\documentclass{jaa}
\usepackage{natbib}
\usepackage{multicol}
\usepackage{contcaption,mathtools}

\usepackage{graphicx}
\usepackage{multirow}
\usepackage{subfigure}
\usepackage{xcolor}
\usepackage{adjustbox}
\usepackage[colorlinks=true,linkcolor=blue,citecolor=blue]{hyperref}

\begin{document}

\title{A new catalogue of head-tail radio galaxies from LoTSS DR1}

\author{Sabyasachi Pal$^{*1}$
Shobha Kumari$^{1}$\\\\
$^{1}$Midnapore City College, Kuturia, Bhadutala, Paschim Medinipur, West Bengal, 721129, India
}
\twocolumn[{
\maketitle
~~~~~~~~~~~~~~~~~\date{Accepted 03 Aug 2022. Revised 19 Jul 2022. In original form 20 May 2022.}\\
\corres{sabya.pal@gmail.com (SP)}\\\\

\begin{abstract}
The unique morphology of head-tail (HT) radio galaxies suggests that radio jets and their intra-cluster medium interact strongly. We conducted a systematic search for HT radio galaxies using the LOFAR Two-metre Sky Survey first data release (LoTSS DR1) at 144 MHz. In this paper, a catalogue of fifty-five new HT radio galaxies is presented, ten of which are narrow-angle tailed sources (NATs) and forty-five of which are wide-angle tailed sources (WATs). NATs are characterised by tails that are bent like a narrow `V' shape with a less than 90 degree opening angle. The opening angle between jets in WAT radio galaxies is greater than ninety degrees, exhibiting wide `C'-like morphologies. We found that thirty out of fifty-five HT radio galaxies reported in this article are associated with known galaxy clusters. Most of the sources presented in the current paper have redshifts $<$ 0.5. The various physical properties and statistical studies of these HT radio galaxies are presented.
\end{abstract}

}]


\section{Introduction}
\label{sec:intro}
The head-tail (HT) morphology of radio galaxies is determined by radio jets that appear to bend in a common direction, resembling a tail, with the luminous host galaxy acting as `head'. This class of radio sources is also widely recognized as {\it C}-shaped radio sources or bent radio sources or Bend-Tail radio sources in general \citep{Ry68, Ru76, Bl00, Pr11, De14} and this type of radio source was first mentioned by \citet{Ry68}. These radio galaxies are classified into two categories based on their degree of bending and luminosity, namely wide-angle tailed (WAT) radio sources and narrow-angle tailed (NAT) radio sources \citep{Ow76}. 3C 465 is known as the prototype for the WAT class radio sources \citep{Ei84, Ei02, Ha05}.

Initially, WATs were thought to be an extension of the HT subclass because of the large opening angles between their tails \citep{Ru76}. WATs now include HT as well as objects with an opening angle greater than $90^{\circ}$ \citep{Ru77, Mi19}. NATs are defined as objects with opening angles of less than $90^{\circ}$ \citep{Mi19}. These types of sources are most commonly found in the cluster of galaxy and thus they can serve as galaxy cluster tracers \citep{Ma10}. In the local Universe (up to $z=1$), \citet{Bl03} detected galaxy clusters by using HT radio galaxies. In addition, \citet{De14} discovered more distant clusters up to $z=2$. WAT sources are thought to be a sub-population of FR-Is \citep{Mi19}. 

According to a recent study on 47 WATs, it is found that these sources are located in the region of the modified Owen-Ledlow diagram where FR-Is are most densely populated \citep{Mis19, Bh21}. The spectroscopic classification indicated that all detected FR-Is were low-excitation radio galaxies (LERGs). However, a multifrequency comparative study of WATs with detected FR-Is and FR-IIs at the same redshifts showed that the multifrequency properties of WATs were remarkably similar to FR-I radio galaxies and had radio powers typical of FR-IIs \citep{Mis19}. Further study is needed to find whether WATs should be regarded as a subset of FR-Is or as a distinct class in their own right.

What governs the formation of these very distinct morphologies is still unknown.
Jets in HT radio galaxies are thought to be caused by two potential mechanisms related to the velocity of the host galaxy in the intra-cluster medium (ICM).
1) The velocity of the host galaxy is unusually higher than expected, resulting in ram pressure on the jets \citep{Mi72, Ru76}, where ram pressure, $P_{ram}$, is defined as
\begin {equation}
P_{ram} \propto \rho v^2
\end {equation}
where $\rho$ is the density of the ICM, and $v$ is the velocity between the galaxy and the ICM \citep{Gu72}. 2) `Cluster weather' is responsible for the formation of these types of radio galaxies \citep{Bu98}.
Dynamical interactions, such as cluster–cluster mergers or group accretion onto clusters, are thought to influence the bending of the jets in the ICM.
It is believed that NATs form as a result of ram pressure on the jets by the host galaxy when the host galaxy acquires a higher velocity (in the order of 600 km s$^{-1}$) than the expected velocity \citep{Ve94} whereas WATs are thought to be formed by `cluster weather' \citep{Kl04}.

NGC 1265 is the most commonly used prototype of a NAT radio source \citep{Ry68, O'D86}. 
Recently, with the help of the Tata Institute of Fundamental Research (TIFR) Giant Metrewave Radio Telescope (GMRT) Sky Survey Alternative Data Release 1 (TGSS ADR 1; \citep{In17}), \citet{Bh21} presented 265 HT radio galaxies (204 are WATs and 61 are NATs). Using the Australia Telescope Large Area Survey of the South Pole Telescope Spitzer Deep Field (ATLAS-SPT), \citet{O'B18} presented a sample of 24 sources with HT morphology. A Multi-frequency study of an interacting narrow-angle tail radio galaxy J0037+18 was done by \citet{Pa19}. With the Australia Telescope Large Area Survey (ATLAS) of the Chandra Deep Field South, \citet{De14} detected 55 HT radio galaxies. With the help of the VLA FIRST survey, \citet{Mis19} and \citet{Sa22} reported 47 and 717 (287 NAT and 430 WAT) HT radio galaxies, respectively.
 \citet{Pi11} catalogued a list of radio sources via automated pattern recognition algorithms using NVSS. From the listed catalogues of \citet{Pi11}, \citet{Yu19} confirmed 412 HT radio galaxies.

In this paper, we report a systematic finding of 55 HT radio galaxies from the high-resolution Low-Frequency Array (LOFAR) Two-metre Sky Survey first data release (LoTSS DR1). In Section \ref{sec:source}, we describe our search methodology, which includes a description of the LoTSS survey, the search strategy, the definition of WATs and NATs, and the respective optical counterpart of each source. In the next section (section \ref{sec:result}), we summarise the result. In the last section (section \ref{sec:discuss}), we made a discussion on the findings.
We have used the following cosmological parameters in this paper from the final full-mission Planck measurements of the cosmic microwave background anisotropies, combining information from the temperature and polarisation maps and the lensing re-construction; $H_0 = 67.4$ km s$^{-1}$ Mpc$^{-1}$, $\Omega_m = 0.315$ and $\Omega_{vac} = 0.685$ \citep{Ag20}. 

\section{Methodology}
\label{sec:source}
\subsection{LoTSS DR1}
LOFAR is currently the largest ground-based radio telescope operating in the low-frequency range of 120--168 MHz. Unlike a single-dish telescope, LOFAR is a multipurpose sensor network, with an innovative computer and network infrastructure that can handle extremely large data volumes. LOFAR Two-metre Sky Survey first data release (LoTSS DR1) covers 424 square degrees of radio sky in the HETDEX spring field region, accounting for about 2 per cent of the total LoTSS coverage \citep{Sh19}. It is an ongoing survey that aims to cover the entire northern sky. 
We used this survey at a frequency of 144 MHz to detect HT radio galaxies in the covered region. The DR1 data release includes right ascension (RA) ranges from 10h45m00s to 15h30m00s and declination (DEC) from 45$^{\circ}$00$^{\prime}$00$^{\prime\prime}$ to 57$^{\circ}$00$^{\prime}$00$^{\prime\prime}$. The survey images are developed by using a fully automated, direction-dependent calibration and imaging pipeline. The survey has a median sensitivity of $S_{144} = 71$ $\mu$Jy beam$^{-1}$ and the point-source completeness of the survey is 90 per cent at an integrated flux density of 0.45 mJy. The angular resolution of the images is 6$^{\prime\prime}$ \citep{Sh19}.

At 1.4 GHz, there are two high-frequency sky surveys: a) the NRAO VLA Sky Survey (NVSS) with a VLA D configuration that covers 82 per cent of the celestial sphere with an rms of $\sim$ 0.45 mJy and an angular resolution of 45$^{\prime\prime}$ \citep{Co98}.
b) Faint Images of the Radio Sky at Twenty-Centimeters (FIRST) with VLA B configuration, has an angular resolution of 5$^{\prime\prime}$ and a typical rms of 0.15 mJy \citep{Be95}.
Due to limited resolution, high-sensitive NVSS is suitable to detect large-scale diffuse emissions but can not reveal finer details in galaxy morphology. The high-resolution FIRST survey, on the other hand, can provide more information about the core but is not good for detecting large-scale diffuse emissions.
LoTSS data with high sensitivity and high resolution provides more information about large-scale diffuse emission as well as detailed information about galaxy morphology. Taking advantage of 50--1000 times better sensitivity and 5--30 times higher resolution in comparison to other low-frequency wide-area surveys (like GLEAM; \citep{Wa15}, TGSS; \citep{In17}, MESS; \citep{He15} and VLSSr; \citep{La14}), LoTSS DR1 could be used to look for a variety of radio sources. Recently, using this survey, \citet{Pa21} listed 33 winged radio galaxies. 459 HT radio galaxies were also reported using this survey by \citet{Mi19}.


\begin{table*}[htb]
\tabularfont
\scriptsize
\begin{footnotesize}
\caption{WAT Radio Sources from LoTSS DR1.}\label{tab:-Head-tail-W} 
\begin{tabular}{cccccccccc}
\topline
Cat. &  Name	& R.A.    & Decl.   & Redshift&$F_{144}$&$F_{1400}$&$\alpha_{1400}^{144}$&   $\log L_{\textrm{rad}}$	   &Others 	\\
Num. &       &(J2000.0)&(J2000.0)&	   ($z$)  & (mJy)    & (mJy)   &		    & (W~Hz$^{-1}$)	     &	Catalog	\\
	&       &         &         &             &          &         &      		    &&	\\
(1)&(2)&(3)&(4)&(5)&(6)&(7)&(8)&(9)&(10)\\\midline
~~W1	& J1049+4619	& 10 49 55.89	& +46 19 20.4	     & 0.35  & ~546	& ~109	& 0.72  & 26.35    &1, 4, 7, 8, 14  \\
~~W2	& J1051+5041 	& 10 51 58.99	& +50 41 59.8  	     & --  		 & ~160	& ~~~19	& 0.95  & --     & 1, 5 \\
~~W3	& J1057+4749	& 10 57 26.97	& +47 49 27.4		 & 0.18  & ~~~96	& ~~~27	&0.57  & 24.94    & 1, 9  \\
~~W4	& J1101+5603	& 11 01 08.50	& +56 03 20.6       & --  		 & ~132	& ~~~17	& 0.92  & --     &  -- \\
~~W5	& J1105+4800	& 11 05 04.47	& +48 00 43.5       & --  		 & ~~~60	& ~~~14	& 0.65  & --     & 1 \\
~~W6	& J1112+5151	& 11 12 21.67	& +51 51 27.7       & --  		 & ~~~49	& ~~~~~8	& 0.81  & --     & 1   \\
~~W7	& J1113+4952	& 11 13 17.60	& +49 52 00.2	    & 0.61 & ~169	& ~~~57	& 0.49  & 26.34  & 1   \\
~~W8	& J1114+4611	& 11 14 16.50	& +46 11 15.1       & -- 		 & ~289	& ~~~26	& 1.08  & --     & 1, 4  \\
~~~W9$^{\ast}$	& J1115+4729	& 11 15 31.64	& +47 29 29.7	   & -- 		 & ~~~63	& ~~~15	& 0.63  & --     & 1  \\
W10	& J1115+4834	& 11 15 41.19	& +48 34 19.4	    & 0.07  		 & ~239	& ~~~64	& 0.59  & 24.47   & 1, 9  \\
W11	& J1120+5308	& 11 20 40.88	& +53 08 09.4	    & -- 		 & ~~~60	& ~~~~~5	& 1.11  & --     & 1 \\ 
W12	& J1121+5421	& 11 21 13.68	& +54 21 34.4	     & 0.21  & ~~~87	& ~~~15	& 0.79  & 25.06    & 1, 9  \\
W13	& J1124+5546	& 11 24 25.08	& +55 46 14.8	     & --  		 & ~~~40	& ~~~~~7	& 0.78  & --     & 1  \\
W14	& J1127+5533	& 11 27 18.40	& +55 33 32.8	     & -- 		 & ~~~22	& ~~~~~5	& 0.66  & --     & 1 \\
W15	& J1147+4917	& 11 47 51.70	& +49 17 31.3	     & --	      & ~~~55	& ~~~~~9	& 0.81  & --     & 1  \\
W16	& ~~J1147+5548$^{+}$	& 11 47 15.68	& +55 48 19.4	   & -- 		 & ~~~41	& ~~~~6	& 0.85  & --     & --  \\
W17	& J1154+4949	& 11 54 12.94	& +49 49 01.0	     & 0.29 & ~~~61	& ~~~~~4	& 1.22  & 25.27    & 1  \\
W18	& J1208+5458	& 12 08 05.93	& +54 58 07.2	     & 0.46  & ~110	& ~~~~~9	& 1.12  & 25.99   & 1 \\
W19	& J1220+5334	& 12 20 50.22	& +53 34 33.8	     & 0.21  & ~239	& ~~~47	& 0.73  & 25.49    & --   \\
W20	& J1224+4905	& 12 24 12.49	& +49 05 51.8	     & --       & ~198	& ~~~~~6	& 1.57  & --     & 1  \\
W21	& J1224+5419	& 12 24 49.89	& +54 19 34.2	     & 0.47  & ~111	& ~~~~~9	& 1.13  & 26.01   & 1  \\  
W22	& J1236+5525	& 12 36 47.43	& +55 25 10.6	     & 0.32 & ~230	& ~~~61	& 0.59  & 25.87   & 1, 4 \\
W23	& J1238+4838	& 12 38 58.26	& +48 38 20.5	     & 0.46 & ~174	& ~~~25	& 0.87 & 26.14   & 1 \\
W24	& J1247+4852	& 12 47 42.73	& +48 52 05.9	     & 0.21  & ~293	& ~~~15	& 1.33  & 25.63   & 3, 16 \\
W25	& J1249+4954	& 12 49 39.23	& +49 54 03.1	     & -- 		 & ~119	& ~~~26	& 0.68  & --      & 1   \\
W26	& J1303+4743	& 13 03 48.51	& +47 43 19.4	     & -- 		 & ~226	& ~~~42	& 0.75  & --      & 1, 10 \\
W27	& J1303+4935	& 13 03 27.70	& +49 35 56.5	     & 0.25  & ~234	& ~~~51	& 0.68  & 25.65    & 1 \\
W28	& J1303+5220	& 13 03 32.48	& +52 20 10.3	     & -- 		 & ~108	& ~~~24	& 0.67  & --      & 10, 11 \\
W29	& J1306+5144	& 13 06 10.82	& +51 44 22.6	     & 0.28  & ~652	& ~180	& 0.58  & 26.19    & 1, 2, 8 \\
W30	& J1315+4841	& 13 15 31.82	& +48 41 11.7	     & -- 		 & ~455	& ~~~75	& 0.81  & --      & 1, 4 \\
W31	& J1325+5617	& 13 25 04.38	& +56 17 42.3  	     & -- 		 & ~107	& ~~~43	& 0.41  & --      & 1 \\
W32	& J1330+4730	& 13 30 32.67	& +47 30 55.6	     & 0.33 & ~~~98	& ~~~32	& 0.50  & 25.54    & 12 \\
~~W33$^{\ast}$	& J1344+5552	& 13 44 44.58	& +55 52 03.3 	   & -- 		 & ~543	& ~166	& 0.52  & --      & --  \\
W34	& ~~J1346+5250$^{+}$	& 13 46 06.92 	& +52 50 17.6    & -- 		 & ~237	& ~~~63	& 0.58  & --      & 1, 17 \\
W35	& J1351+5216	& 13 51 50.83	& +52 16 29.7	     & 0.17 & ~~~87	& ~~~14	& 0.82  & 24.89     & 1      \\
W36	& J1359+4905	& 13 59 44.53	& +49 05 33.7		 & -- 		 & ~124	& ~~~23	& 0.75  & --      & 1   \\
W37	& J1426+5547	& 14 26 06.46	& +55 47 26.5		 & -- 		 & ~~~89	& ~~~15	& 0.80  & --      & 1 \\
W38	& J1428+5520	& 14 28 42.42	& +55 20 51.5	     & 0.29  & ~210	& ~~~51	& 0.63  & 25.38    & 12  \\
~~W39$^{\ast}$	& J1431+4743	& 14 31 50.54	& +47 43 57.1	      & -- 		 & ~255	& ~~~64	& 0.61  & --      & 1 \\
W40	& J1434+4951	& 14 34 41.95	& +49 51 09.7	     & 0.20 & ~221	& ~~~49	& 0.67  & 25.38     &	--  \\
W41	& J1439+5631	& 14 39 19.32	& +56 31 39.4	     & -- 		 & ~121	& ~~~31	& 0.61  & --      & 12, 15, 16  \\
W42	& J1446+4841	& 14 46 06.73	& +48 41 40.0	     & 0.38 & ~205	& ~~~27	& 0.91  & 26.03    & 1  \\
W43	& J1459+4947	& 14 59 42.61	& +49 47 17.5		 & 0.17  & ~630	& ~184	& 0.55  & 25.70    & 1, 2  \\
W44	& J1459+5334	& 14 59 35.96	& +53 34 19.8	     & 0.08  & ~262	& ~~~13	& 1.35  & 24.66     & 1  \\
~W45$^{\ast}$	& J1523+5255	& 15 23 27.06	& +52 55 07.4	       & -- 		 & ~234	& ~~~50	& 0.68  & --      & 1, 16 \\
\hline
\end{tabular}
\tablenotes{1. NVSS \citep{Co98}; 2. VLSS \citep{Co07}; 3. 5C \citep{Ke66, Po68, Po69, Wi70, Pe75, Wa77, Pe78, Sc81, Be82, Be88}; 4. 6C \citep{Ba85}, \citep{Ha88, Ha90, Ha91, Ha93a, Ha93b}; 5. 7C \citep{Mc90, Ko94, Wa96, Ve98}; 6. PKS \citep{Bo64}; 7. TXS \citep{Do96}; 8. 87GB \citep{Gr91}; 9. ASK \citep{Sa11}; 10. 2MASX \citep{Sk06}; 11. GALEXASC \citep{Ag05}; 12. GALEXMSC \citep{Ag05}; 13. NGC \citep{Dr88}; 14. B3 \citep{Fi85}; 15. WISE \citep{Ch11, Re11}; 16. 2MASS \citep{Sk06}}
NOTE: `$\ast$' Sources with no optical counterparts\\
`$+$' possible radio relic sources
\end{footnotesize}
\end{table*}

\begin{table*}[htb]
\tabularfont
\scriptsize
\begin{footnotesize}
\caption{NAT Radio Sources from LoTSS DR1.}\label{tab:-Head-tail-N} 
\begin{tabular}{cccccccccc}
\topline
Cat. &  Name	& R.A.    & Decl.   & Redshift&$F_{144}$&$F_{1400}$&$\alpha_{1400}^{144}$&   $\log L_{\textrm{rad}}$	   &Others 	\\
Num. &       &(J2000.0)&(J2000.0)&	   ($z$)  & (mJy)    & (mJy)   &		    & (W~Hz$^{-1}$)	     &	Catalog	\\
	&       &         &         &             &          &         &      		    &&	\\
(1)&(2)&(3)&(4)&(5)&(6)&(7)&(8)&(9)&(10)\\\midline
~N1	& J1132+5459	& 11 32 01.98	& +54 59 51.9	     & 0.55 & ~~~51	& ~~~~9	& 0.78  & 25.77    & 1   \\
~N2	& J1147+4805	& 11 47 48.43	& +48 05 49.1	     & 0.70  & ~111	& ~~34	& 0.53  & 26.30  & 1  \\
~N3	& J1207+4805	& 12 07 19.42	& +48 05 58.7	     & 0.35 & ~488	& ~~84	& 0.79  & 26.31    & 1, 2  \\
~~N4$^{\ast}$	& J1325+5544	& 13 25 48.42	& +55 44 14.7       & --		 & ~231	& ~~35	& 0.83  & --      & 1, 16  \\
~~N5$^{\ast}$	& J1329+5246	& 13 29 09.51 	& +52 46 19.9  	   & -- 		 & ~135	& ~~19	& 0.86  & --      & 1  \\ 
N6	& ~~J1347+5022$^{+}$	& 13 47 01.03	& +50 22 17.9 & -- 		 & ~107	& ~~13	& 0.93  & --      & 1, 17  \\
~~N7$^{\ast}$	& J1431+5518	& 14 31 16.33	& +55 18 49.9 	   & -- 		 & ~~~66	& ~~~~8	& 0.93  & --      & 1  \\
~N8	& J1505+4706	& 15 05 38.47	& +47 06 22.4		 & 0.27  & ~264	& ~163	& 0.22  & 25.73   & 1, 16 \\
~N9	& J1506+5311	& 15 06 03.10	& +53 11 09.3	     & 0.14 & ~299	& ~~~66	& 0.68  & 25.21     & 1   \\
~~N10	& ~~J1510+5146$^{+}$	& 15 10 03.48	& +51 46 28.1	   & -- 		 & ~198	& ~~~57	& 0.55  & --      & 1, 14   \\
\hline
\end{tabular}
\tablenotes{1. NVSS \citep{Co98}; 2. VLSS \citep{Co07}; 3. 5C \citep{Ke66, Po68, Po69, Wi70, Pe75, Wa77, Pe78, Sc81, Be82, Be88}; 4. 6C \citep{Ba85}, \citep{Ha88, Ha90, Ha91, Ha93a, Ha93b}; 5. 7C \citep{Mc90, Ko94, Wa96, Ve98}; 6. PKS \citep{Bo64}; 7. TXS \citep{Do96}; 8. 87GB \citep{Gr91}; 9. ASK \citep{Sa11}; 10. 2MASX \citep{Sk06}; 11. GALEXASC \citep{Ag05}; 12. GALEXMSC \citep{Ag05}; 13. NGC \citep{Dr88}; 14. B3 \citep{Fi85}; 15. WISE \citep{Ch11, Re11}; 16. 2MASS \citep{Sk06}}
NOTE: `$\ast$' Sources with no optical counterparts\\
`$+$' possible radio relic sources
\end{footnotesize}
\end{table*}

\subsection{Search strategy}
\label{subsec:search strategy}
LoTSS DR1 includes a total of 325,694 sources with a 5$\sigma$ detection limit. The source density in LoTSS is a factor of 10 times higher than the most sensitive existing very wide-area radio-continuum surveys \citep{Sh19}. 
Using this survey, we aimed to find HT radio galaxies. To identify a list of HT radio galaxies, we filtered the sources in the catalogue with angular sizes $\ge12^{\prime\prime}$ (i.e. at least twice the convolution beam size), which yielded 18,500 sources. We visually inspected the fields of all of these 18500 sources in search of a new finding for HT radio galaxies. We categorized the list of sources on the basis of bending angle. We removed sources with larger bending angles ($>$160$^{\circ}$) and listed only those sources whose bending angles are $\leq160^{\circ}$. 

We cross-matched the list of sources presented in this paper with previously reported HT radio galaxies by \citet{Mi19, Sa22}. From our catalogue, we removed all the HT radio galaxies that were previously reported by \citet{Mi19} and we also removed one HT radio galaxy (J1242+5021), as this source is already listed by \citet{Sa22}. Out of fifty-five listed sources in this current paper, \citet{Mi19} catalogued thirty-six sources without classifying them as HT radio galaxies.

\subsection{Optical counterparts}
\begin{figure*}
\vbox{
\centerline{
\includegraphics[height=5.6cm,width=5.6cm]{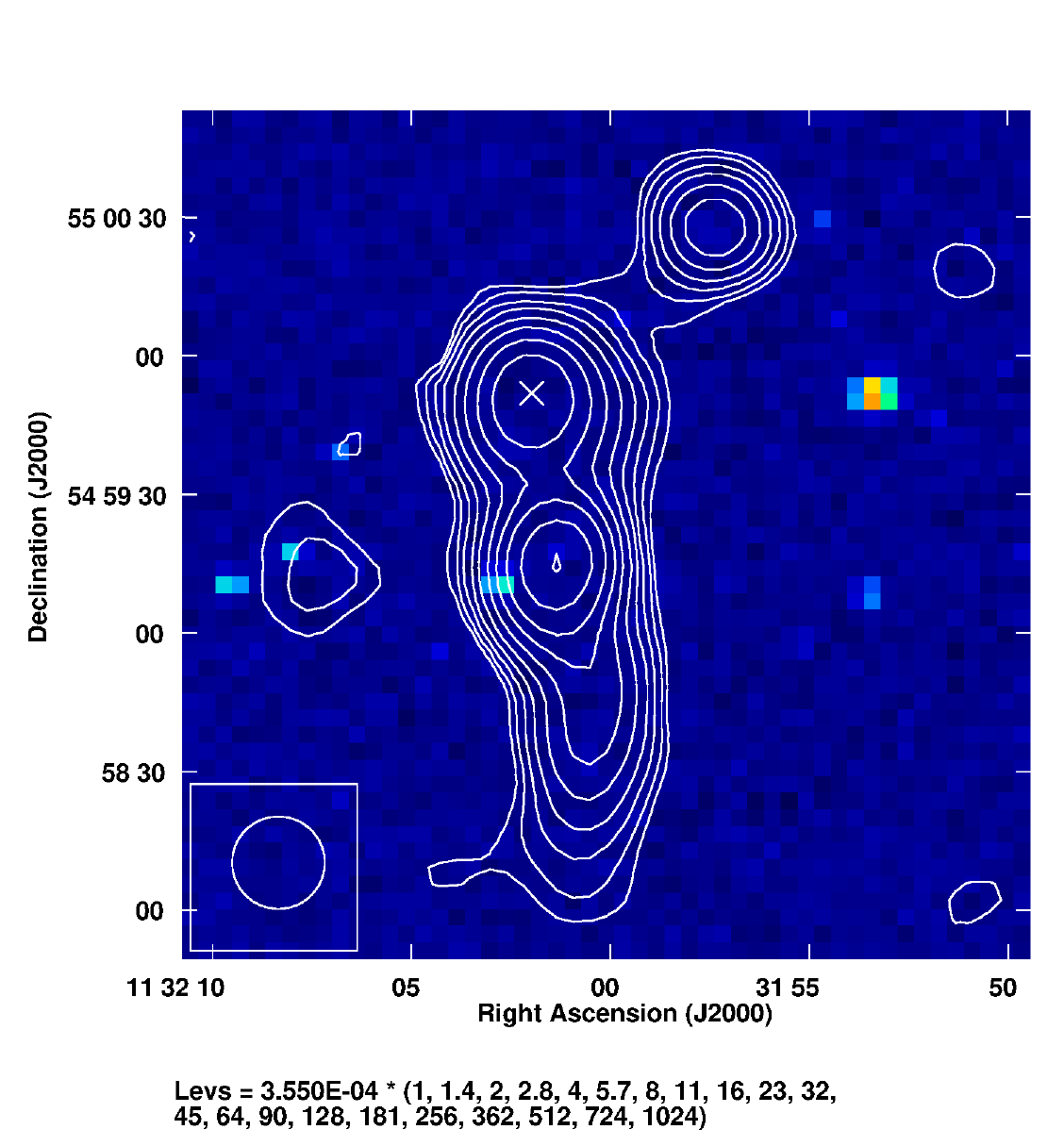}
\includegraphics[height=5.6cm,width=5.6cm]{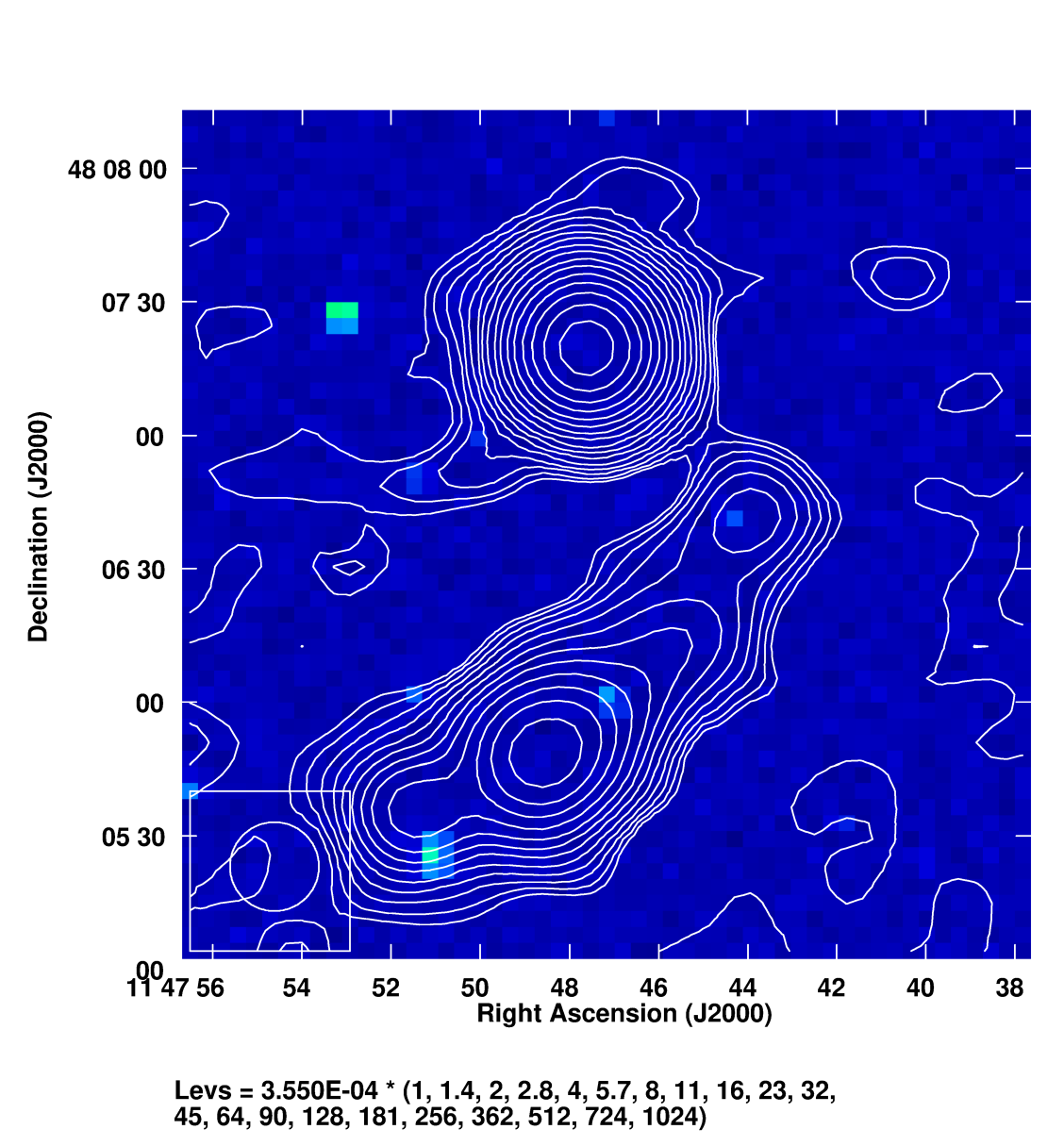}
\includegraphics[height=5.6cm,width=5.6cm]{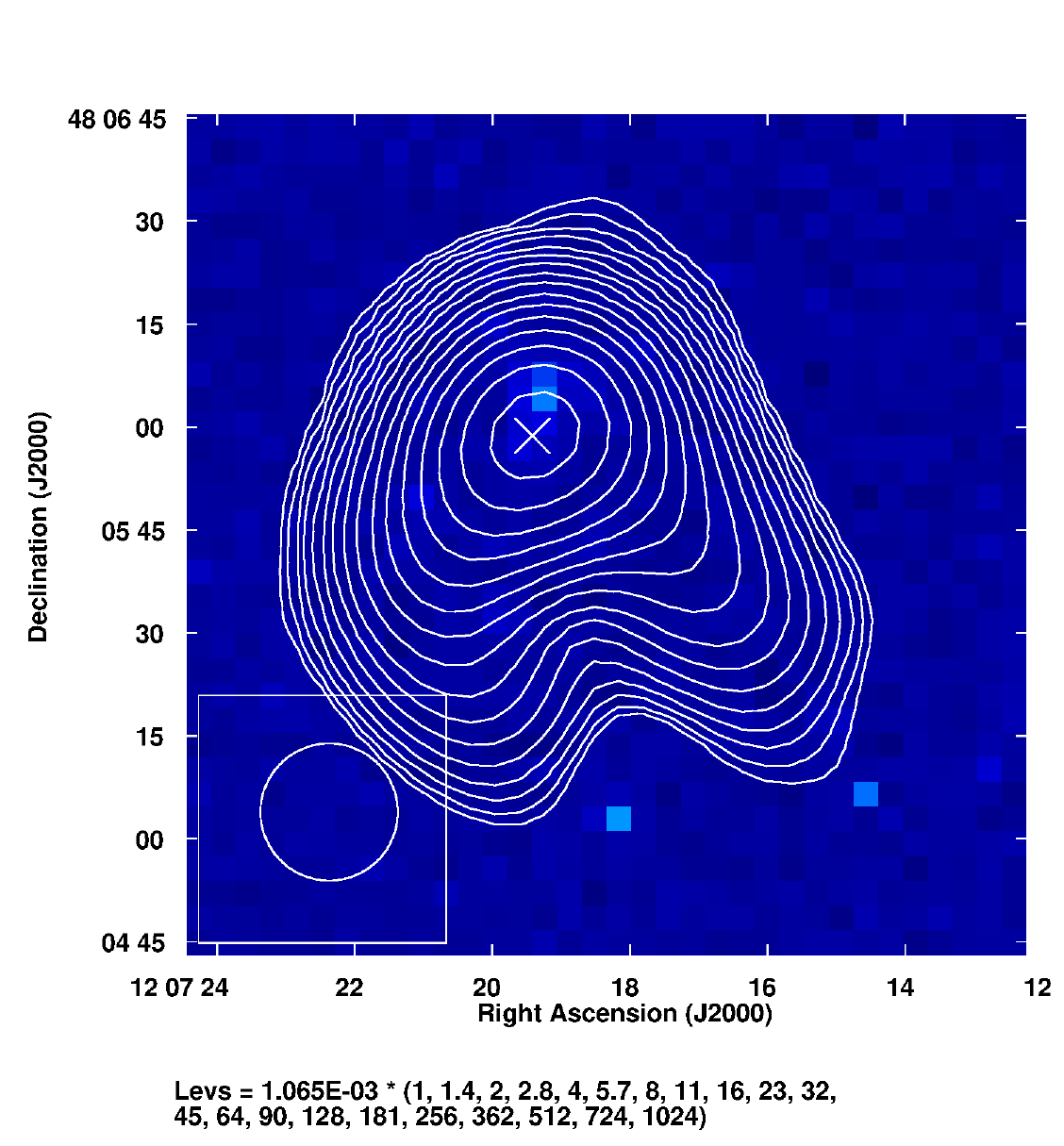}
}
}
\vbox{
\centerline{
\includegraphics[height=5.6cm,width=5.6cm]{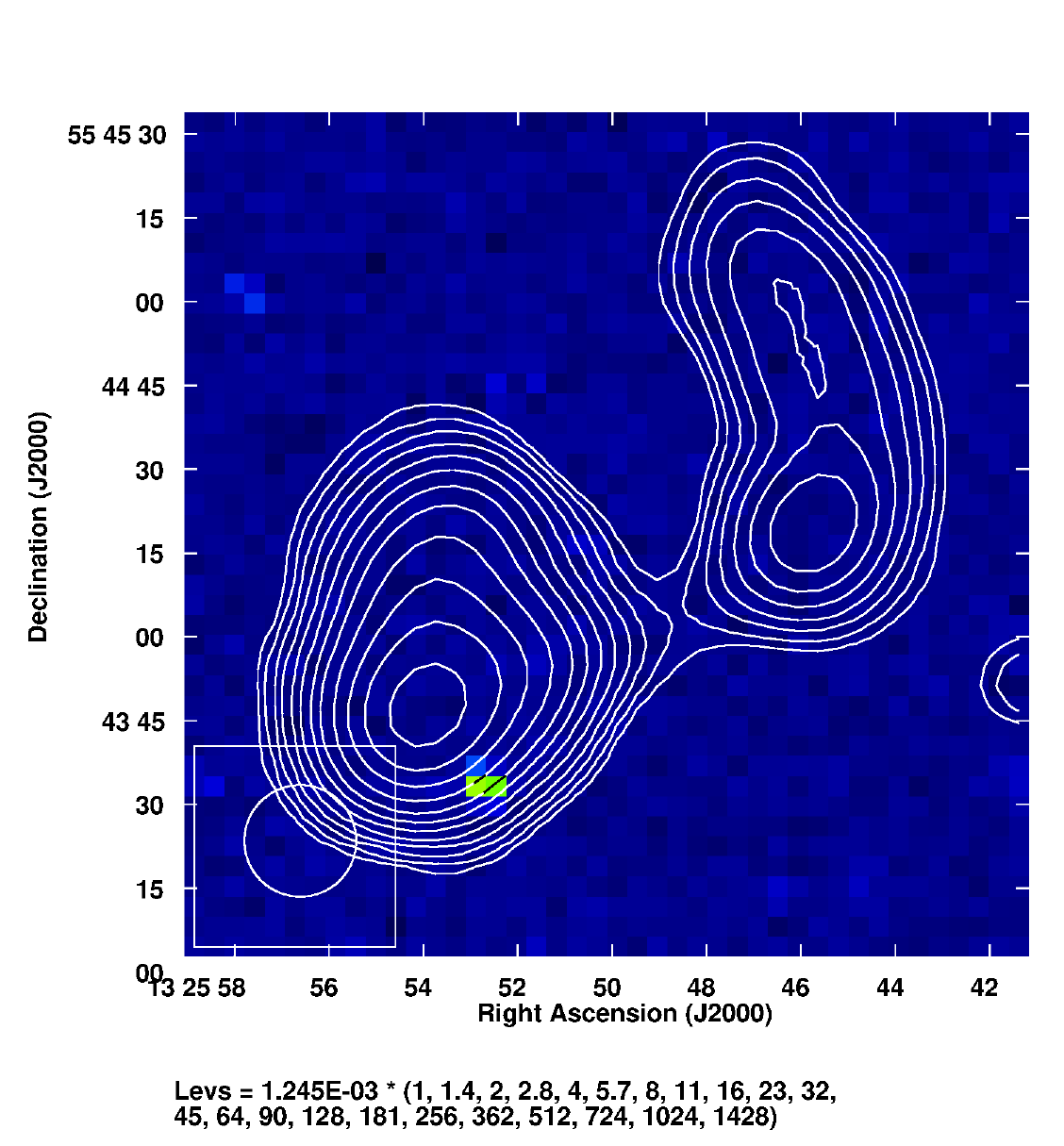}
\includegraphics[height=5.6cm,width=5.6cm]{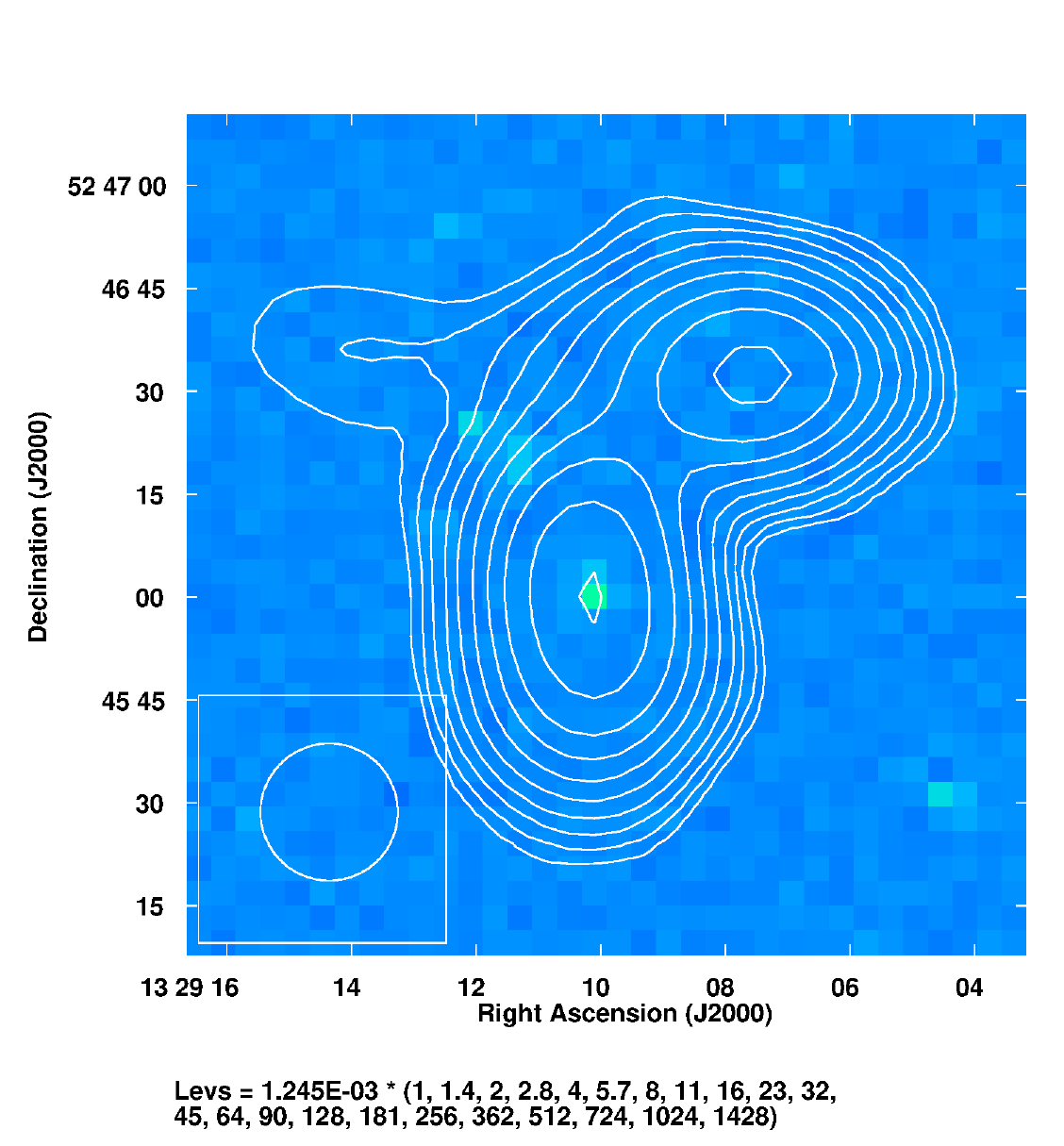}
\includegraphics[height=5.6cm,width=5.6cm]{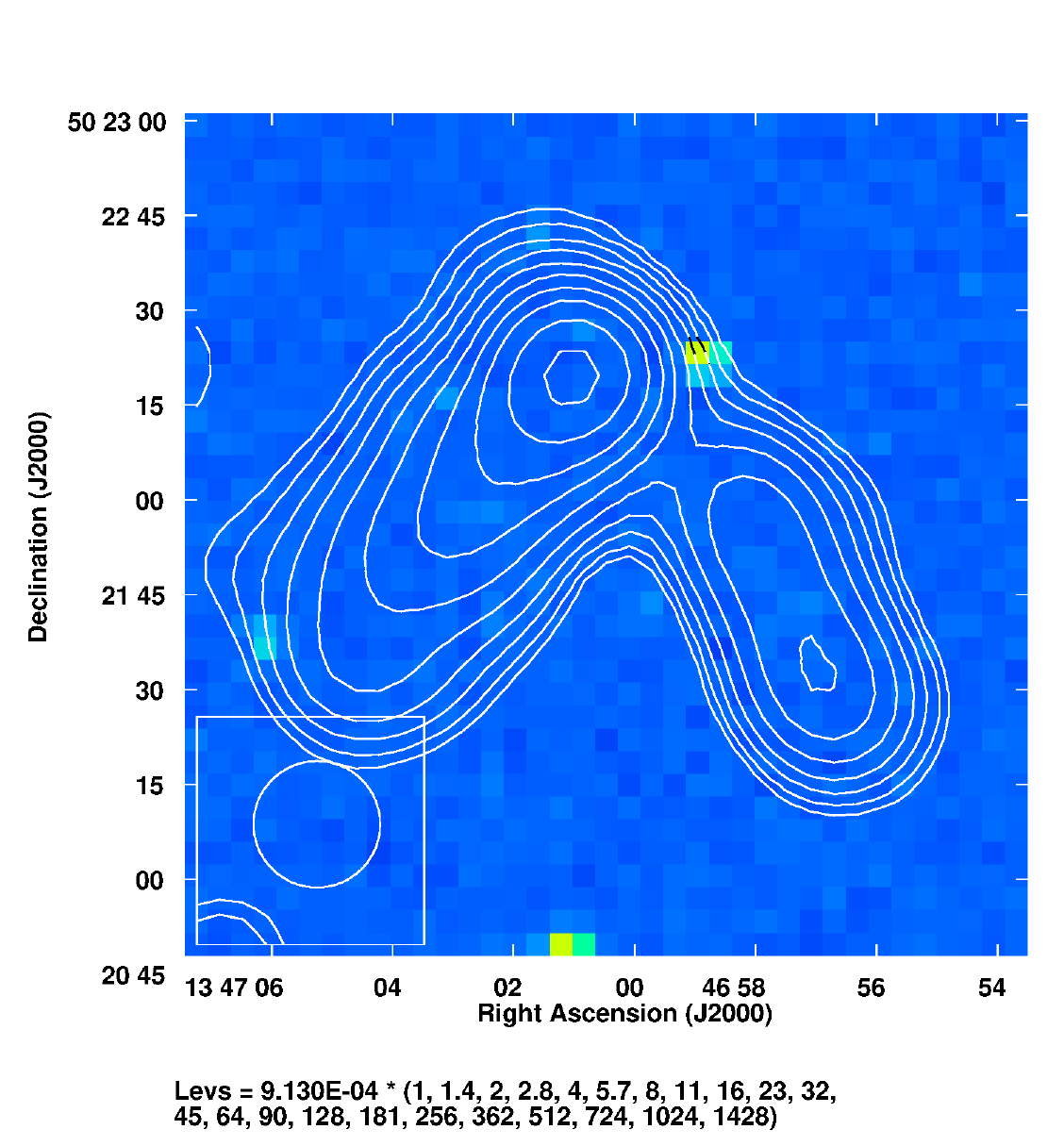}
}
}
\vbox{
\centerline{
\includegraphics[height=5.6cm,width=5.6cm]{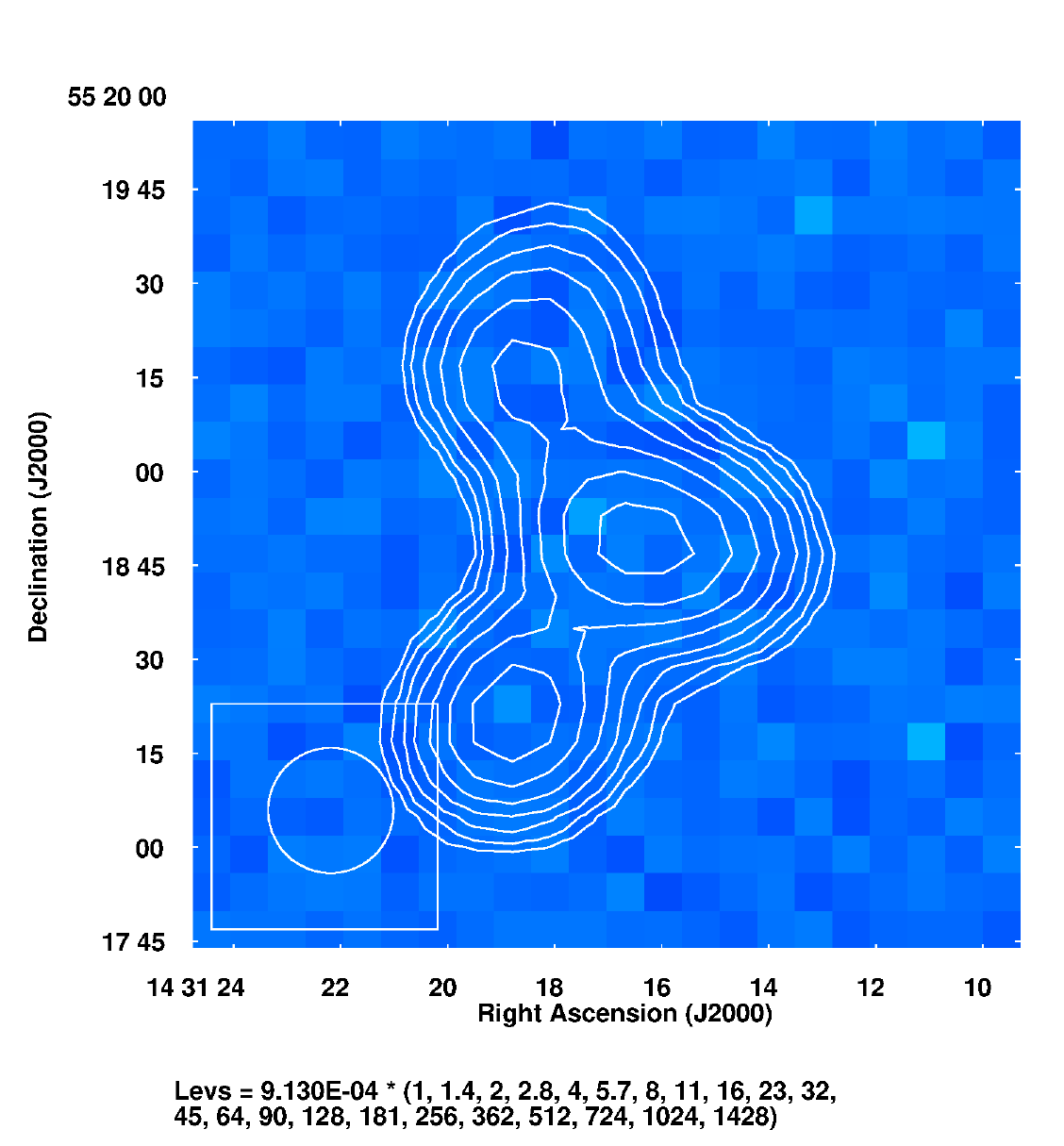}
\includegraphics[height=5.6cm,width=5.6cm]{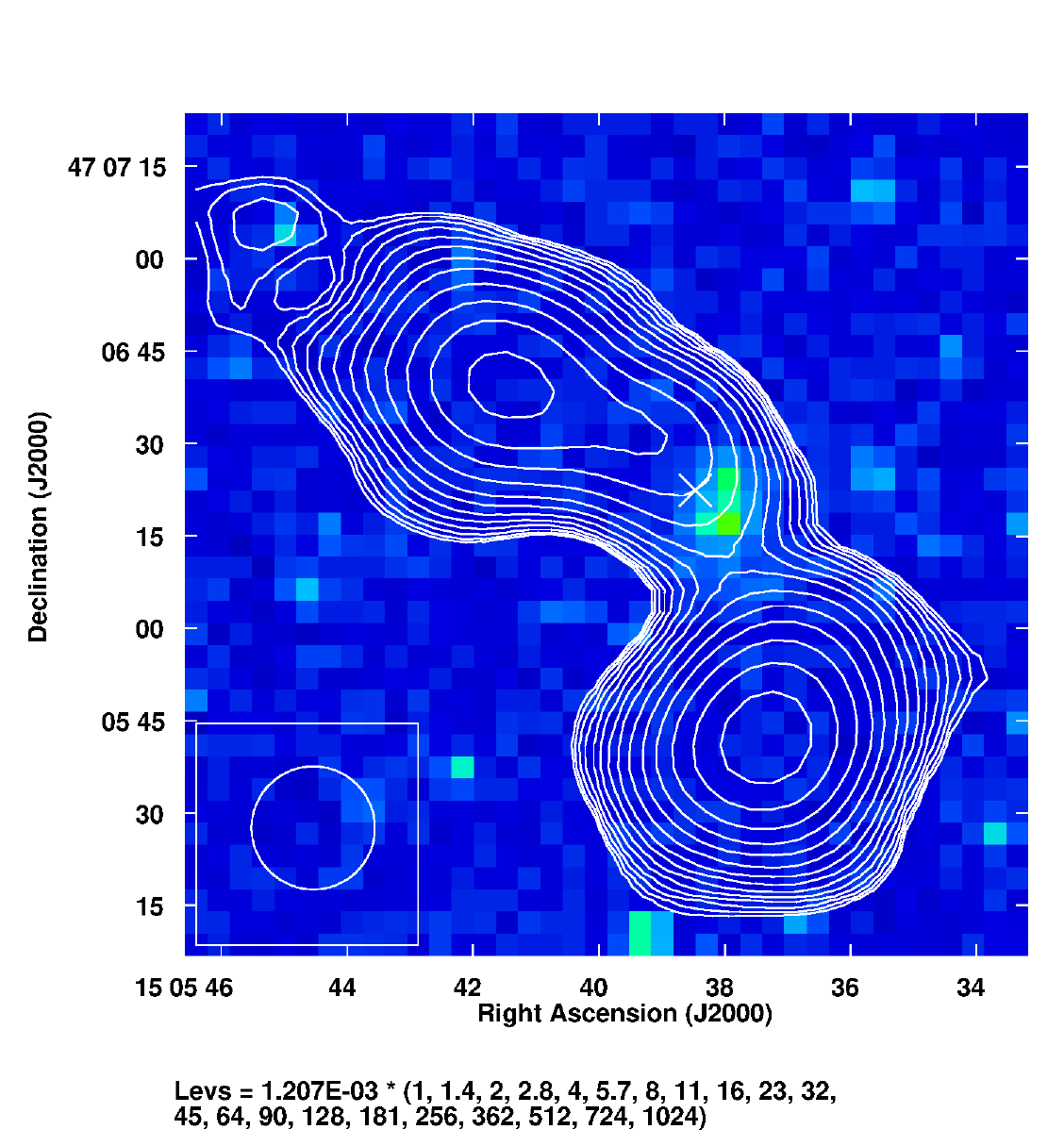}
\includegraphics[height=5.6cm,width=5.6cm]{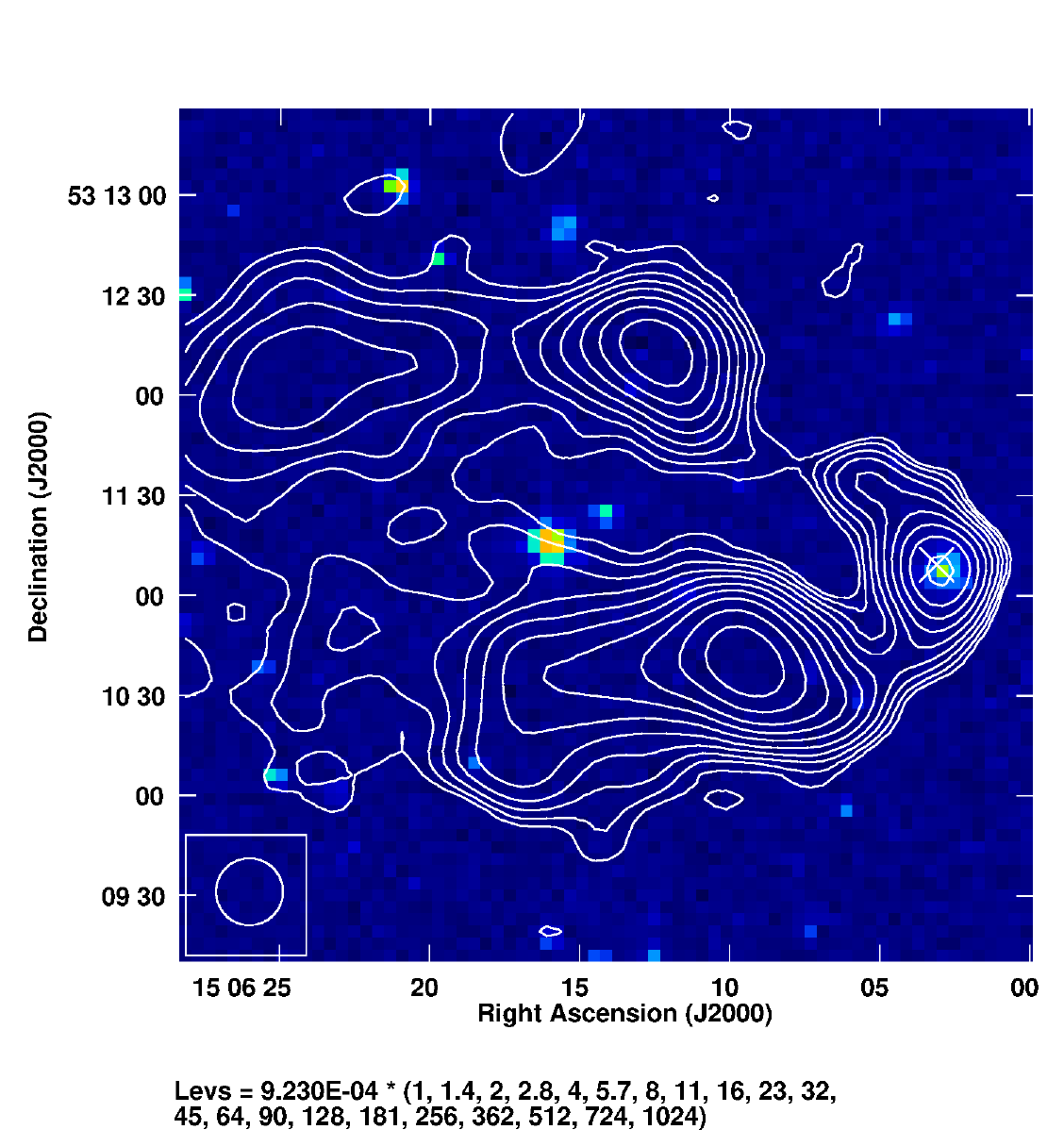}
}
}
\vbox{
\centerline{
\includegraphics[height=5.6cm,width=5.6cm]{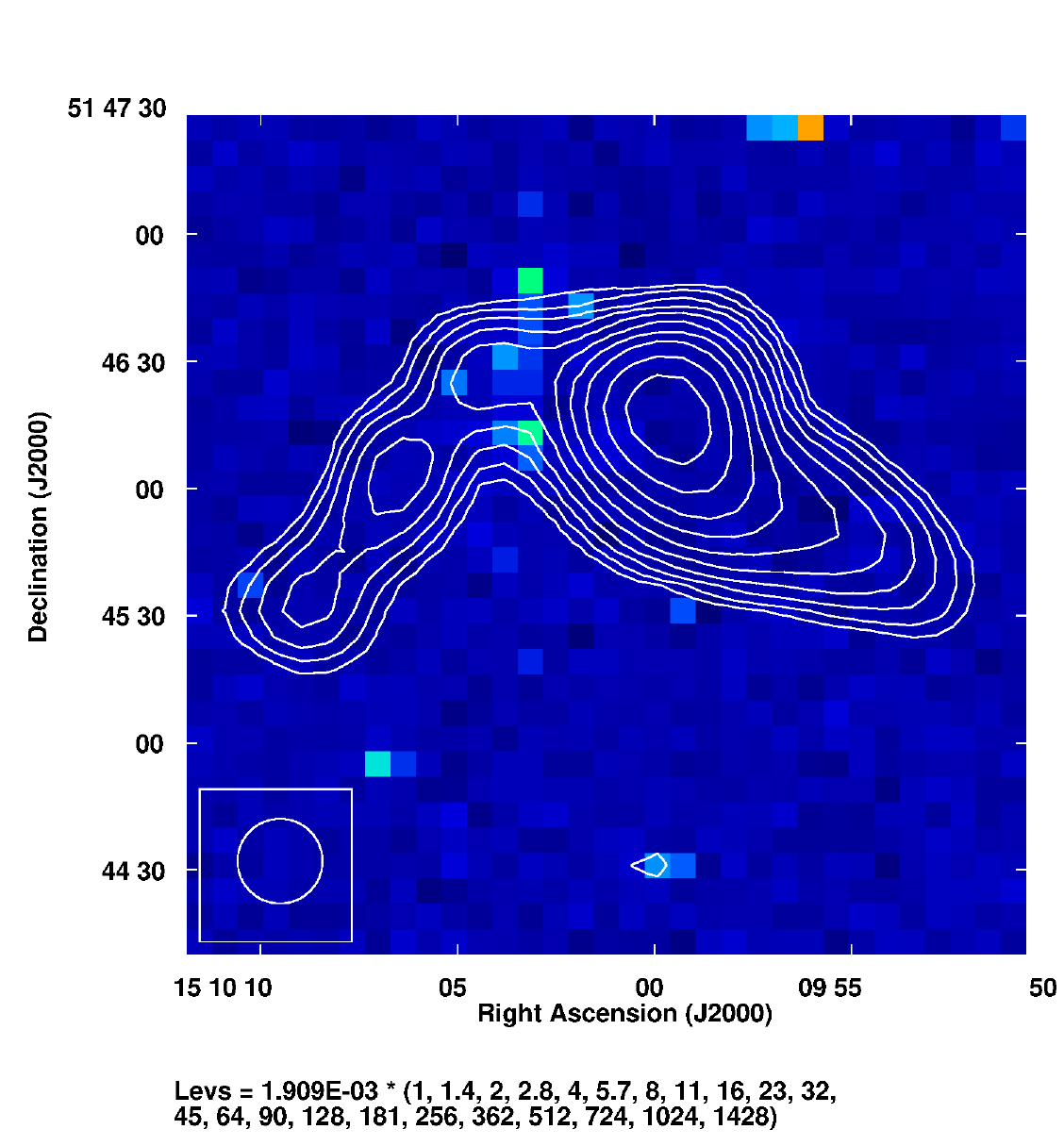}
}
}
\caption{LOFAR images of NAT radio galaxies (contours) overlaid on DSS2 red images (colour). Here white cross mark represent the optical counterpart for the sources, when available.}
\label{fig:HT-NAT}
\end{figure*}

In the current paper, we searched for optical counterparts for all the reported HT galaxies using the Sloan Digital Sky Survey (SDSS) data catalogue \citep{Gu06}. The optical counterparts of these galaxies are marked with $\times$ in Figs. \ref{fig:HT} and \ref{fig:HT-NAT}. Out of 55 newly discovered HT radio galaxies, optical counterparts are found for 44 HT radio galaxies. For some HT radio galaxies, their optical counterparts are not detected. In those cases, we used our best-guess eye estimated (EE) position as the position of HT radio galaxies. Sources with no optical counterparts are marked with $\ast$ in the table \ref{tab:-Head-tail-W} and \ref{tab:-Head-tail-N}.

\subsection{Definition of the NAT and WAT sources}
\label{subsec:definition}
We identified HT radio galaxies by manually inspecting all extended sources ($\ge12^{\prime\prime}$) in the LoTSS DR1 survey. HT radio galaxies exhibit significant jet bending apart from the typical linear trajectory.
The HT radio galaxies are classified into two sub-classes based on the angle of bending between the two radio jets: wide angle tail (WAT) and narrow-angle tail (NAT).
WATs are defined as sources with bending angle $\ge90^{\circ}$ while sources with an angle of less than $90^{\circ}$ are termed NAT sources. The bending angle is measured by drawing two straight lines parallel to the two opposite sides of radio jets emanating from the core of the galaxy (optical counterpart). 
It should be noted that a particular WAT source may appear as a NAT source on our radio map due to the projection effect or the orientation to the plane of the sky.

\vspace{2em}

\begin{table*}[htb]
\tabularfont
\scriptsize
\begin{footnotesize}
\caption{LoTSS WATs in galaxy clusters.}\label{tab:-Head-tail2W} 
\begin{tabular}{ccccccccccc}
\topline
Source&Source&Cluster Name&Redshift&Comoving&Angular&$D_{cl}$& $r_{500}$&$R_{L}$&$M_{500}$&$N_{500}$ \\
Id&Name&    &($z$)	 & Dist.   & Sep.  &        &        &          &        &               \\ 
&   &    &   & (Mpc)   &(arcsec) &(kpc)  & (Mpc)  &          &($\times 10^{14}M_\odot$) &     	\\
	&       &         &         &             &          &         &      		    &&	\\
(1)&(2)&(3)&(4)&(5)&(6)&(7)&(8)&(9)&(10)&(11)\\\midline
~1	& J1049+4619	&WHL J104955.4+461912		&0.3508		&1426.1 	&~~10.38	&~~48.57	&0.73  &30.58 &1.71  &13  \\
~2	& J1057+4749	&MSPM 05580			&0.0873	&~~378.9		&156.24	&~~48.61  & --  & --    & --  & --  \\
~3	& J1112+5151	&WHL J111224.7+515414		&0.5082		&1978.8		&167.64	&936.80	&0.95  &69.43 &4.16  &29 \\
~4	& J1115+4834	&SDSS-C4-DR3 3408		&0.0740		&~~323.3		&~~~~2.70	&~~~~3.84	&--&--&--&--\\
~5	& J1121+5421	&WHL J112113.6+542133		&0.2122		&~~895.1 	&~~~~1.32	&~~~~4.44	&0.68  &20.68	&1.12   &11  \\
~6	& J1147+5548	&WHL J114703.7+554715		&0.5650		&1916.4		&~~~~2.00	&712.27	&0.63  &26.71  &1.48  &~~9 \\
~7	& J1154+4949	&WHL J115412.9+494858 		&0.2873		&1188.1		&~~~~4.02	&~~16.65	&0.67  &21.80  &1.19 &10 \\
~8	& J1220+5334	&WHL J122051.2+533438 		&0.2076		& ~~876.8		&~~~~8.94	&~~29.64	&0.81  &28.83  &1.61 &18 \\
~9	& J1224+4905	&WHL J122418.6+490550 		&0.1012		& ~~439.2		&~~59.88	&112.15	&0.64  &16.49 &0.88   &~~9  \\
10	& J1224+5419	&WHL J122450.7+541929		&0.4672		&1839.9 		&~~~~6.06	&~~33.04	&0.98  &71.72  &4.31 &29 \\
11	& J1236+5525	&WHL J123647.3+552511 		&0.3231		&1323.4	&~~~~1.26	&~~~~5.62	&0.80  &32.86 &1.85  &16  \\
12	& J1247+4852	&WHL J124743.2+485156   	&0.2095		&~~884.4		&~~10.74	&~~35.85	&0.94  &55.86  &3.29 &37  \\
13	& J1249+4954	&NSC J124941+495335		&~~0.2542$^{\star}$	&1060.5		&~~34.92	&133.44	&--&--&--&--\\
14	& J1303+4743	&GMBCG J195.99926+47.72834	&~~0.3770$^{\star}$	&1521.8		&116.34	&567.24	&-- & --  & --    & -- \\
15	& J1303+4935	&WHL J130327.7+493558  		&0.2547		&1062.5		&~~~~1.44	&~~~~5.51	&0.66  &17.22 &0.92   &~~8  \\
16	& J1306+5144	&WHL J130612.2+514407		&0.2773		&1149.8	&~~20.22	&~~81.85	&0.98  &53.93 &3.16  &28 \\
17	& J1315+4841	&WHL J131527.6+484025 		&0.5120		&1991.5		&~~62.64	&256.38	&1.03  &81.90  &4.97 &39  \\
18	& J1330+4730	&WHL J133032.6+473053  		&0.3261		& 1334.6		&~~~~2.70	&~~12.10	&0.72  &22.42  &1.23 &11  \\
19	& J1351+5216	&WHL J135156.1+521542		&0.1653		&~~705.8		&~~67.92	&189.85	&0.66  &20.34  &1.10  &~~9 \\
20	& J1359+4905	&WHL J135949.8+490516		&~~0.1728$^{\star}$	&~~736.4		&~~54.24	&156.87	&--&--&--&--\\
21	& J1346+5250	&MaxBCG J206.50836+52.82084	&~~0.1458$^{\star}$	&~~625.7		&~~76.80	&194.51 & --  & --    & --  & -- \\			
22	& J1428+5520	&WHL J142843.4+552045		&0.2893		&1195.7		&~~11.22	&~~46.68	&0.66  &19.76 &1.07  &10  \\
23	& J1434+4951	&GMBCG J218.67368+49.85134 	&~~0.2040$^{\star}$	& ~~862.4		&~~~~5.52	&~~18.07	&--  & --  & --    & --   \\
24	& J1459+4947	&WHL J145943.2+494716 		&0.1705		&~~727.0		&~~~~6.36	&~~18.20	&1.12  &83.76  &5.09 &50 \\
25	& J1459+5334	&SDSS-C4-DR3 3613		&0.0750		&~~327.6		&~~~~4.14	&~~~~5.97	&--  & --  & --    & --  \\
\hline
\end{tabular}
\end{footnotesize}\\
NOTE: `$\star$' the photometric redshifts\\

\end{table*}

\begin{table*}[htb]
\tabularfont
\scriptsize
\begin{footnotesize}
\caption{LoTSS NATs in galaxy cluster.}\label{tab:-Head-tail2N} 
\begin{tabular}{ccccccccccc}
\topline
Source&Source&Cluster Name&Redshift&Comoving&Angular&$D_{cl}$& $r_{500}$&$R_{L}$&$M_{500}$&$N_{500}$ \\
Id&Name&    &($z$)	 & Dist.   & Sep.  &        &        &          &        &               \\ 
&   &    &   & (Mpc)   &(arcsec) &(kpc)  & (Mpc)  &          &($\times 10^{14}M_\odot$) &     	\\
	&       &         &         &             &          &         &      		    &&	\\
(1)&(2)&(3)&(4)&(5)&(6)&(7)&(8)&(9)&(10)&(11)\\\midline
~1& J1207+4805	&WHL J120719.4+480605 		&0.3532		&1435.0		&~~~6.18	&~~29.03	&0.68  &16.96  &0.91  &~~8 \\
~2	& J1347+5022	&WHL J134708.2+502222		&0.5746		& 2196.7		&~68.70	&410.59	&0.70  &33.36  &1.88 &10 \\			
~3	& J1505+4706	&WHL J150538.1+470617 		&0.2648		&1101.6		&~~~6.48	&~~27.35	&1.01  &74.29  &4.47 &43 \\
~4	& J1506+5311	&WHL J150602.9+531108 		&0.1414		&~~607.5		&~~~2.64	&~~~~6.82	&0.81  &32.89  &1.85 &20 \\
~5& J1510+5146	&WHL J151003.4+514612		&0.2073		&~~875.6 		&~16.56	&~~54.85	&0.84  &30.94  &1.74 &10 \\	
\hline
\end{tabular}

\end{footnotesize}
\end{table*}

\section{Result}
\label{sec:result}
\subsection{Source catalogue}
\label{subsec:source catalogue}
We discovered fifty-five new HT radio galaxies using LoTSS DR1, among which 45 are WAT sources and 10 are NAT sources. In Tables \ref{tab:-Head-tail-W} and \ref{tab:-Head-tail-N}, we tabulated a list of 45 WATs and 10 NATs. Columns (3) and (4) showed the Right Ascension (J2000.0) and Declination (J2000.0) of HT radio galaxies. We used the best location of the optical host galaxy from SDSS as the location of the HT source when the optical host galaxy is available. 

 The redshift of the host galaxy is mentioned in column (5), when available. In column (6), we measured flux density at 144 MHz in mJy ($F_{144}$) using LoTSS DR1. The flux density at 1400 MHz, measured by NVSS, is mentioned in column (7). The spectral indices between 144 MHz and 1400 MHz are mentioned in column (8) (discussed in detail in Section \ref{subsec:specind}) and the radio luminosity in column (9) (discussed in detail in Section \ref{subsec:lum}).

\subsection{Angle between two jets of HT radio galaxies}
\label{subsec:angle}
\begin{figure}
	\centering
\includegraphics[width=8.5cm,angle=0,origin=c]{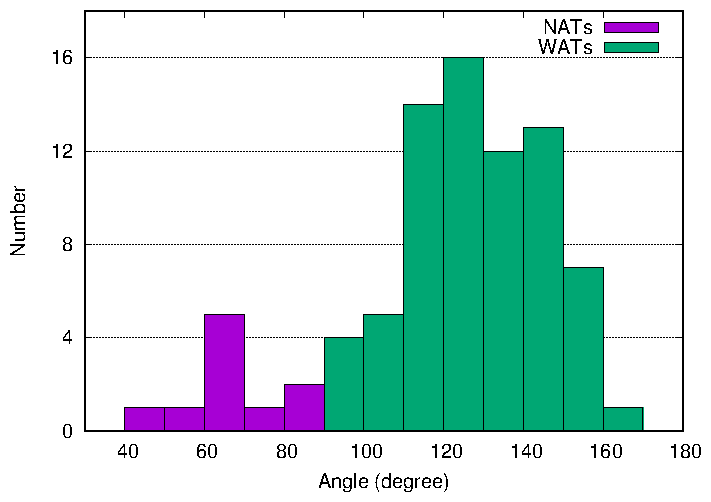}
\caption{A histogram showing the distibution of bending angle of WATs and NATs presented in the current paper.}
\label{fig:angle}
\end{figure}

The opening angle of jets is measured for the WAT and NAT sources presented in the current paper. The opening angle is measured using the peak points in both the lobes and the core of the galaxy. 
The extreme bending NAT radio galaxy is J1506+5311 with a bending angle of $40^{\circ}$ and the lowest bending WAT radio galaxy is J1523+5255 with a bending angle of $160^{\circ}$. In Fig. \ref{fig:angle}, a histogram with the opening angle distribution of WAT and NAT sources is shown. 
The histogram showed a peak between the angles of $120^{\circ}$ to $130^{\circ}$. Because of the complex structure of some of the sources, the bending angle could not be measured. It should be noted that sources with very small bending are not catalogued, so there might be a possibility of bias in this angle distribution diagram.

\subsection{Spectral index}
\label{subsec:specind}
\begin{figure}
	\centering
\includegraphics[width=8.5cm,angle=0]{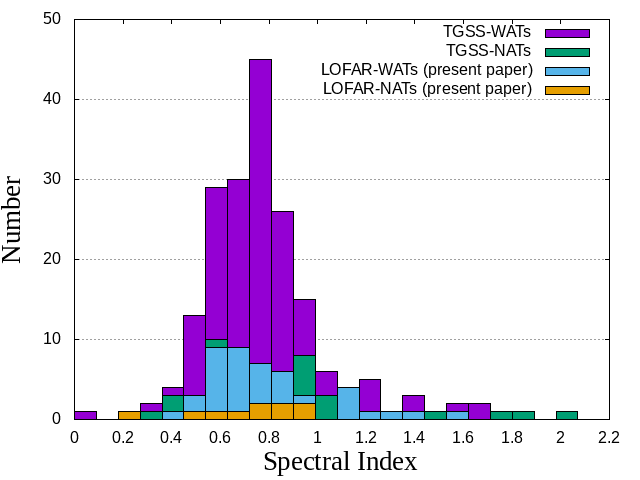}
\caption{Histogram showing the spectral index ($\alpha_{144}^{1400}$) distribution of WAT and NAT radio galaxies presented in the current paper.}
\label{fig:spindex_histgrm}
\end{figure}

The two-point spectral index ($\alpha_{1400}^{144}$) of newly discovered HT radio galaxies is measured using NVSS (at 1400 MHz) and LoTSS (at 144 MHz), under the assumption of $S \propto \nu^{-\alpha}$, where $\alpha$ is the spectral index and $S$ is the radiative flux density at a given frequency ($\nu$). In Fig. \ref{fig:spindex_histgrm}, a histogram showing the spectral index ($\alpha_{1400}^{144}$) distribution of 45 WATs and 10 NATs, reported in the current paper, is shown. We also included 204 WATS and 61 NATs from \citet{Bh21} in the histogram.

The histogram showed a peak near $\alpha_{1400}^{144}$ = 0.80. The total span of $\alpha_{1400}^{144}$ is from 0.22 to 0.79 for NAT sources and from 0.41 to 1.57 for WAT sources. For NAT sources, J1505+4706 has the lowest spectral index with $\alpha_{1400}^{144}$ = 0.22 and J1207+4805 has the highest spectral index with $\alpha_{1400}^{144}$ = 0.79. For WAT sources, J1325+5617 has the lowest spectral index with $\alpha_{1400}^{144}$ = 0.41 and J1224+4905 has the highest spectral index with $\alpha_{1400}^{144}$ = 1.57. None of the HT radio galaxies has an inverted spectrum.

\begin{figure*}
\vbox{
\centerline{
\includegraphics[height=5.6cm,width=5.6cm]{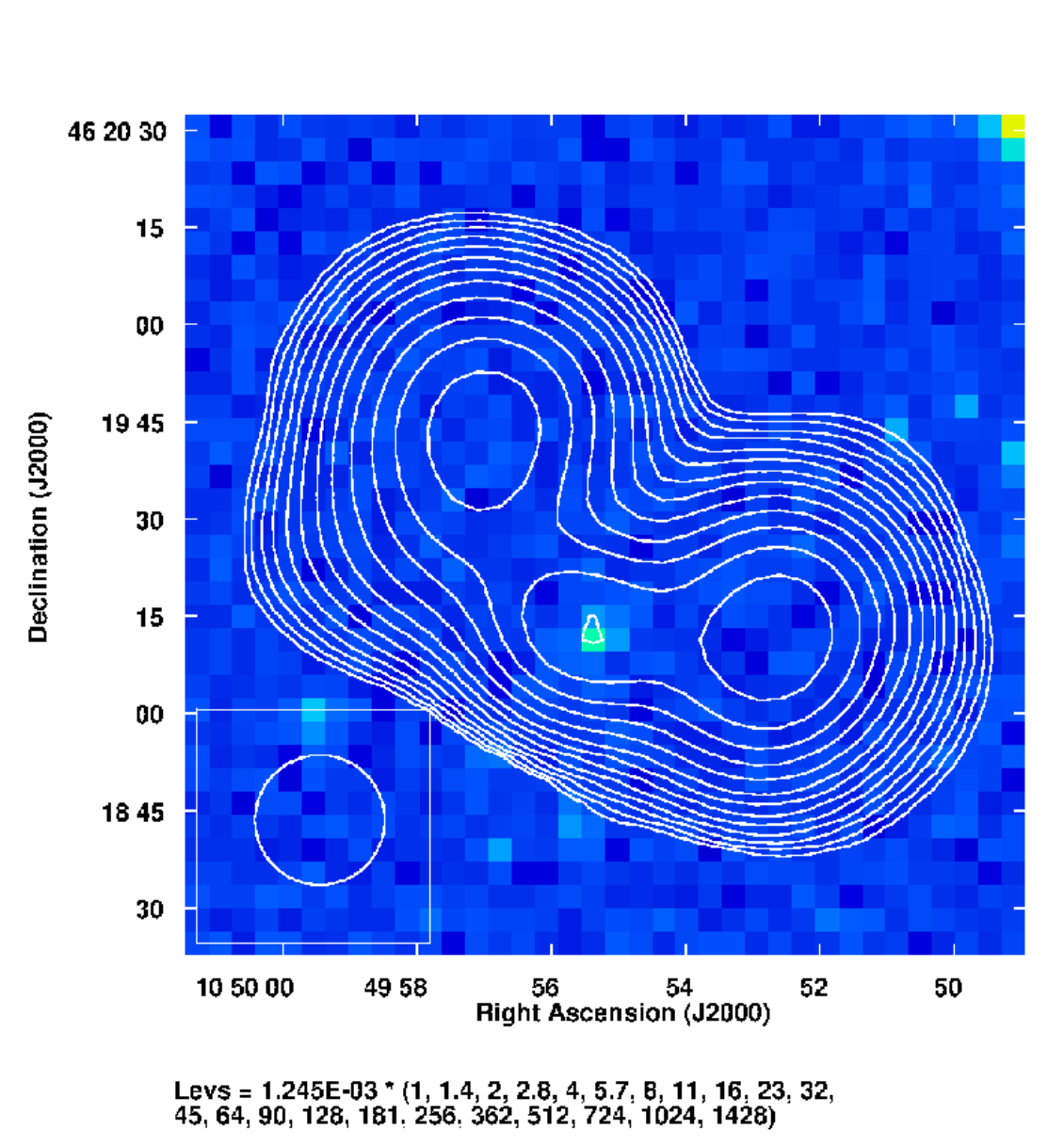}
\includegraphics[height=5.6cm,width=5.6cm]{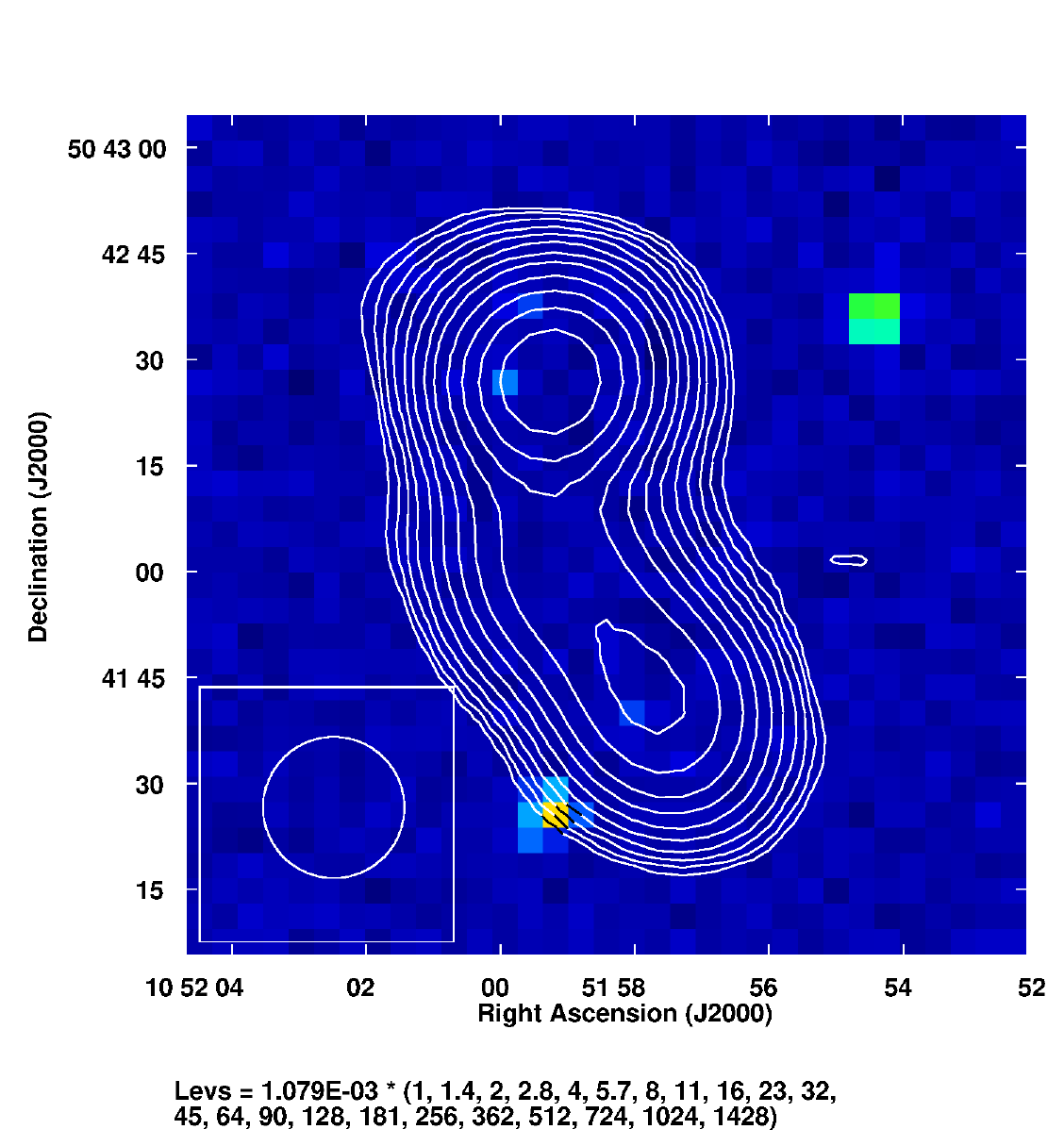}
\includegraphics[height=5.6cm,width=5.6cm]{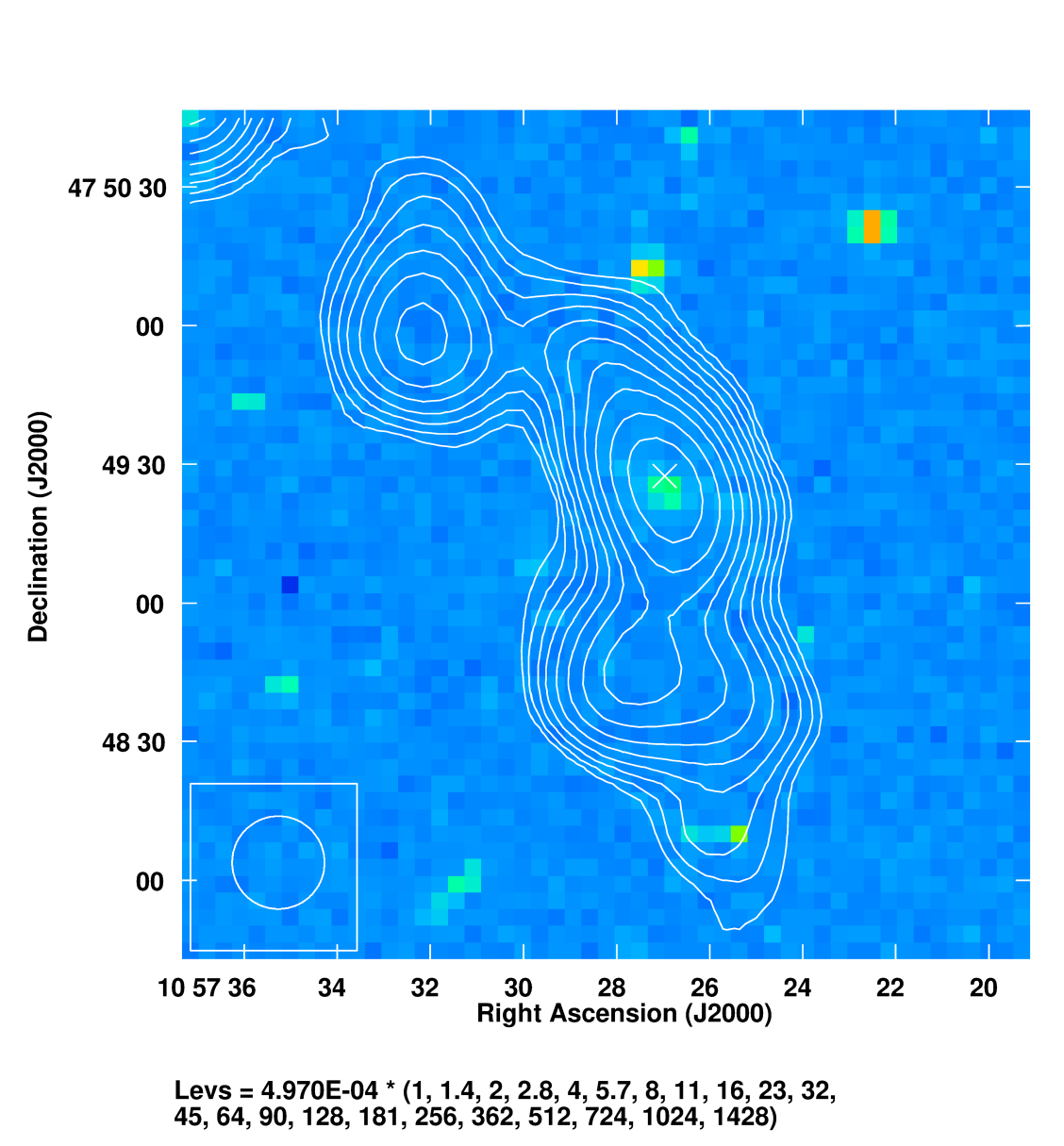}
}
}
\vbox{
\centerline{
\includegraphics[height=5.6cm,width=5.6cm]{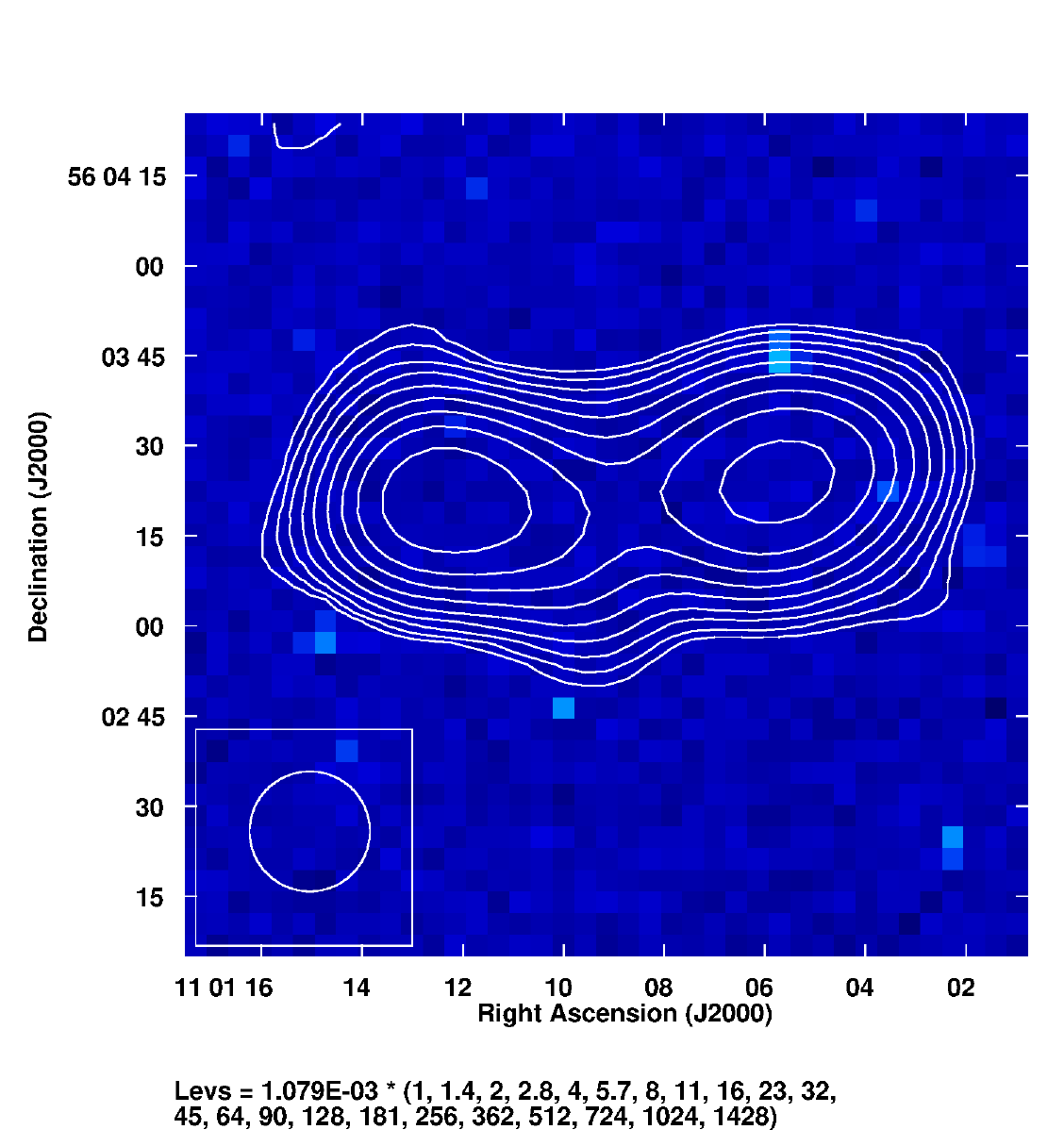}
\includegraphics[height=5.6cm,width=5.6cm]{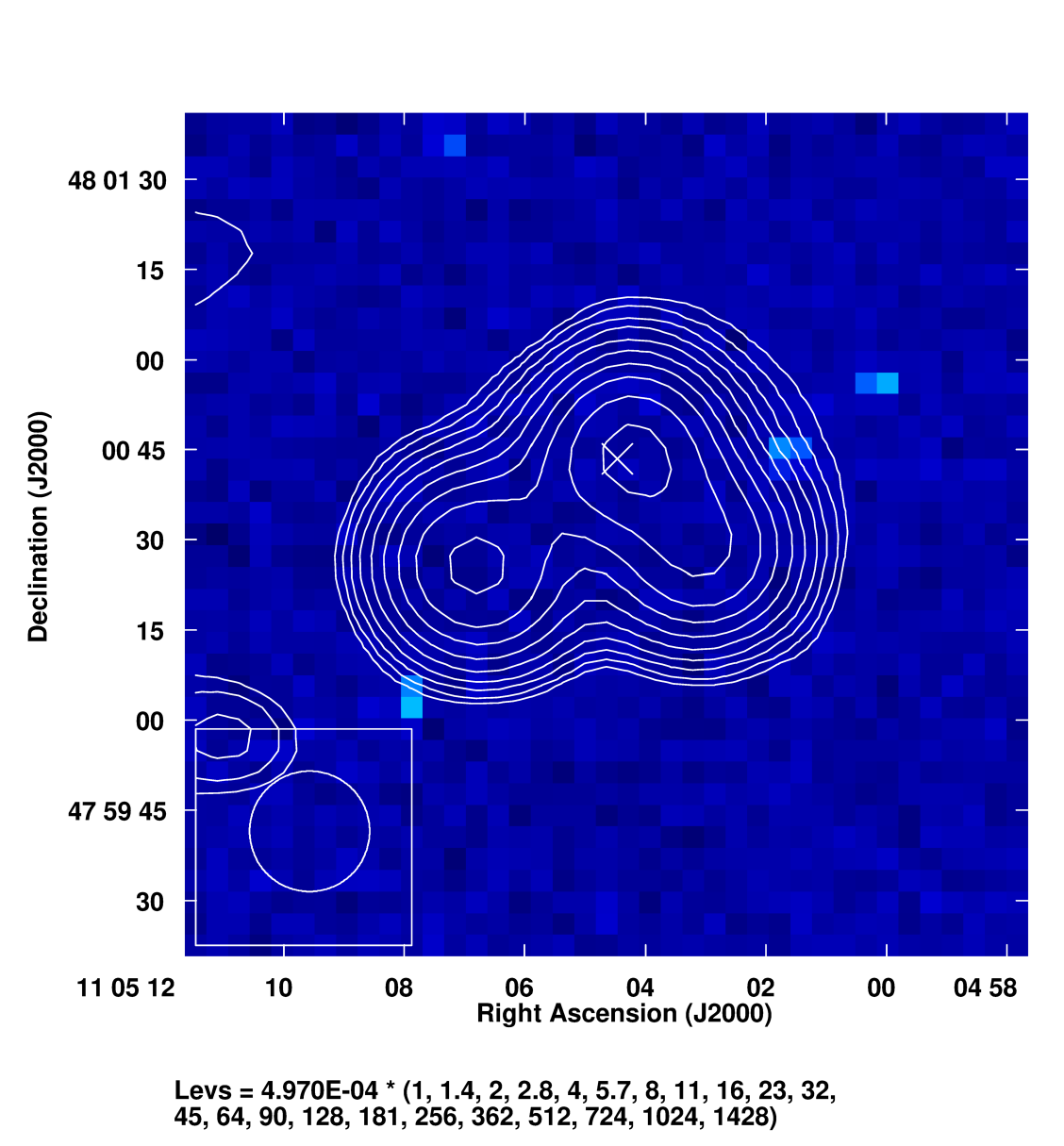}
\includegraphics[height=5.6cm,width=5.6cm]{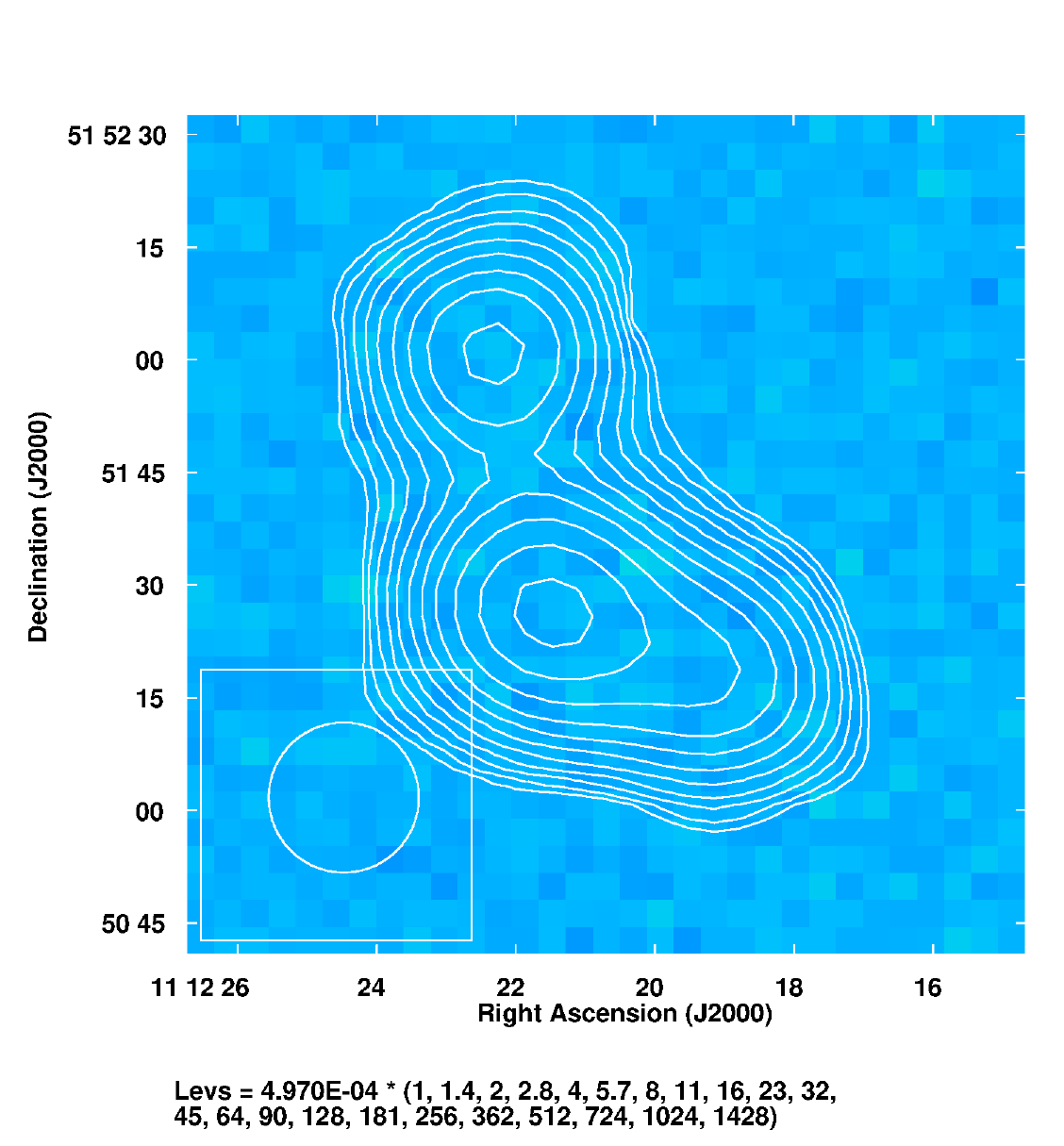}
}
}
\vbox{
\centerline{
\includegraphics[height=5.6cm,width=5.6cm]{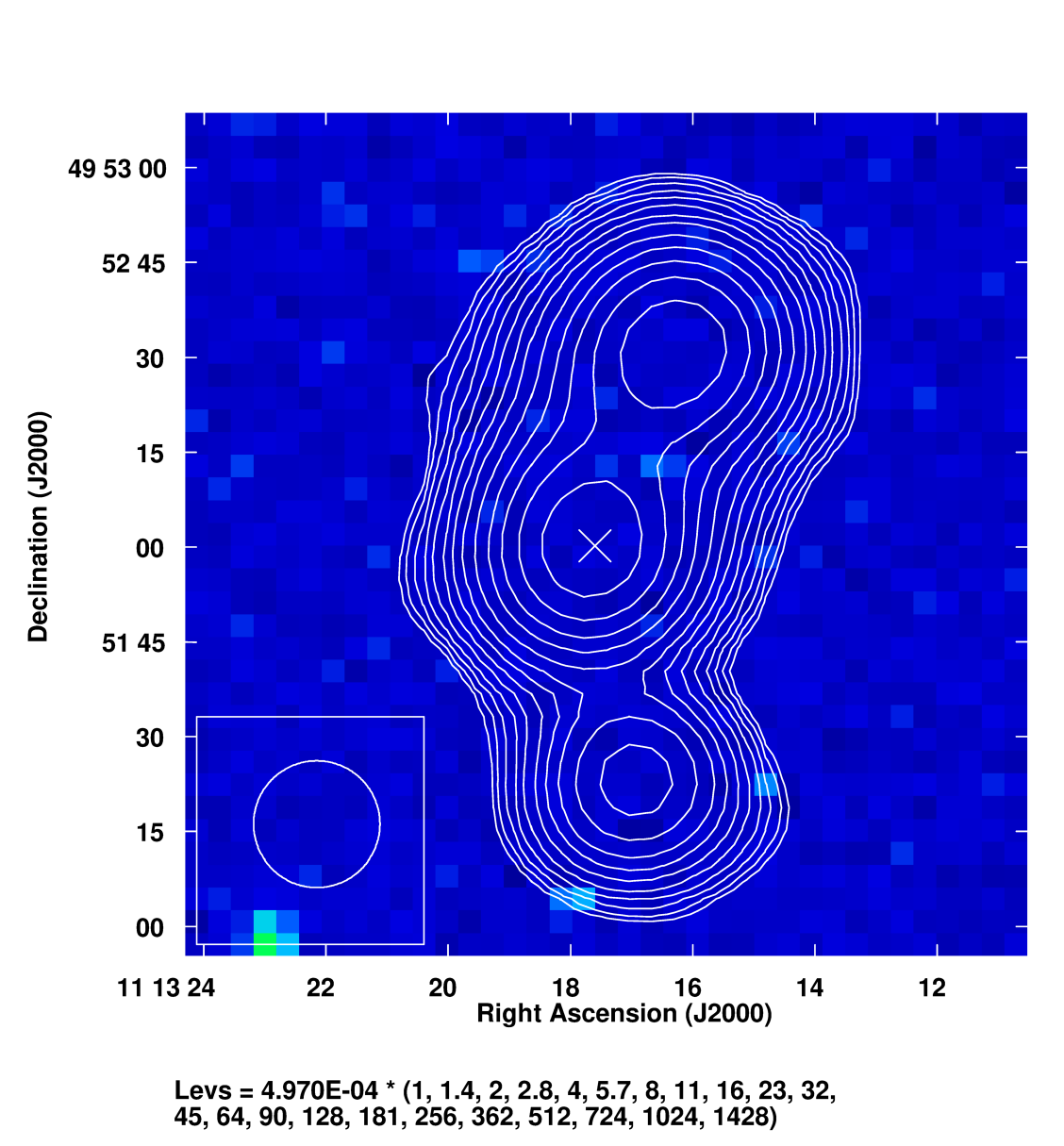}
\includegraphics[height=5.6cm,width=5.6cm]{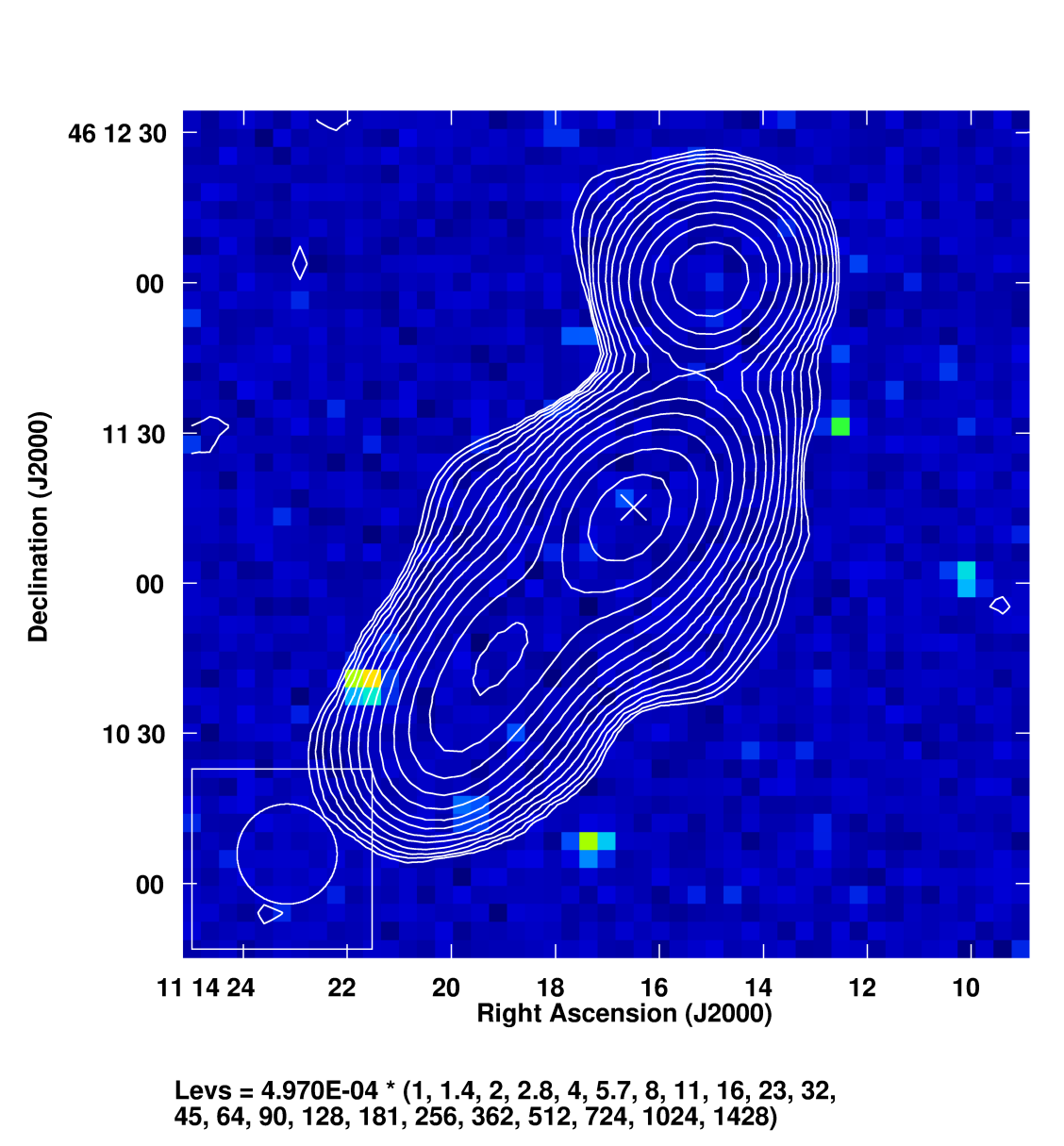}
\includegraphics[height=5.6cm,width=5.6cm]{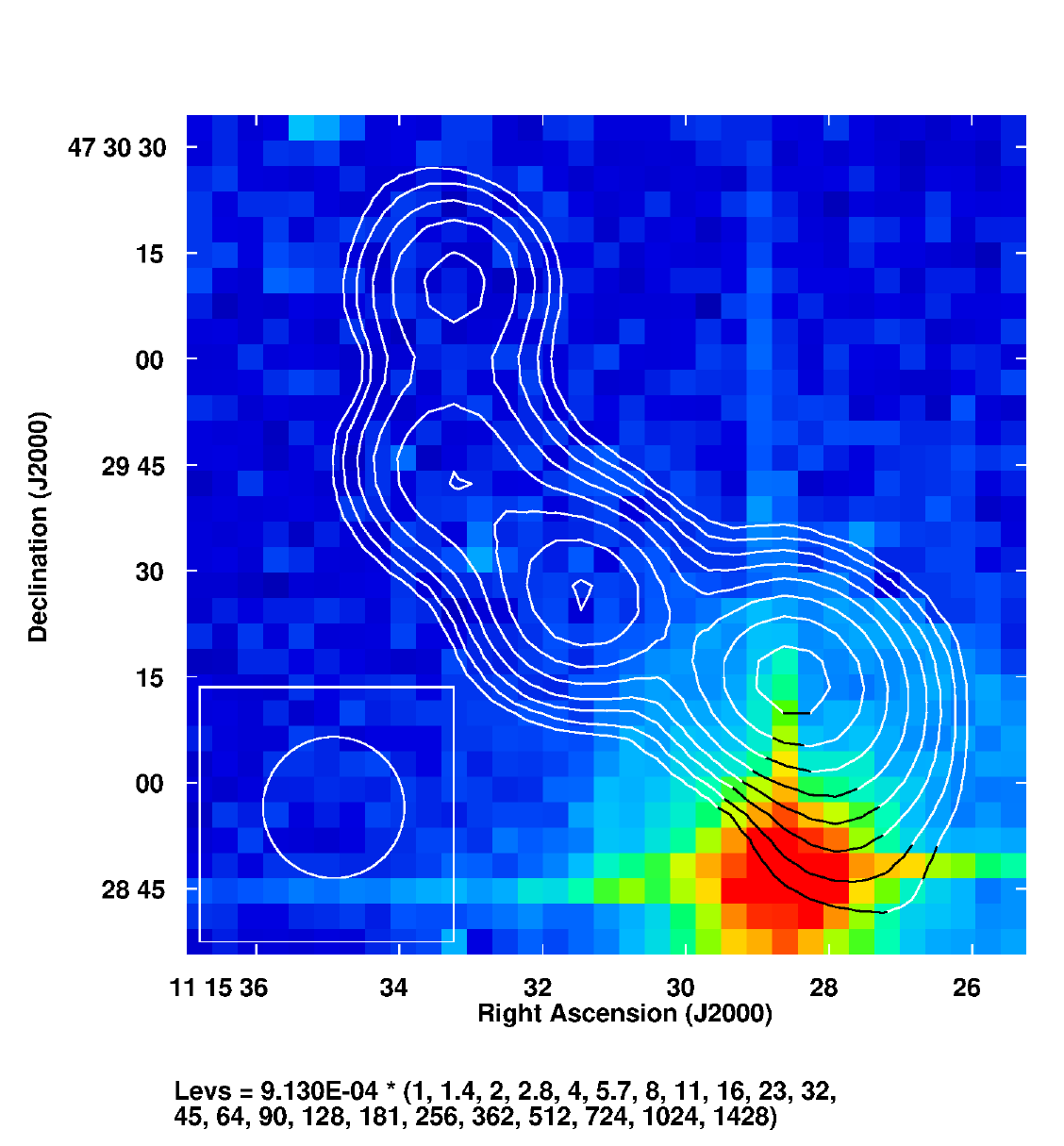}
}
}
\vbox{
\centerline{
\includegraphics[height=5.6cm,width=5.6cm]{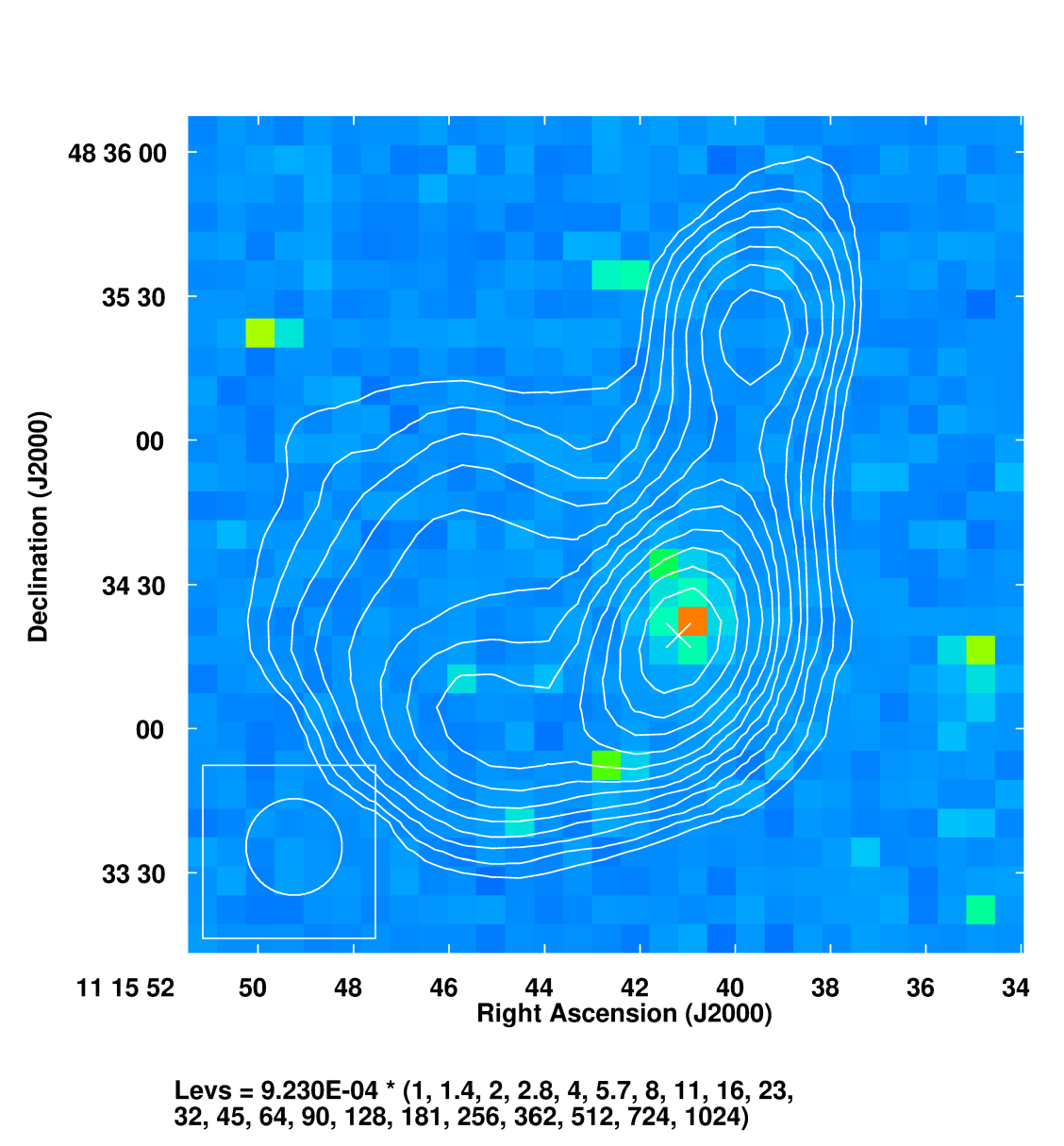}
\includegraphics[height=5.6cm,width=5.6cm]{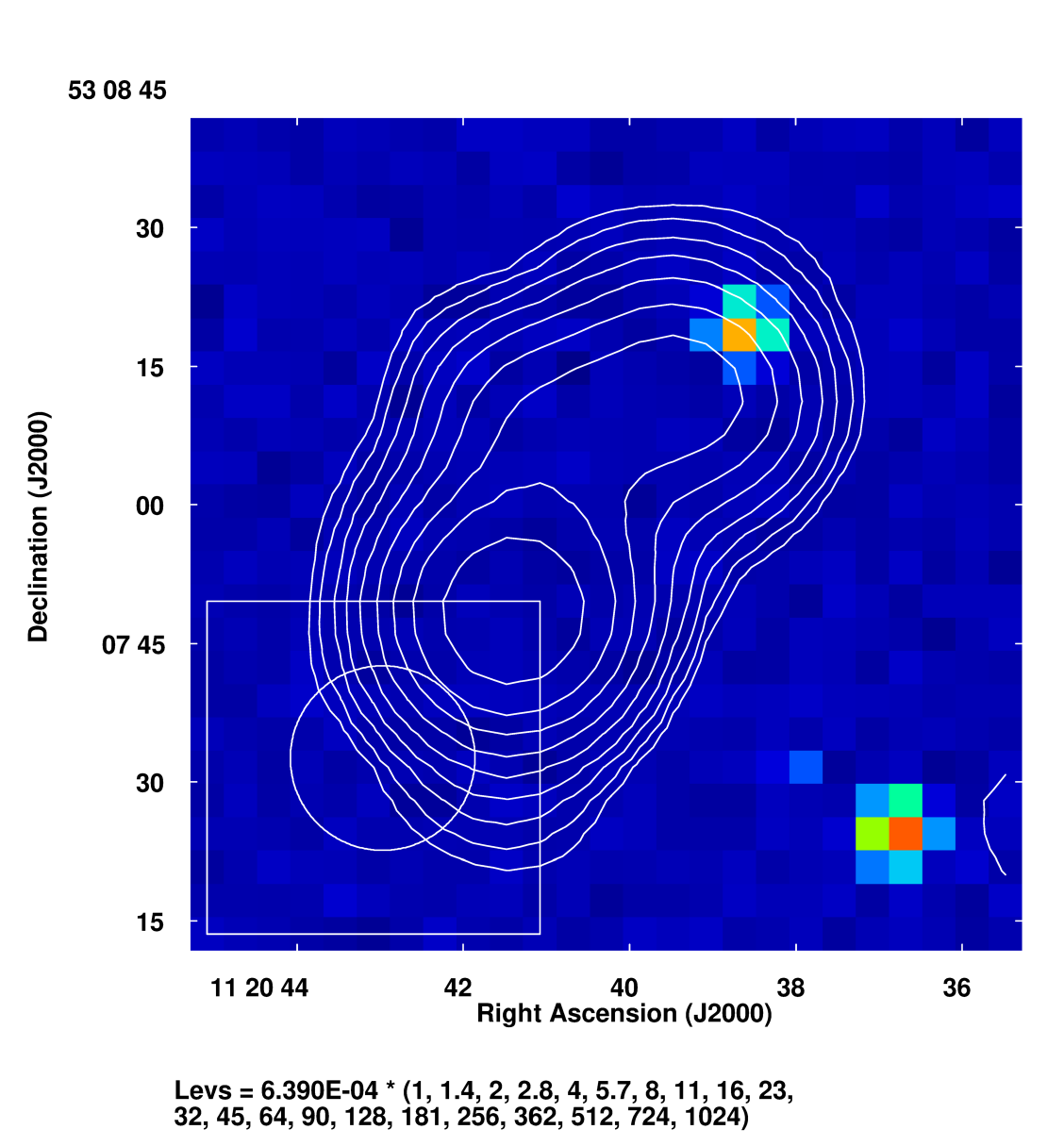}
\includegraphics[height=5.6cm,width=5.6cm]{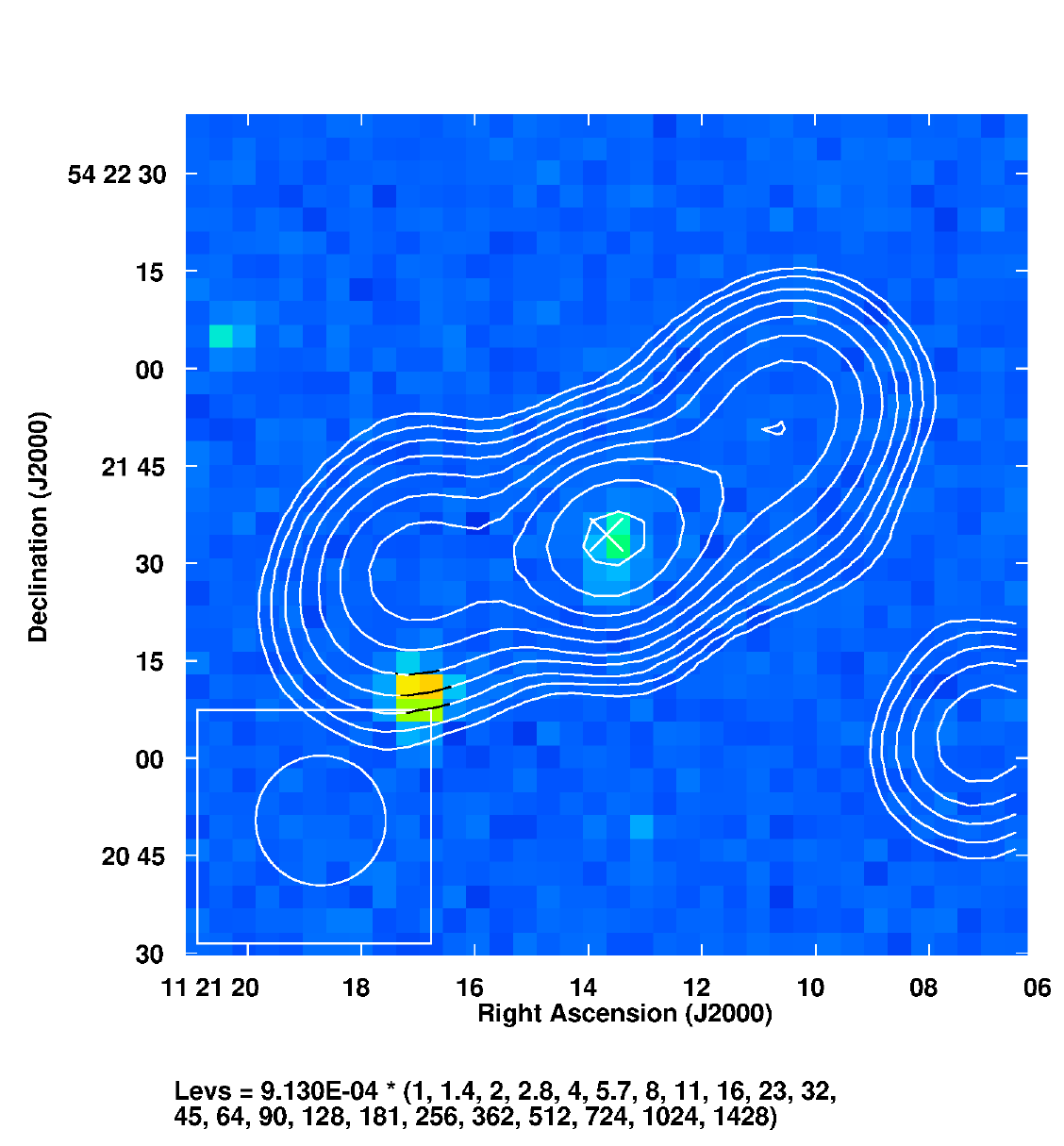}
}
}
\caption{LOFAR images of WAT radio galaxies (contours) overlaid on DSS2 red images (colour). Here white cross mark represent the optical counterpart for the sources, when available.}
\label{fig:HT}
\end{figure*}

\begin{figure*}
\vbox{
\centerline{
\includegraphics[height=5.6cm,width=5.6cm]{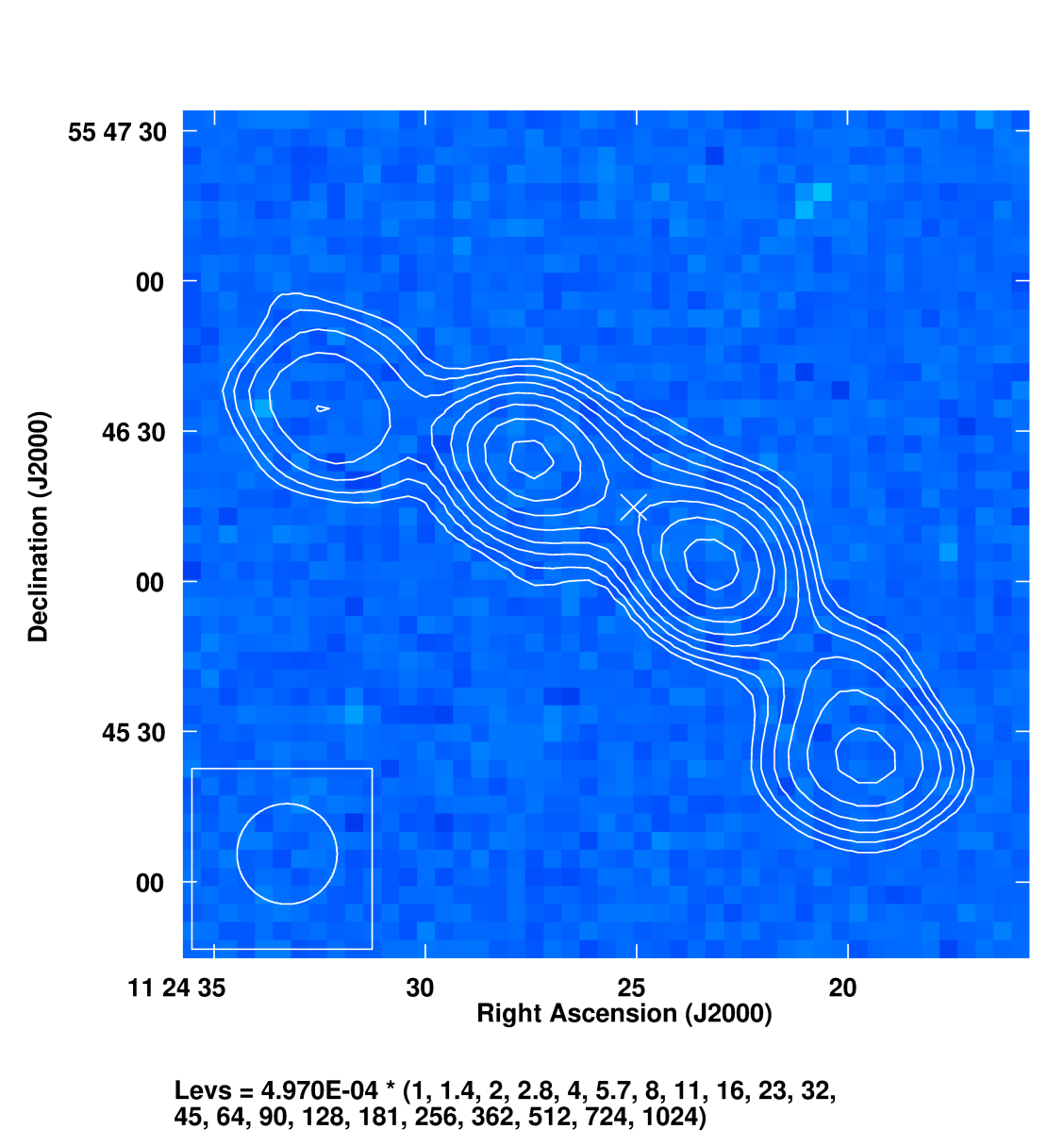}
\includegraphics[height=5.6cm,width=5.6cm]{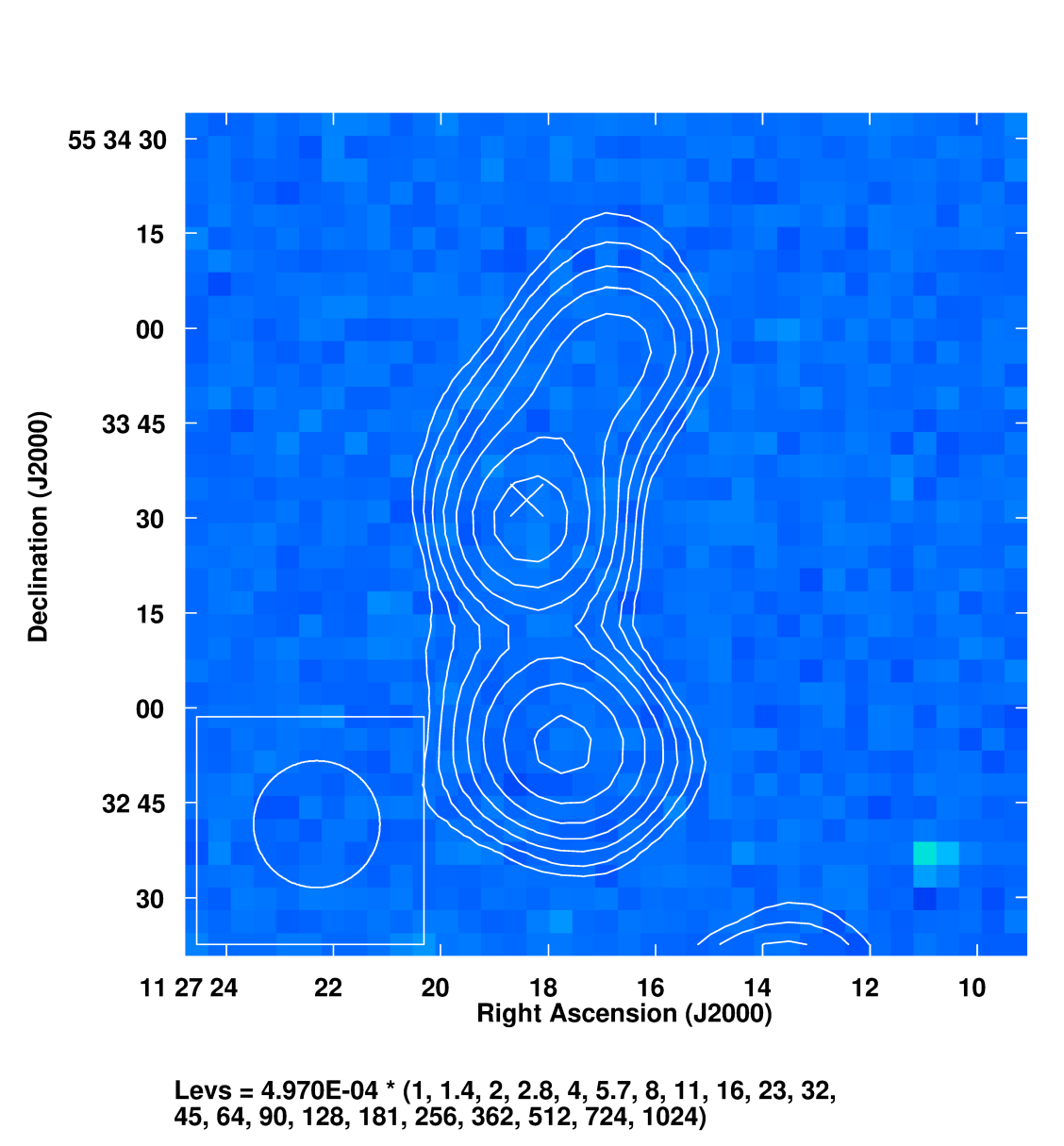}
\includegraphics[height=5.6cm,width=5.6cm]{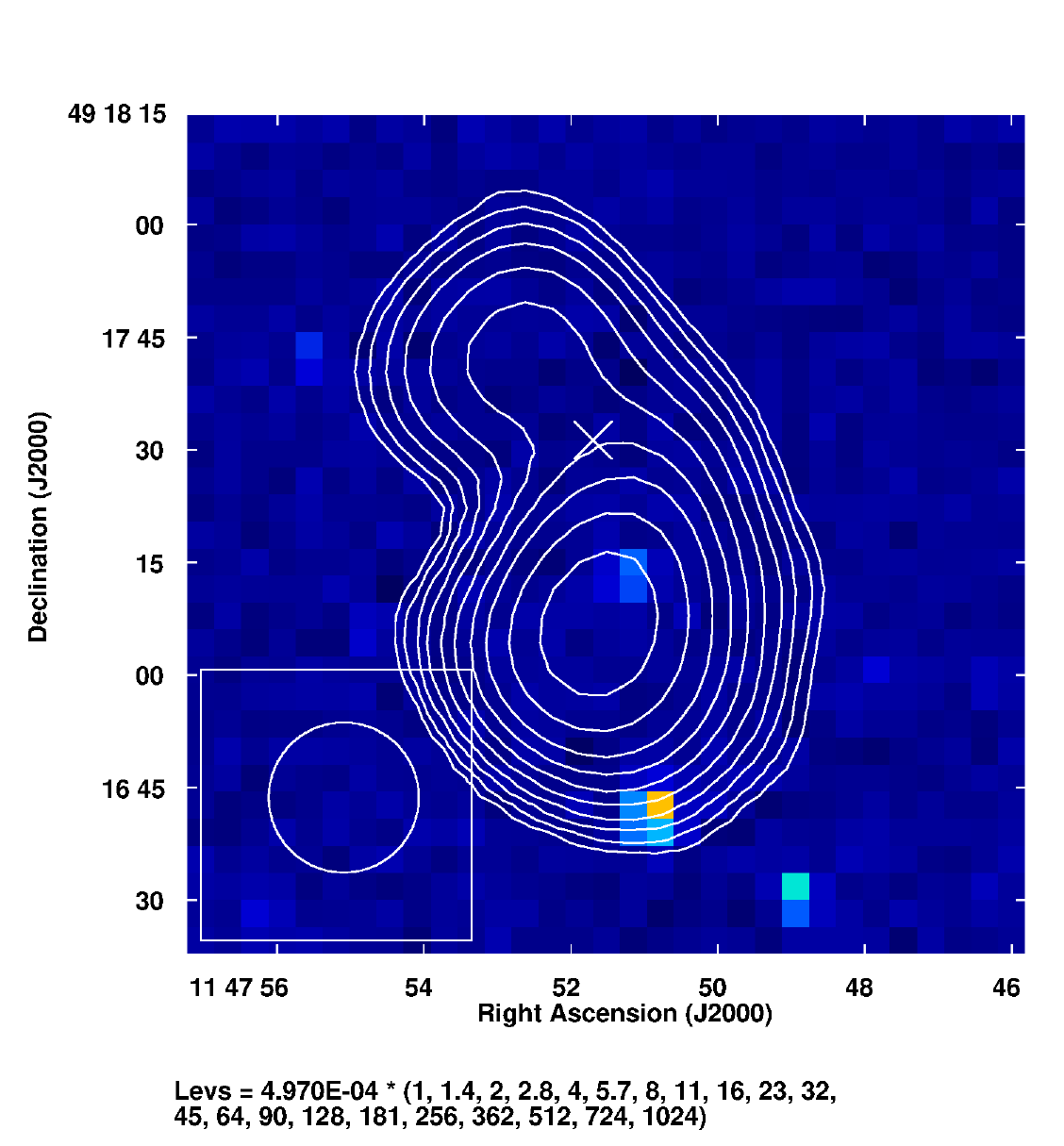}	
}
}
\vbox{
\centerline{
\includegraphics[height=5.6cm,width=5.6cm]{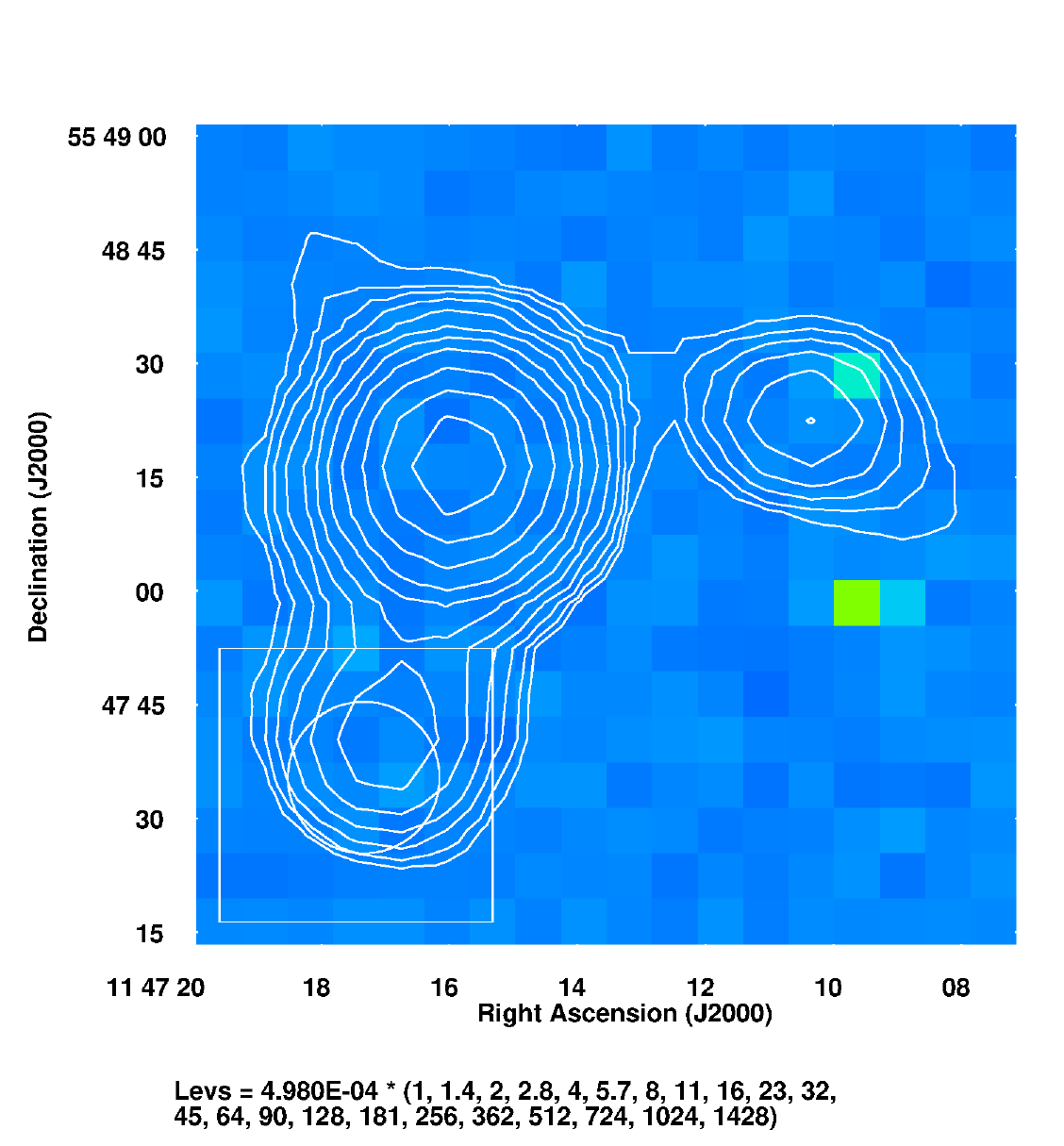}
\includegraphics[height=5.6cm,width=5.6cm]{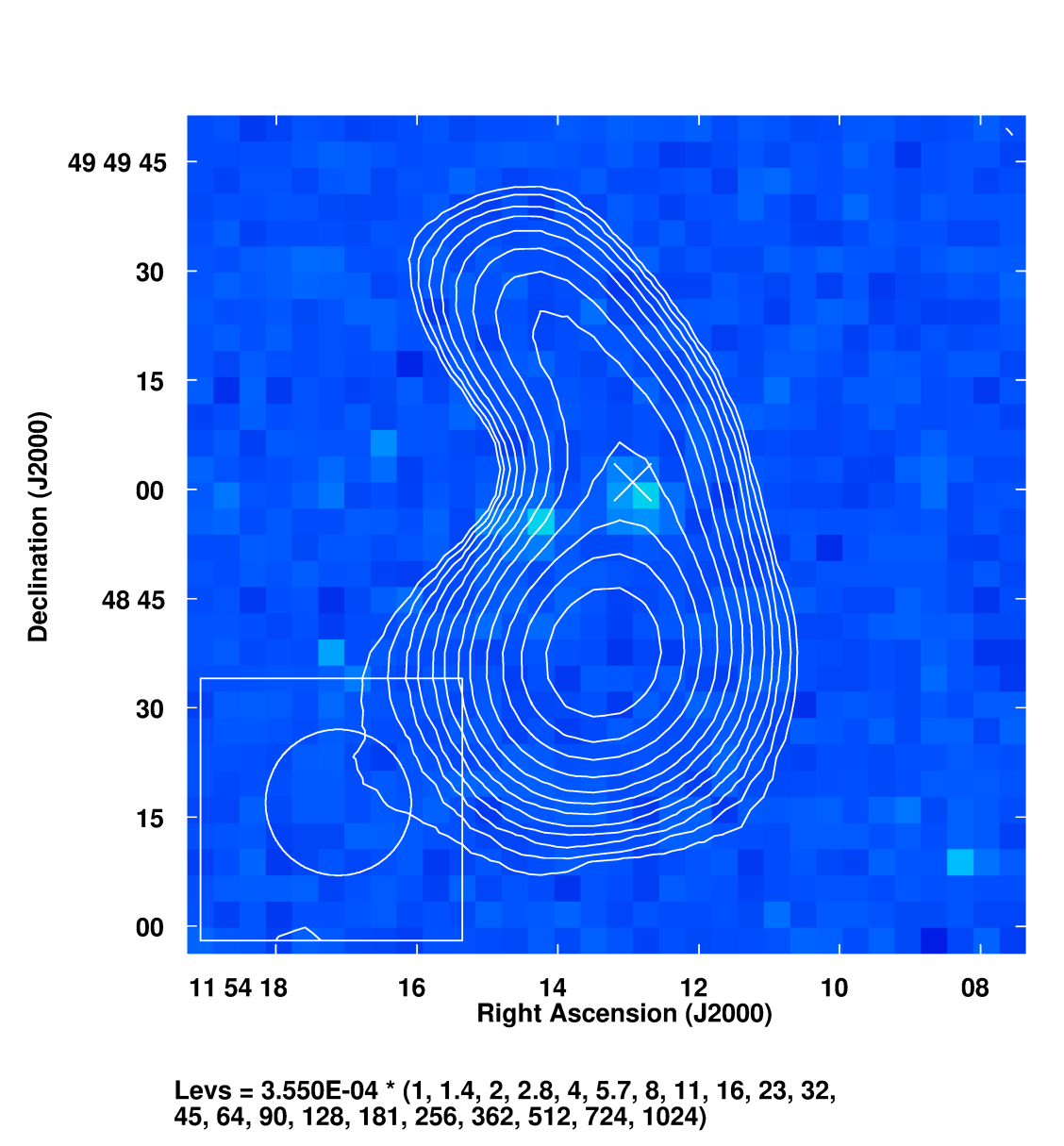}
\includegraphics[height=5.6cm,width=5.6cm]{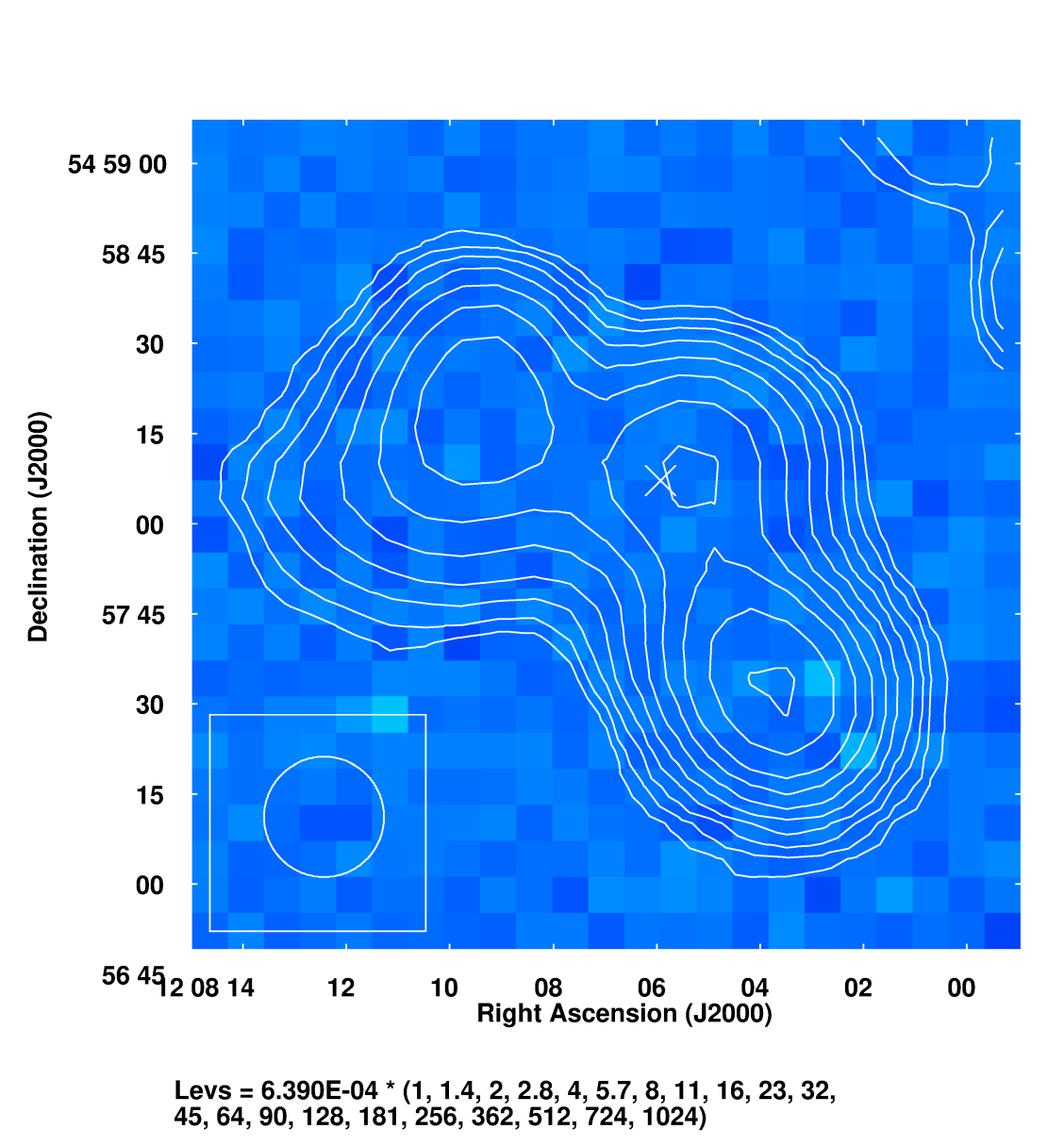}
}
}
\vbox{
\centerline{
\includegraphics[height=5.6cm,width=5.6cm]{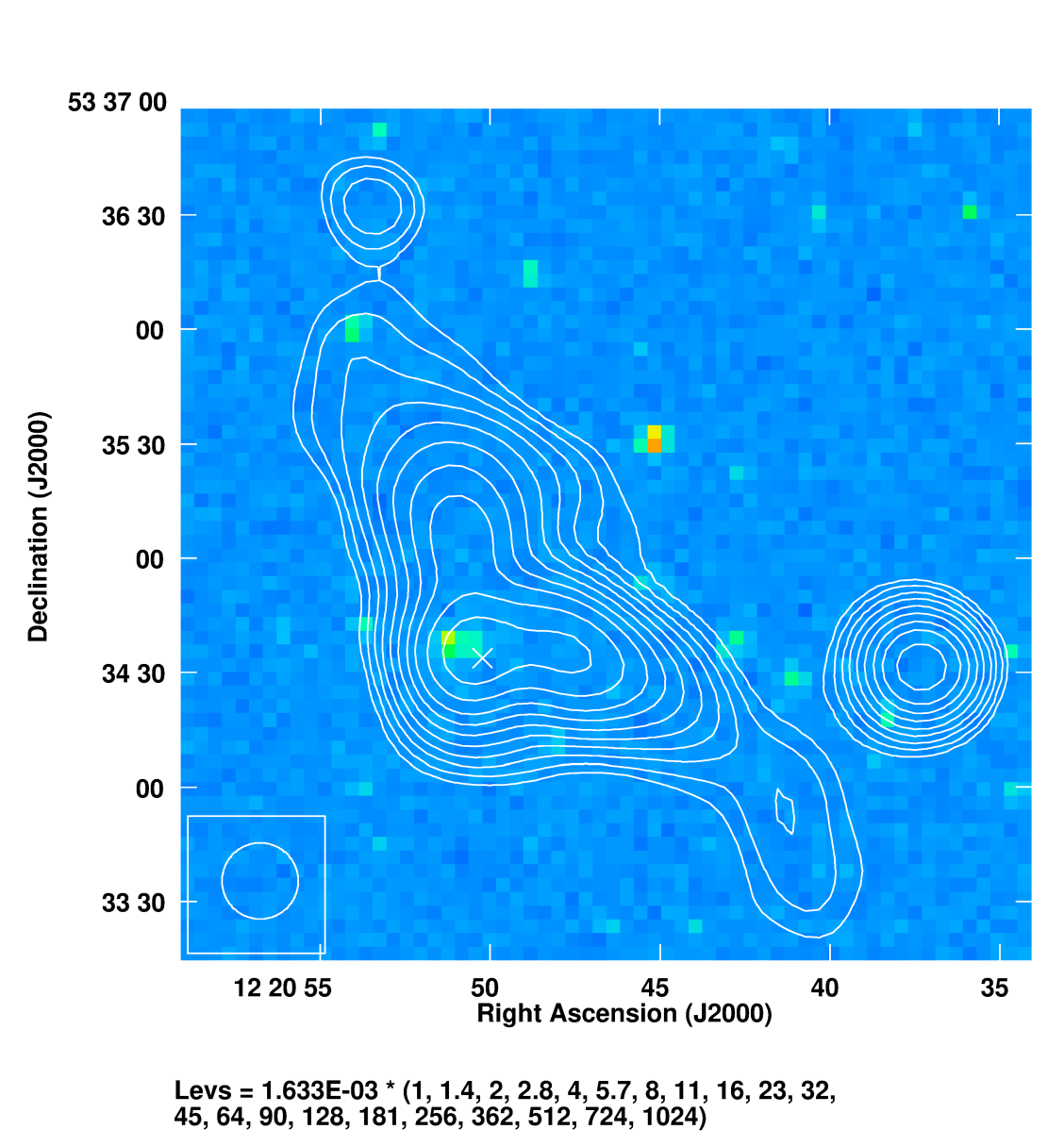}
\includegraphics[height=5.6cm,width=5.6cm]{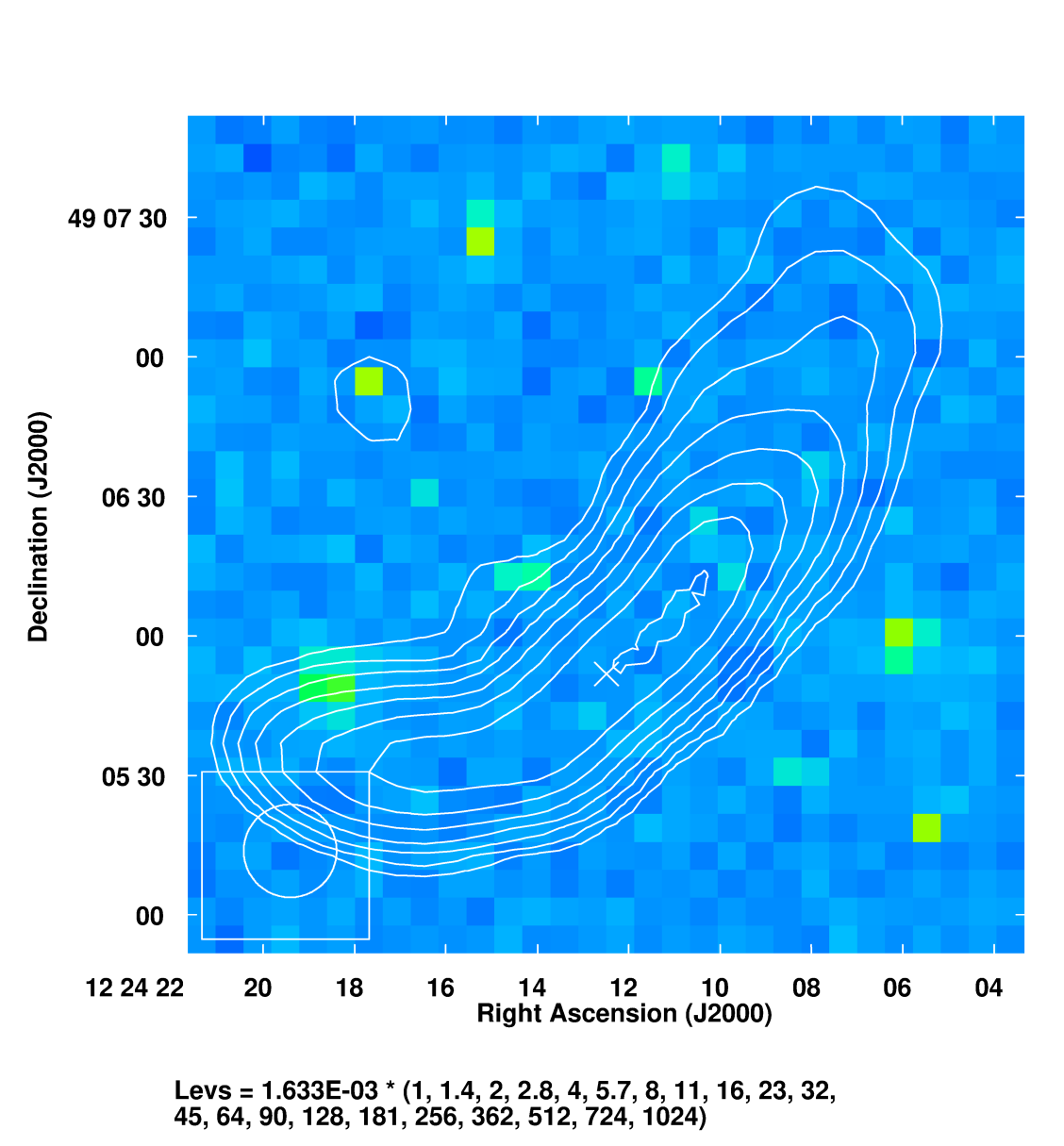}
\includegraphics[height=5.6cm,width=5.6cm]{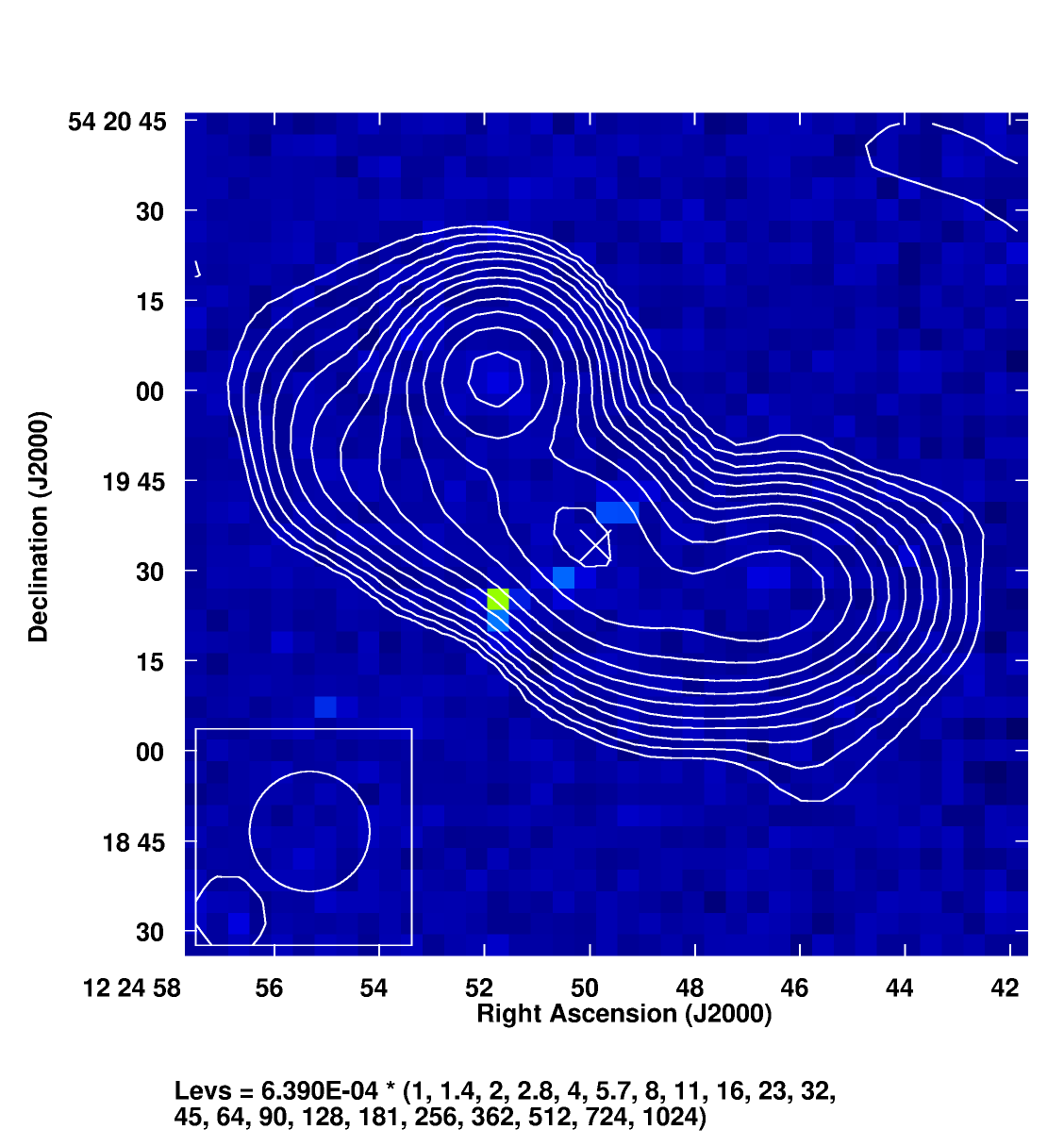}
}
}
\vbox{
\centerline{
\includegraphics[height=5.6cm,width=5.6cm]{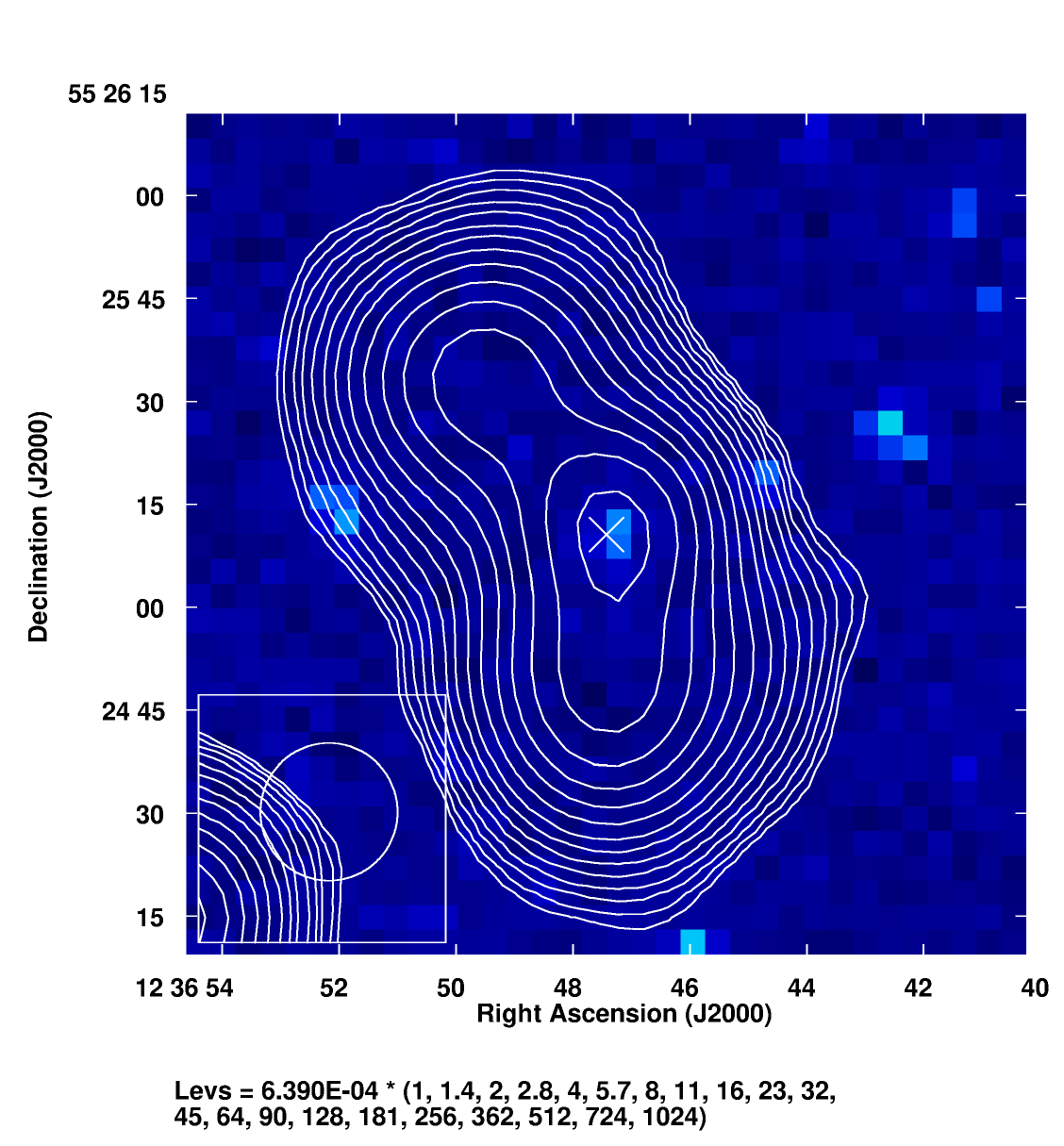}
\includegraphics[height=5.6cm,width=5.6cm]{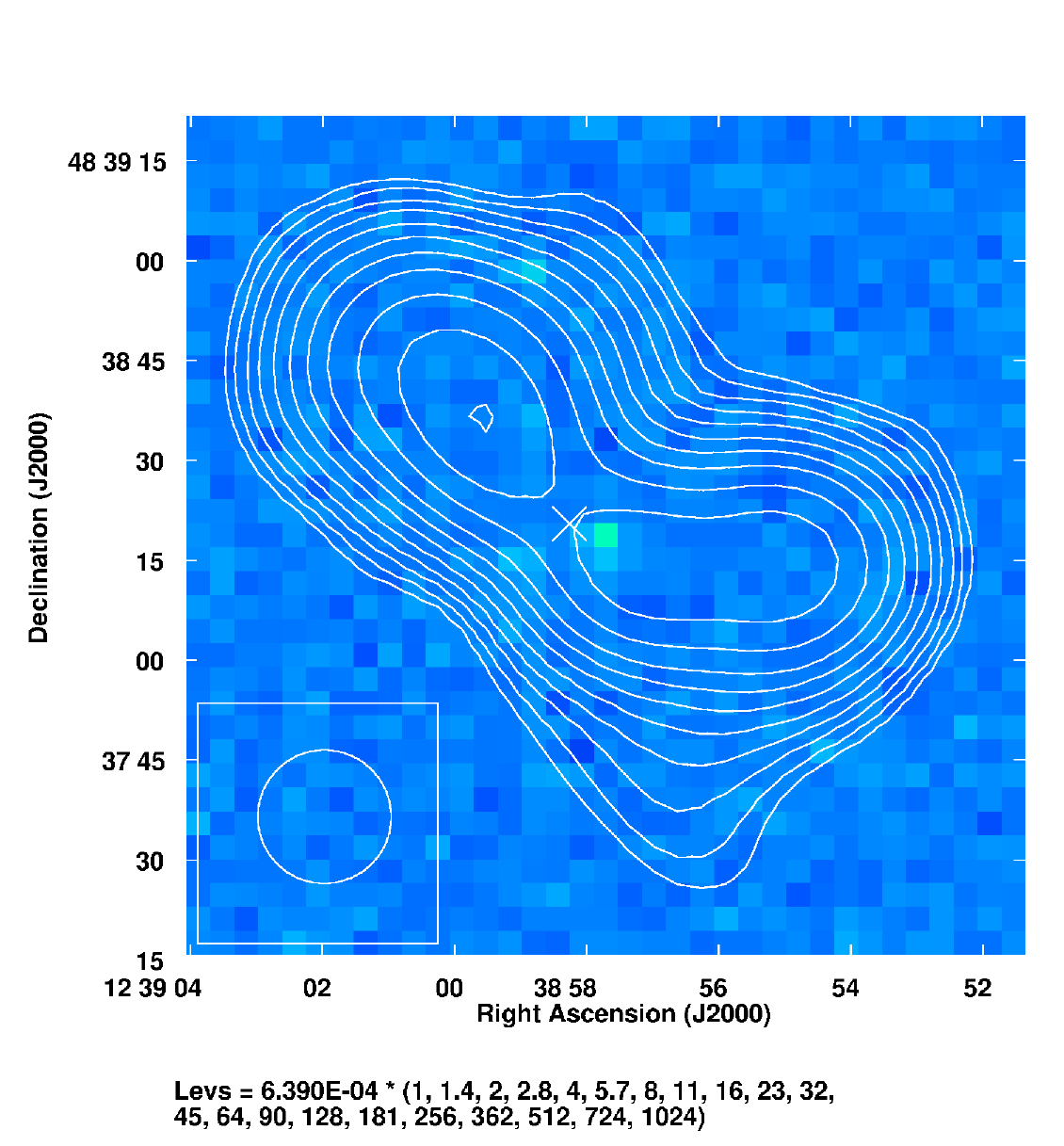}
\includegraphics[height=5.6cm,width=5.6cm]{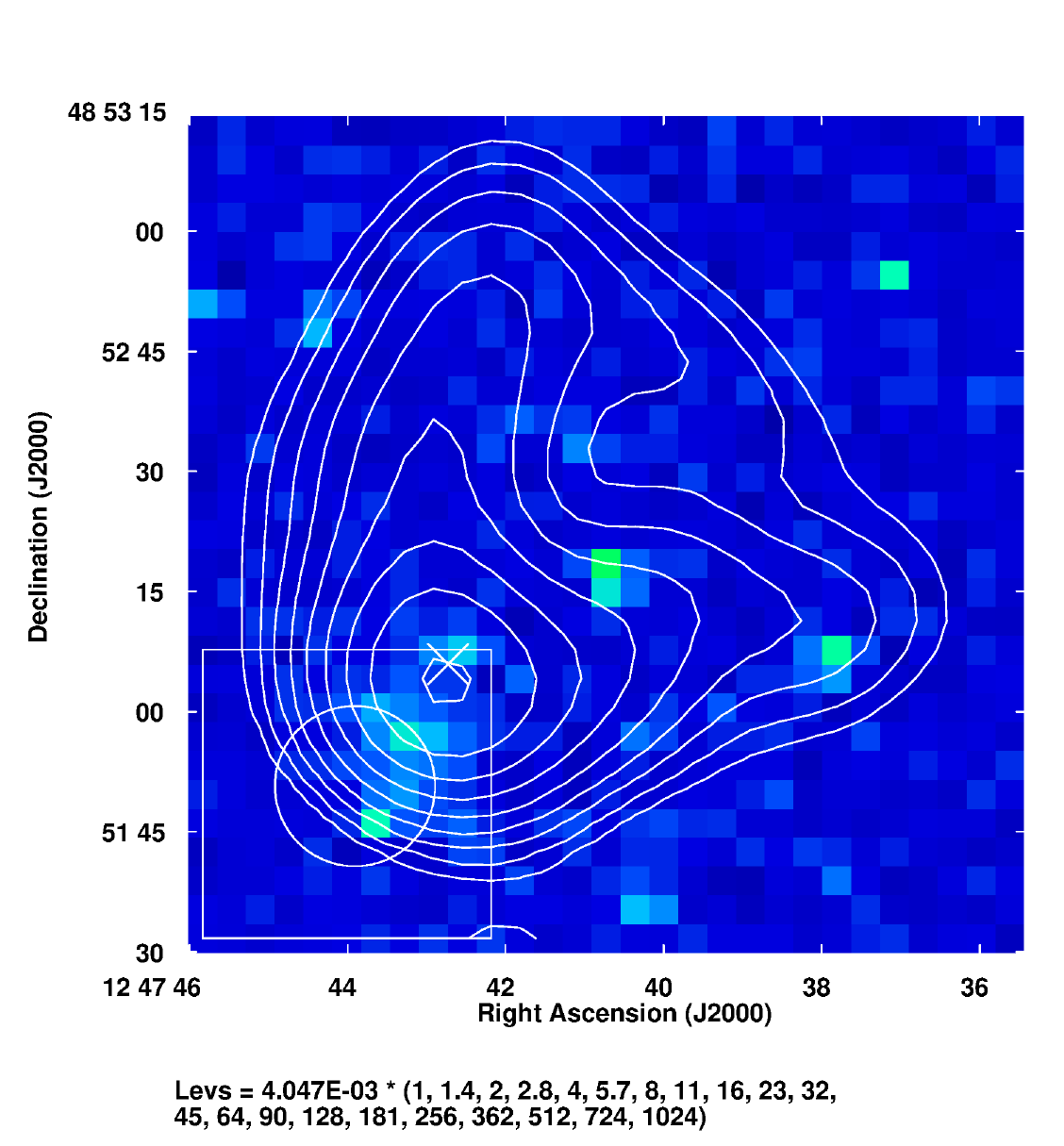}
}
}
\contcaption{LOFAR images of WAT radio galaxies (contours) overlaid on DSS2 red images (colour). Here white cross mark represent the optical counterpart for the sources, when available.}
\end{figure*}

\begin{figure*}
\vbox{
\centerline{
\includegraphics[height=5.6cm,width=5.6cm]{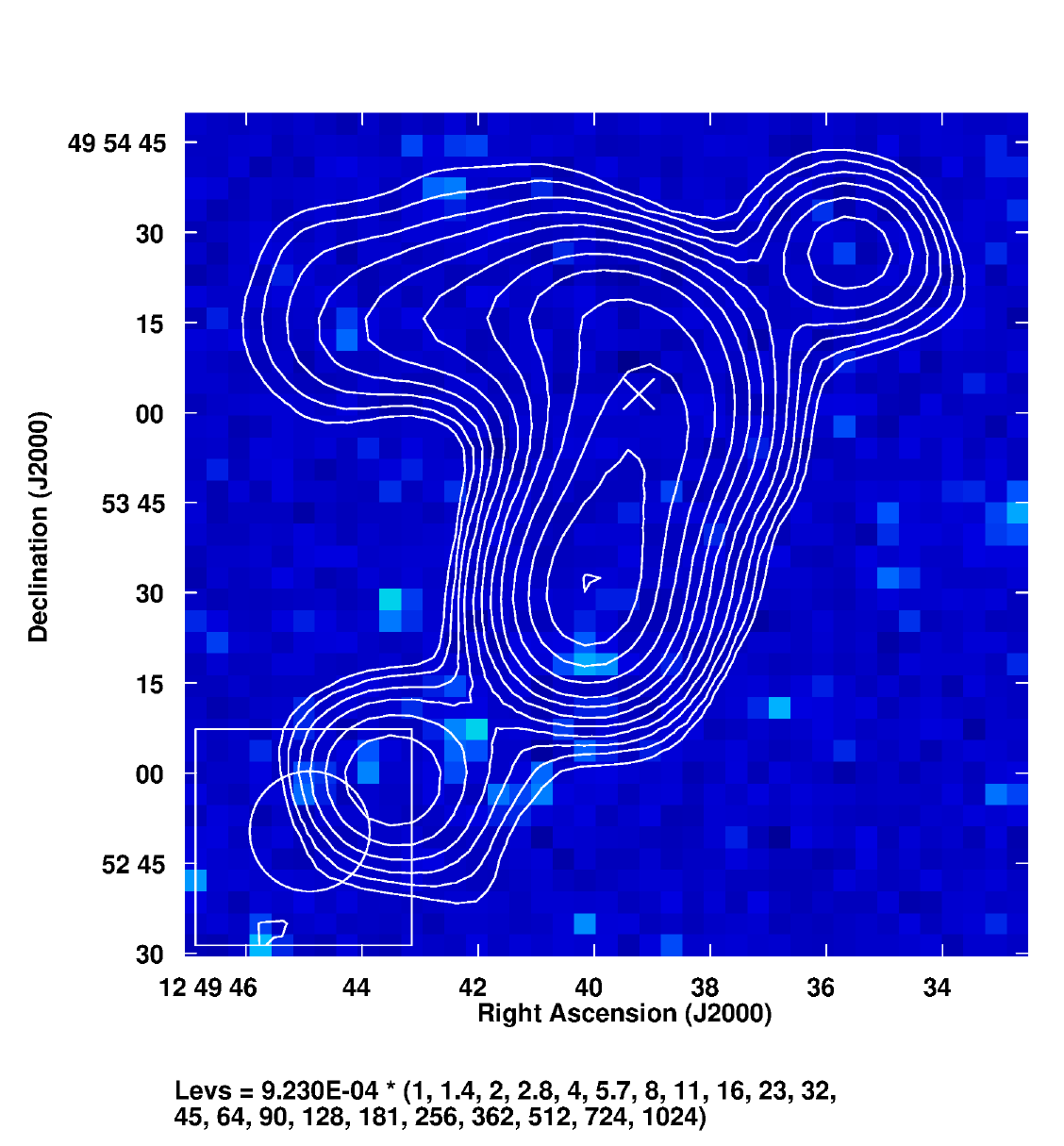}
\includegraphics[height=5.6cm,width=5.6cm]{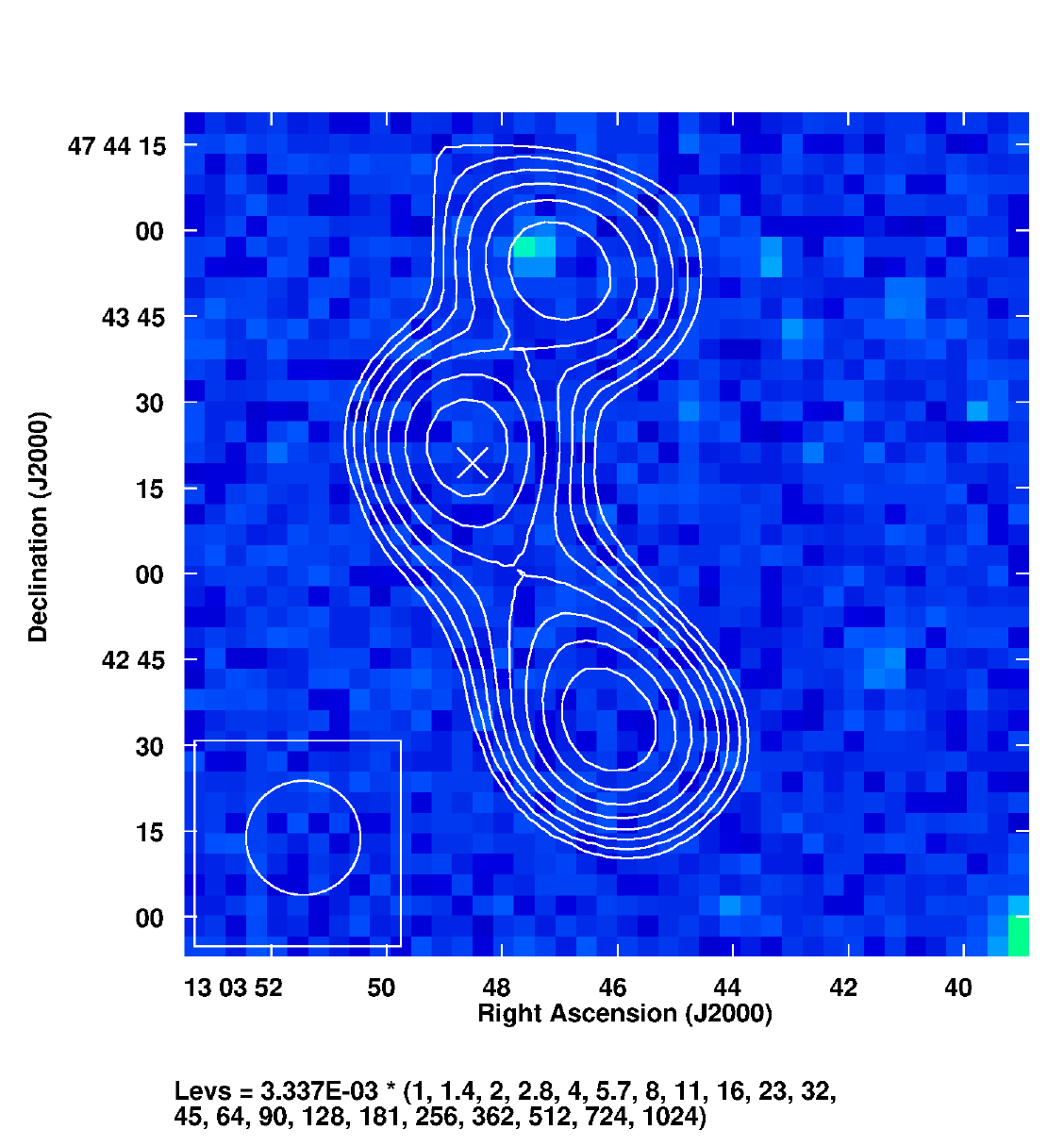}
\includegraphics[height=5.6cm,width=5.6cm]{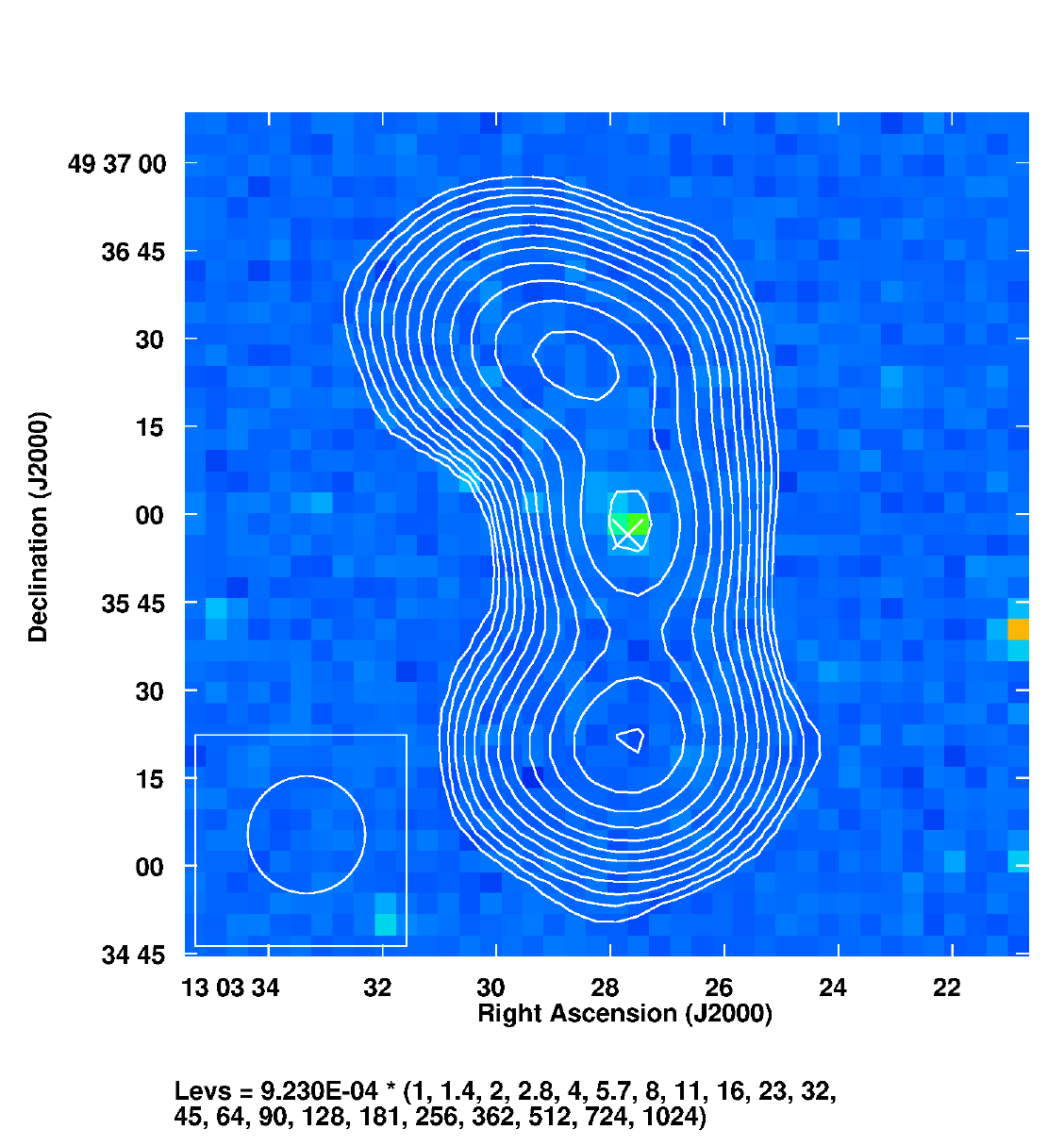}
}
}
\vbox{
\centerline{
\includegraphics[height=5.6cm,width=5.6cm]{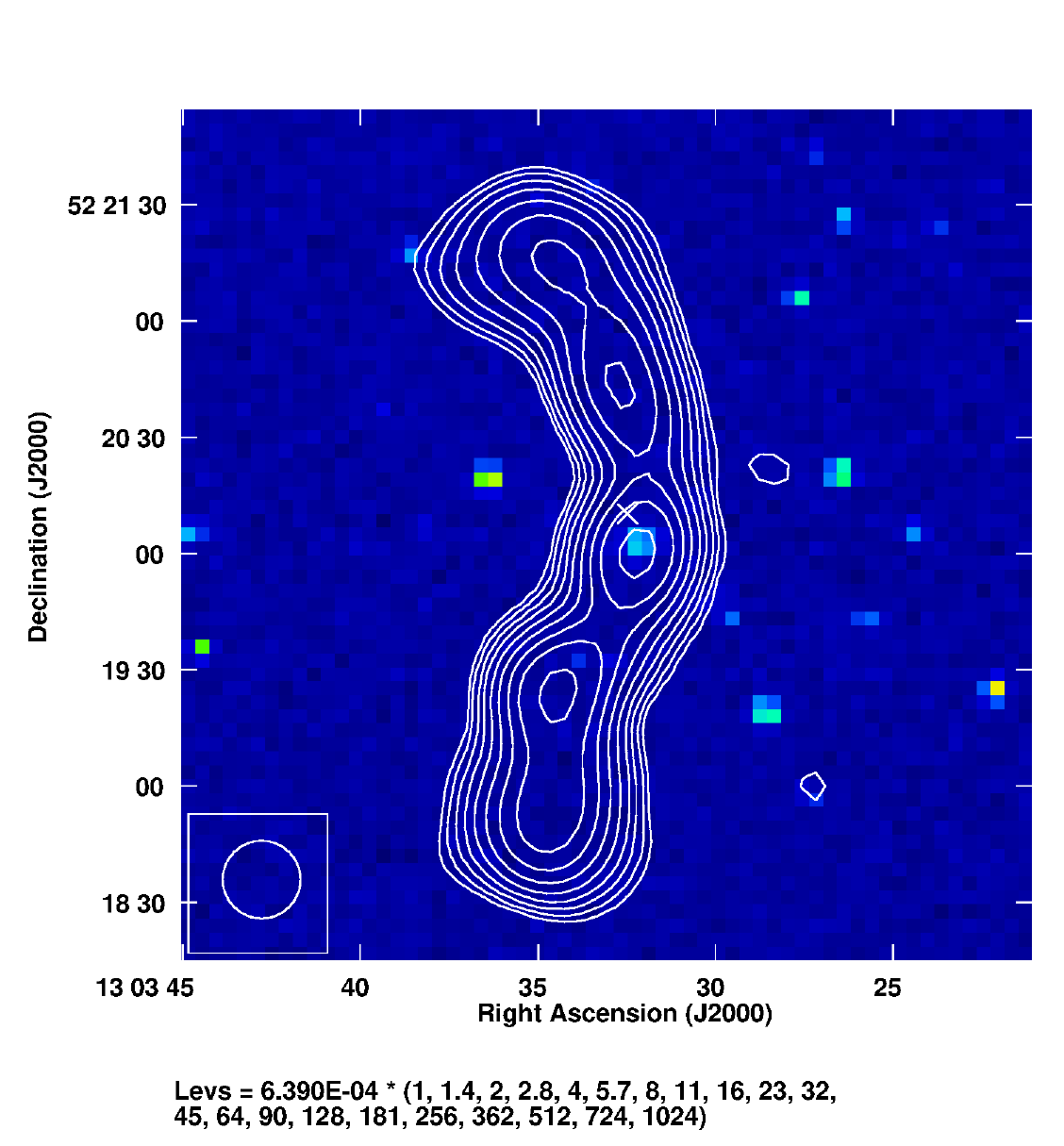}
\includegraphics[height=5.6cm,width=5.6cm]{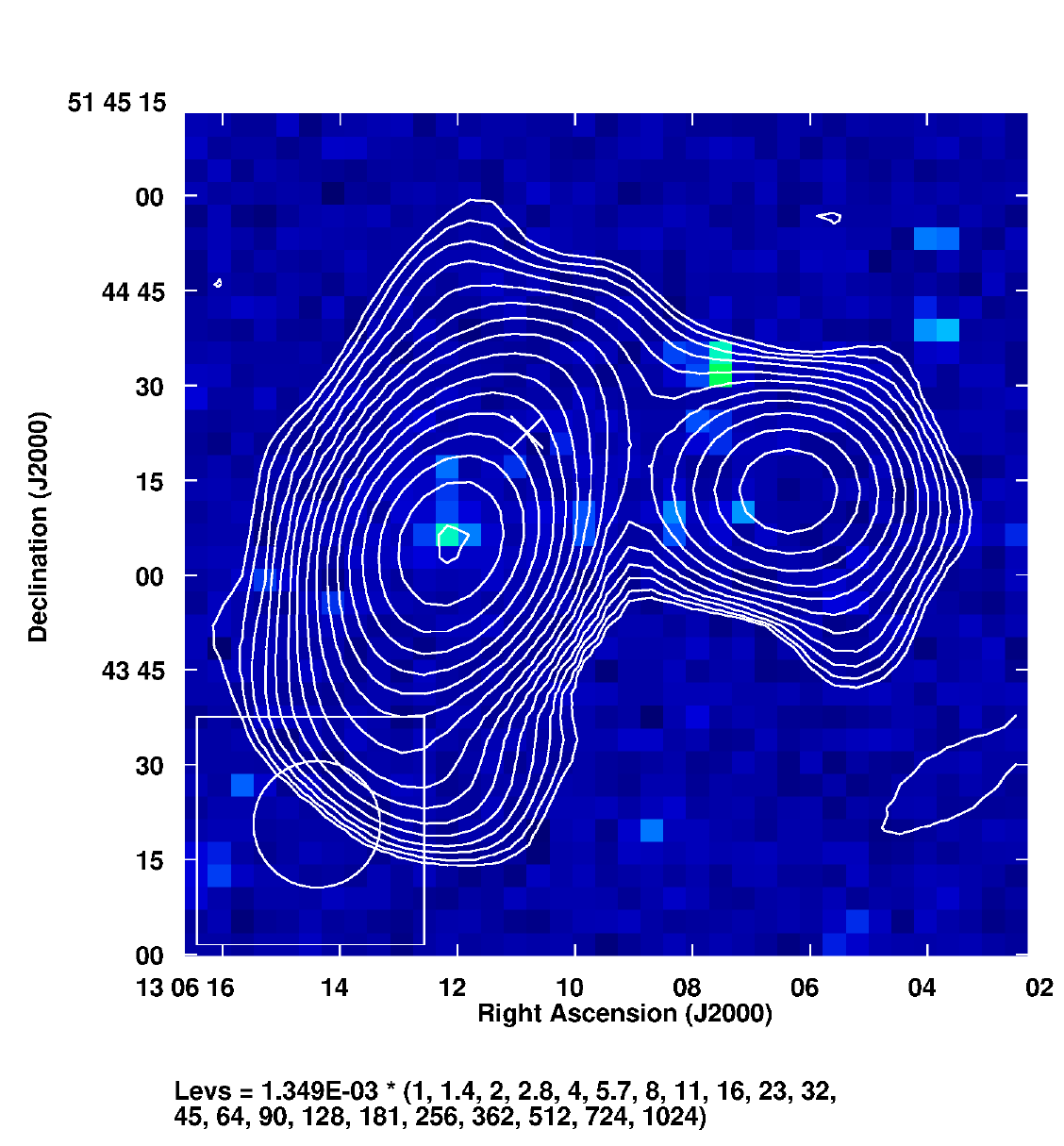}
\includegraphics[height=5.6cm,width=5.6cm]{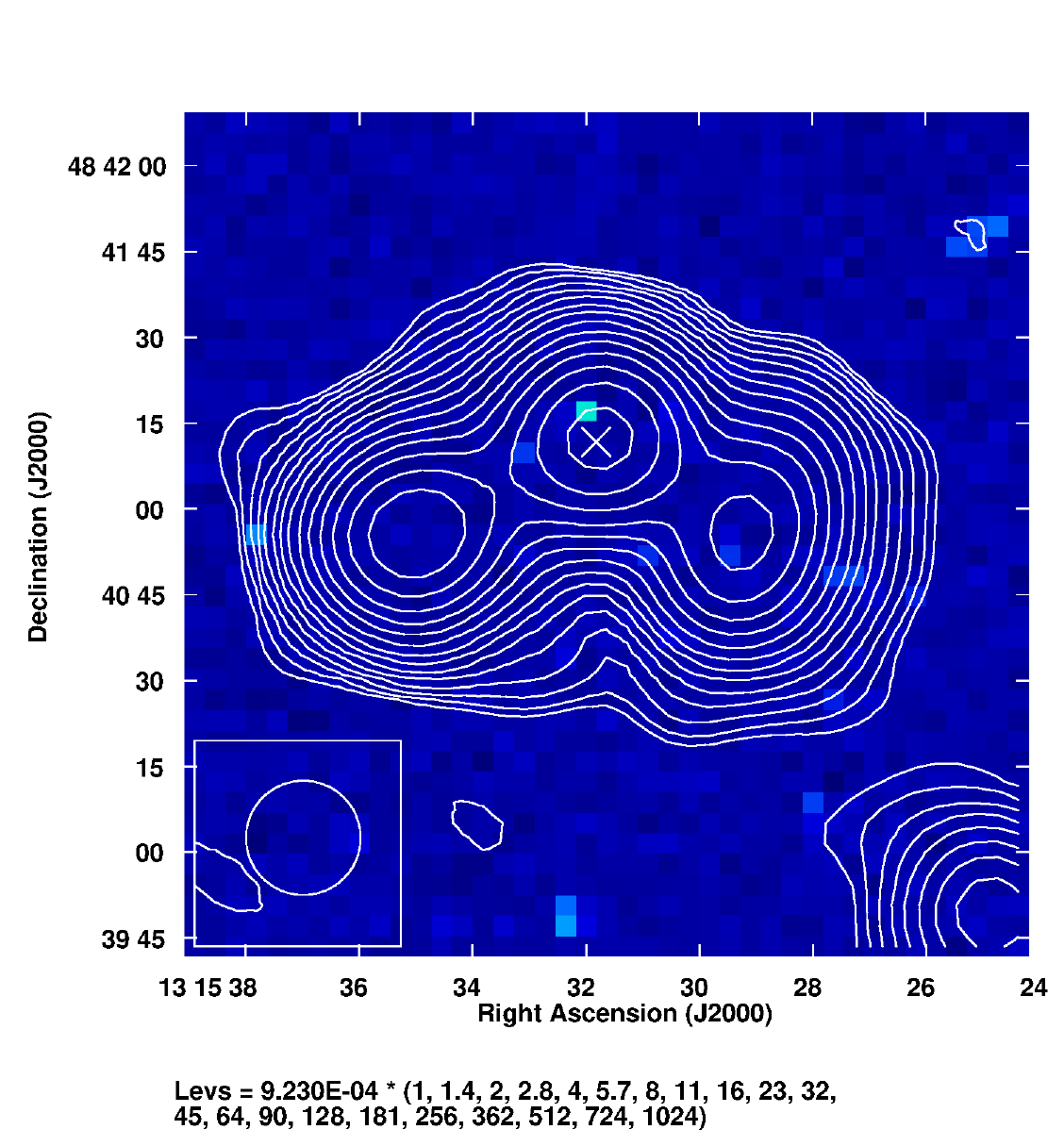}
}
}
\vbox{
\centerline{
\includegraphics[height=5.6cm,width=5.6cm]{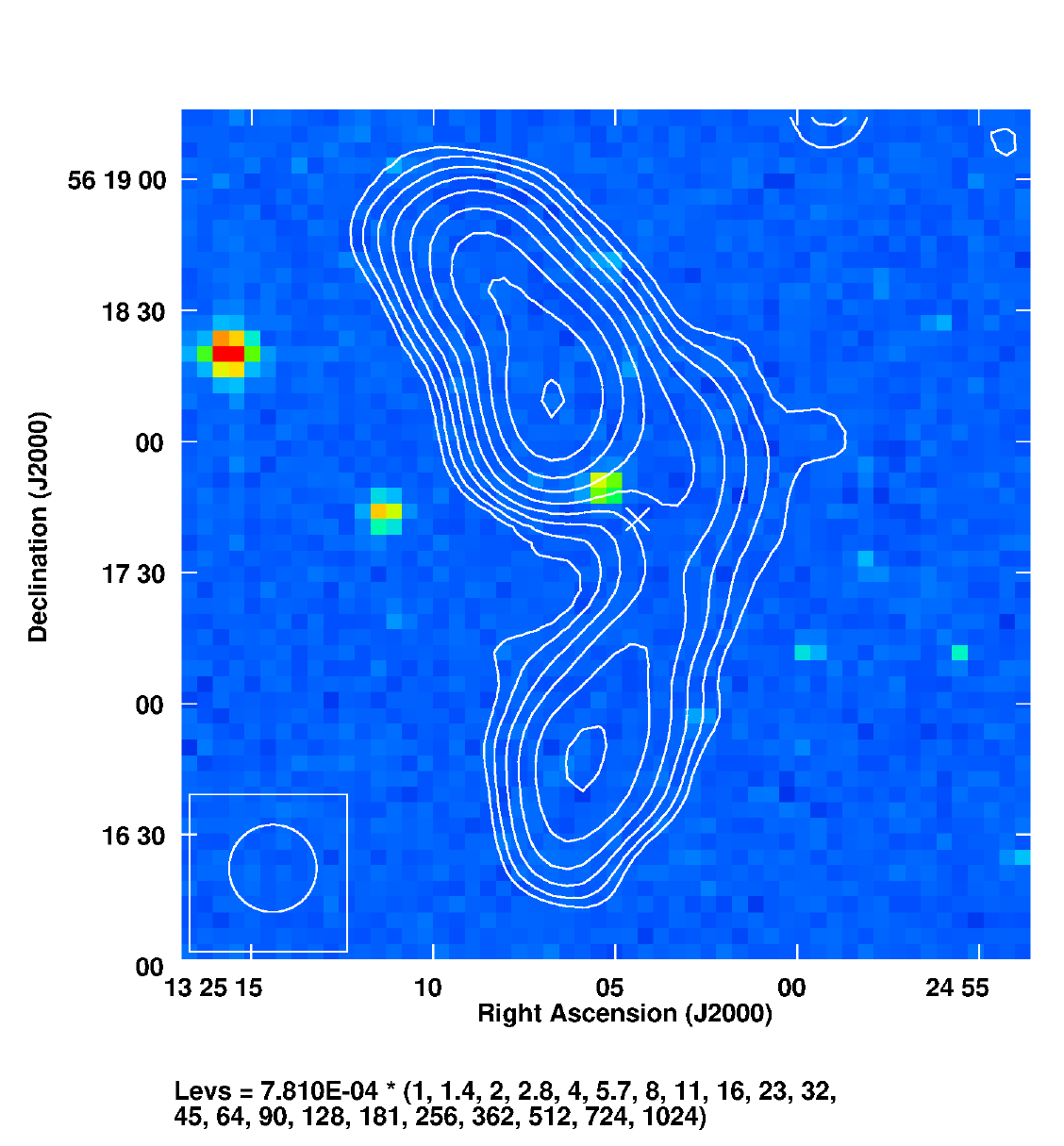}
\includegraphics[height=5.6cm,width=5.6cm]{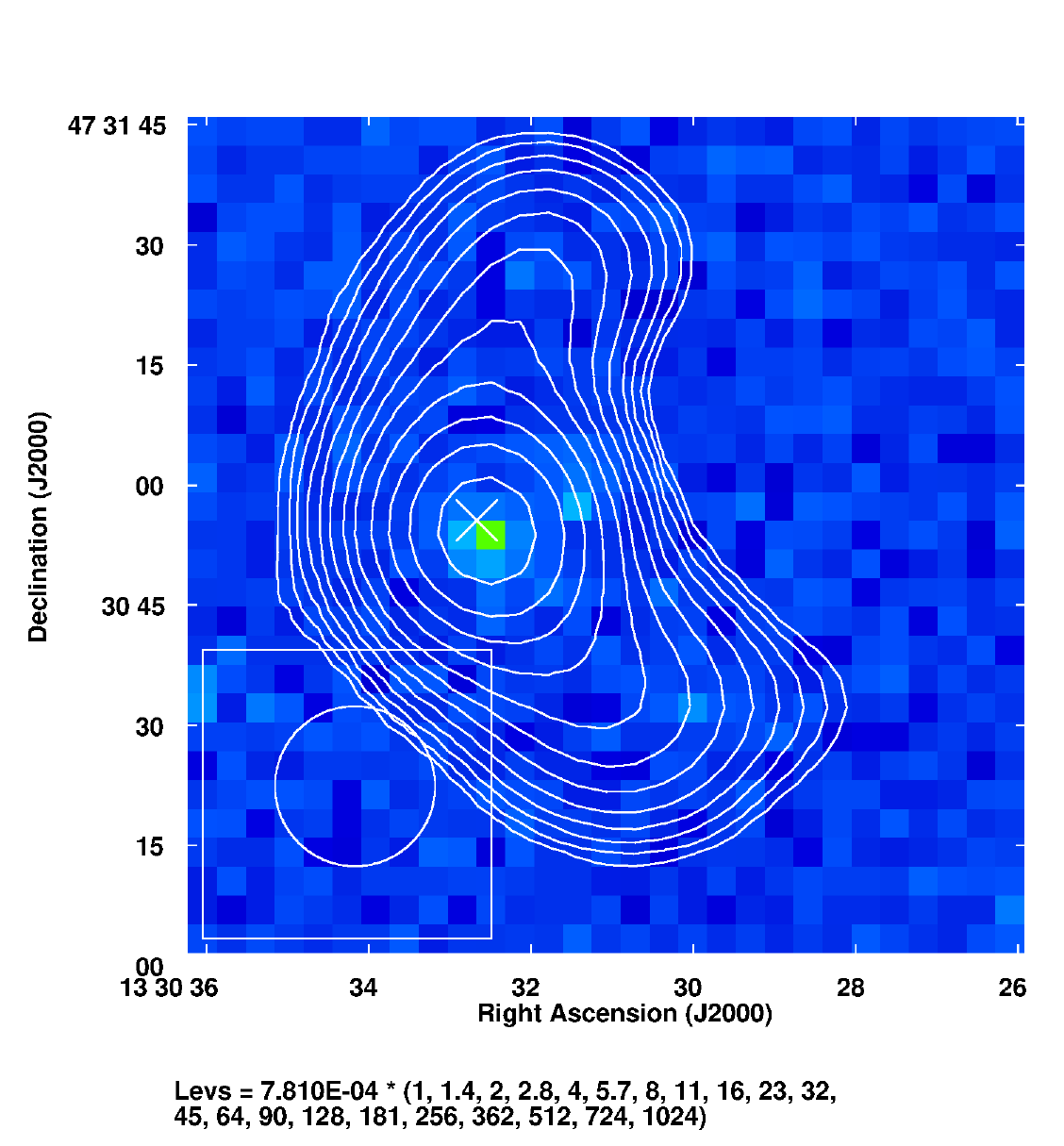}
\includegraphics[height=5.6cm,width=5.6cm]{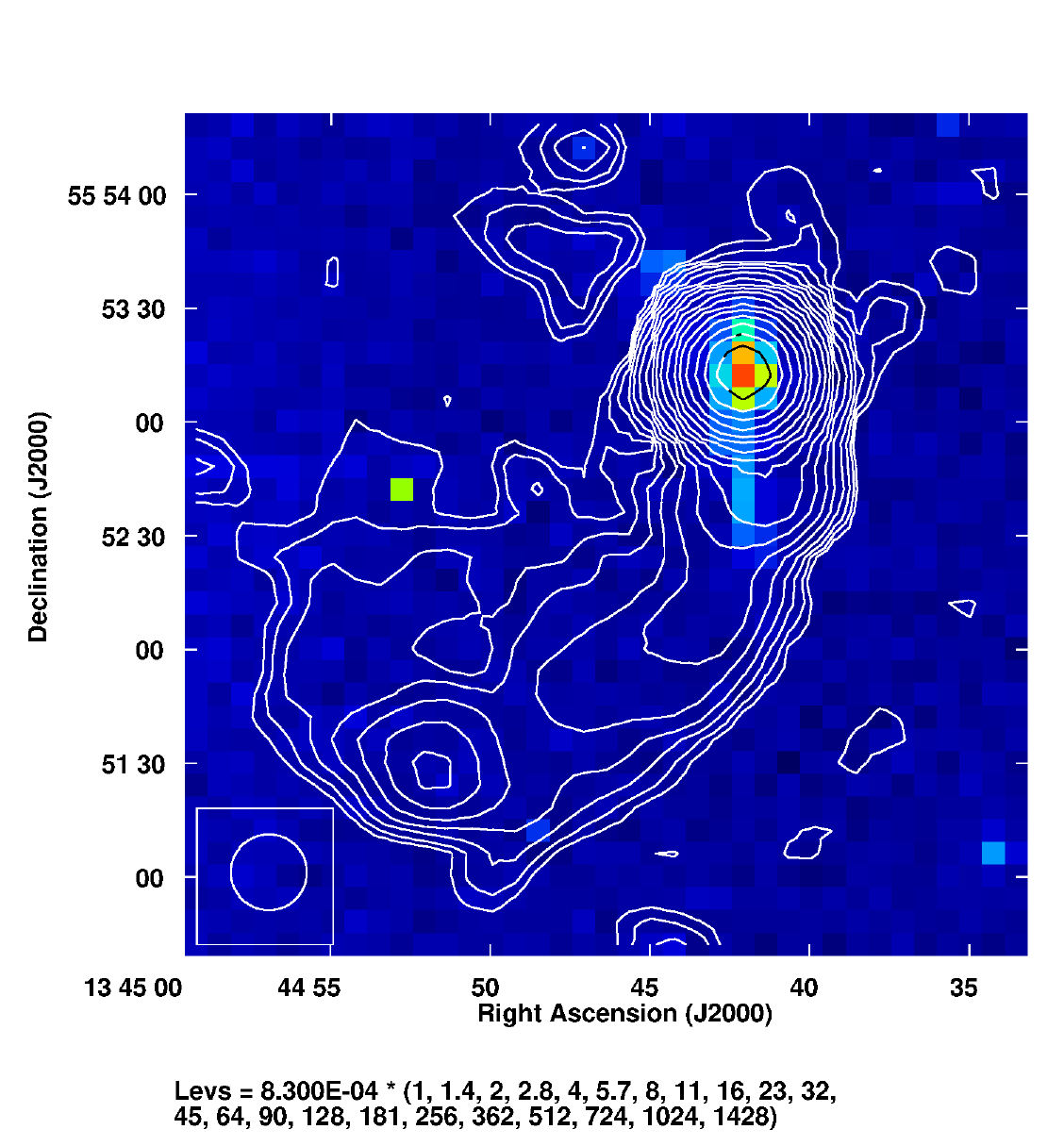}
}
}
\vbox{
\centerline{
\includegraphics[height=5.6cm,width=5.6cm]{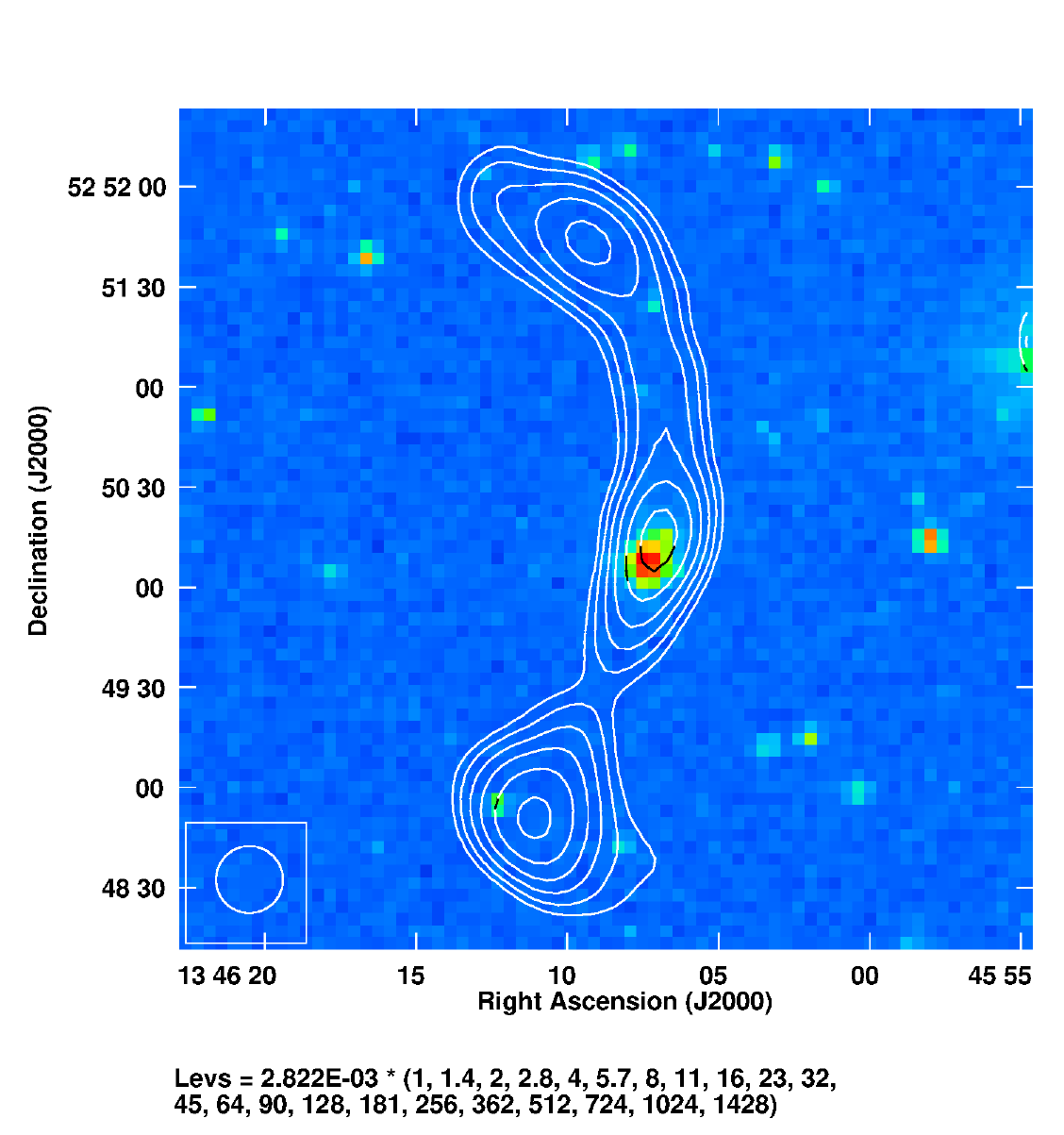}
\includegraphics[height=5.6cm,width=5.6cm]{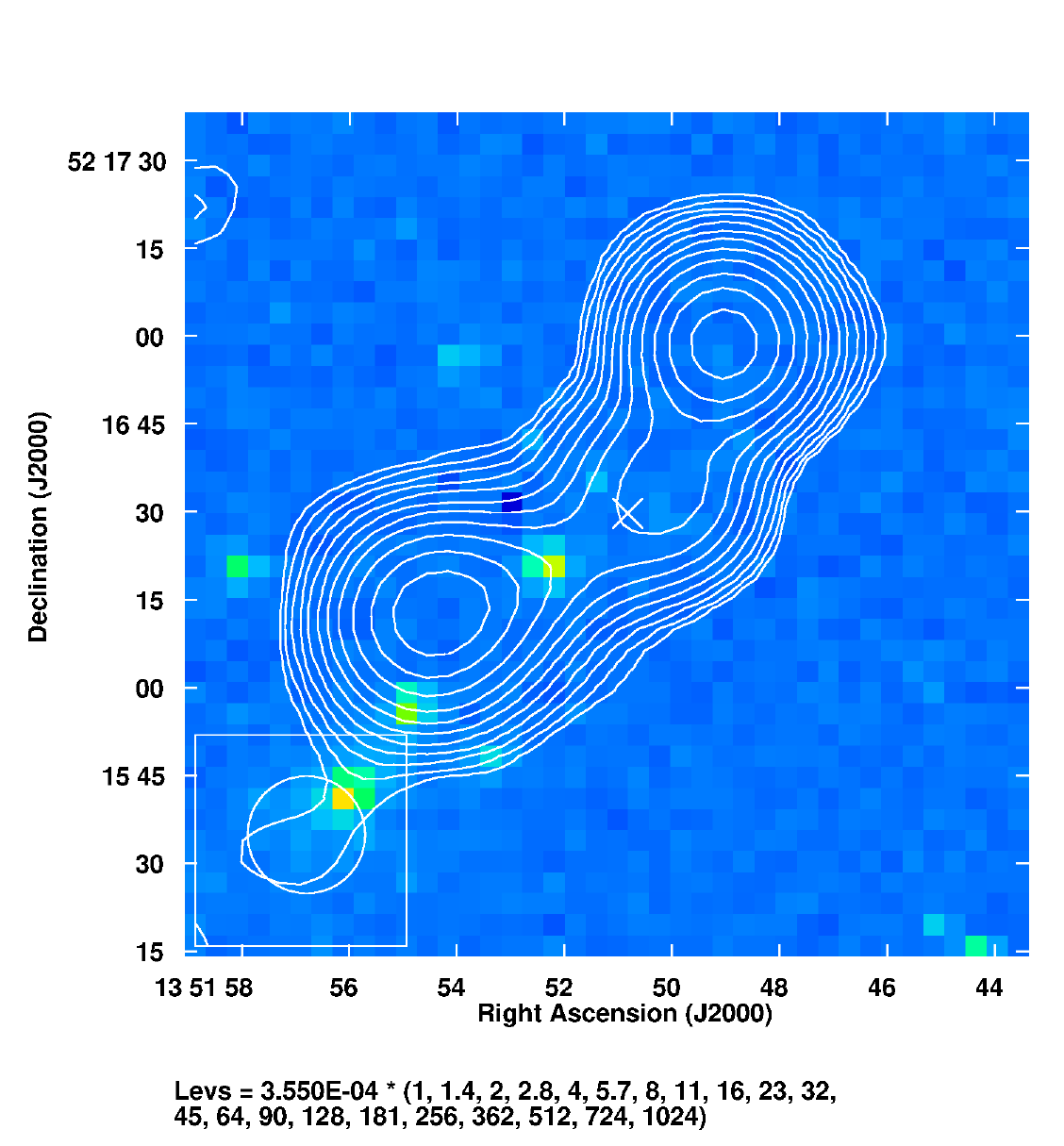}
\includegraphics[height=5.6cm,width=5.6cm]{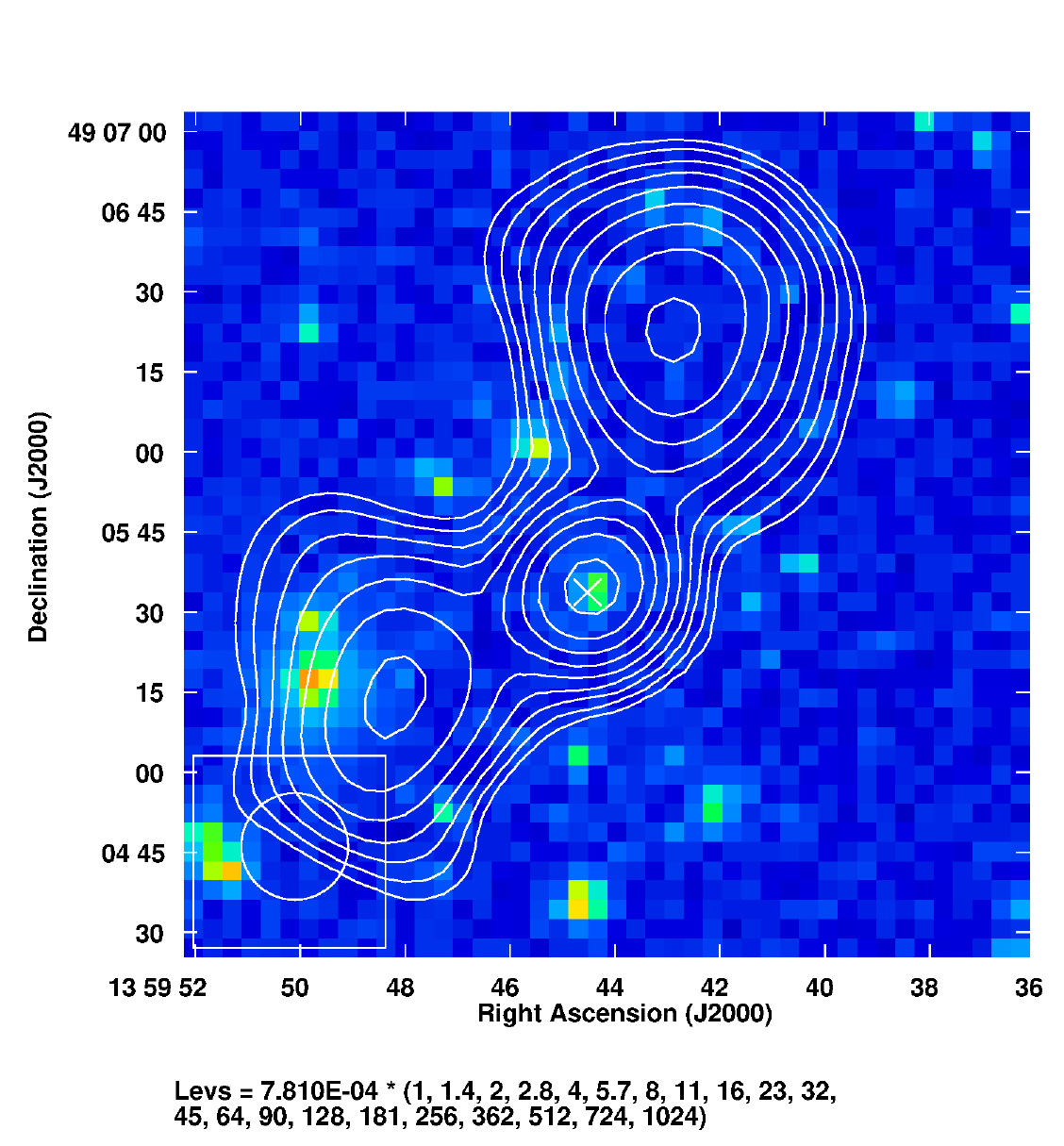}
}
}
\contcaption{LOFAR images of WAT radio galaxies (contours) overlaid on DSS2 red images (colour). Here white cross mark represent the optical counterpart for the sources, when available.}
\end{figure*}

\begin{figure*}
\vbox{
\centerline{
\includegraphics[height=5.6cm,width=5.6cm]{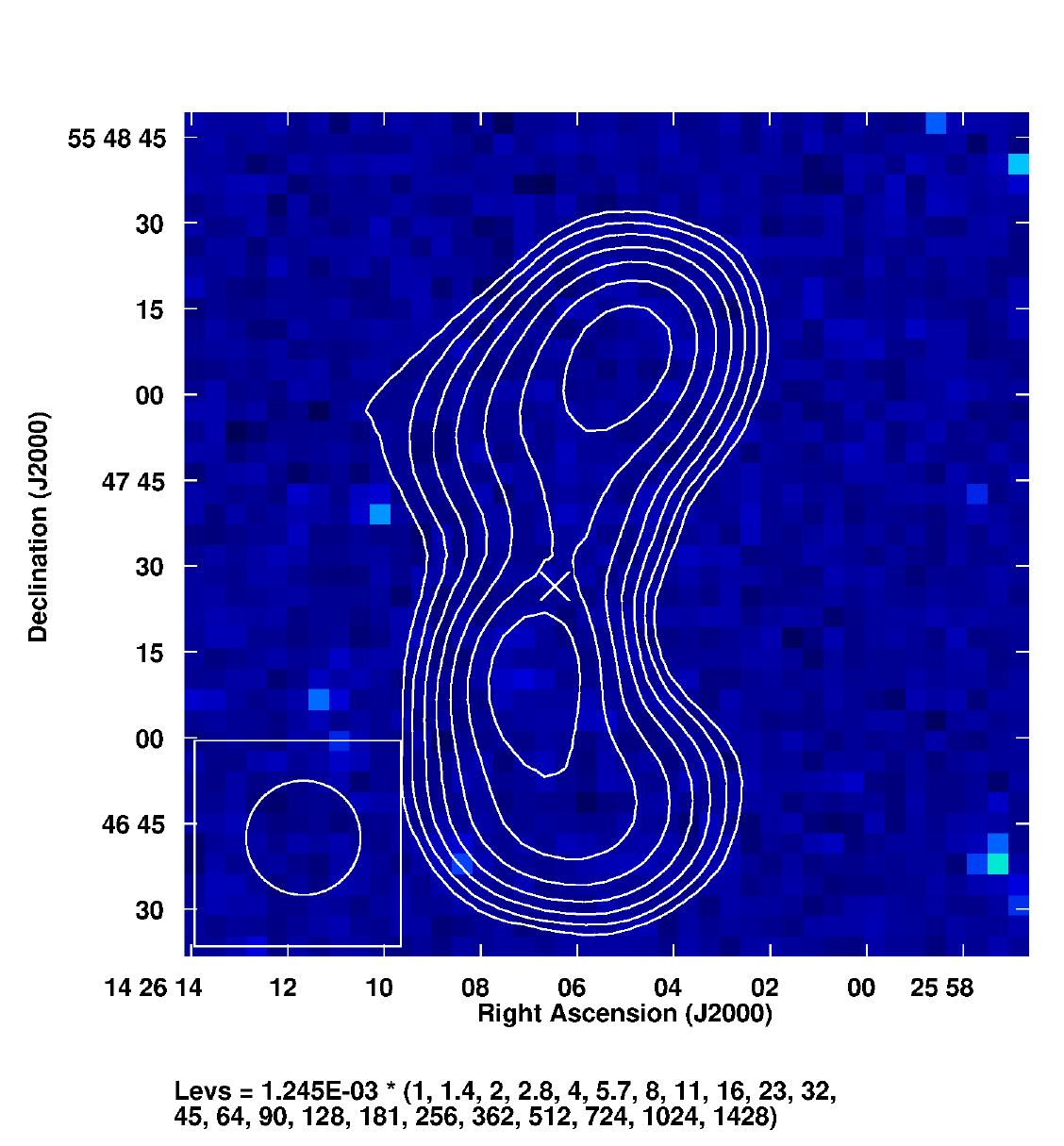}
\includegraphics[height=5.6cm,width=5.6cm]{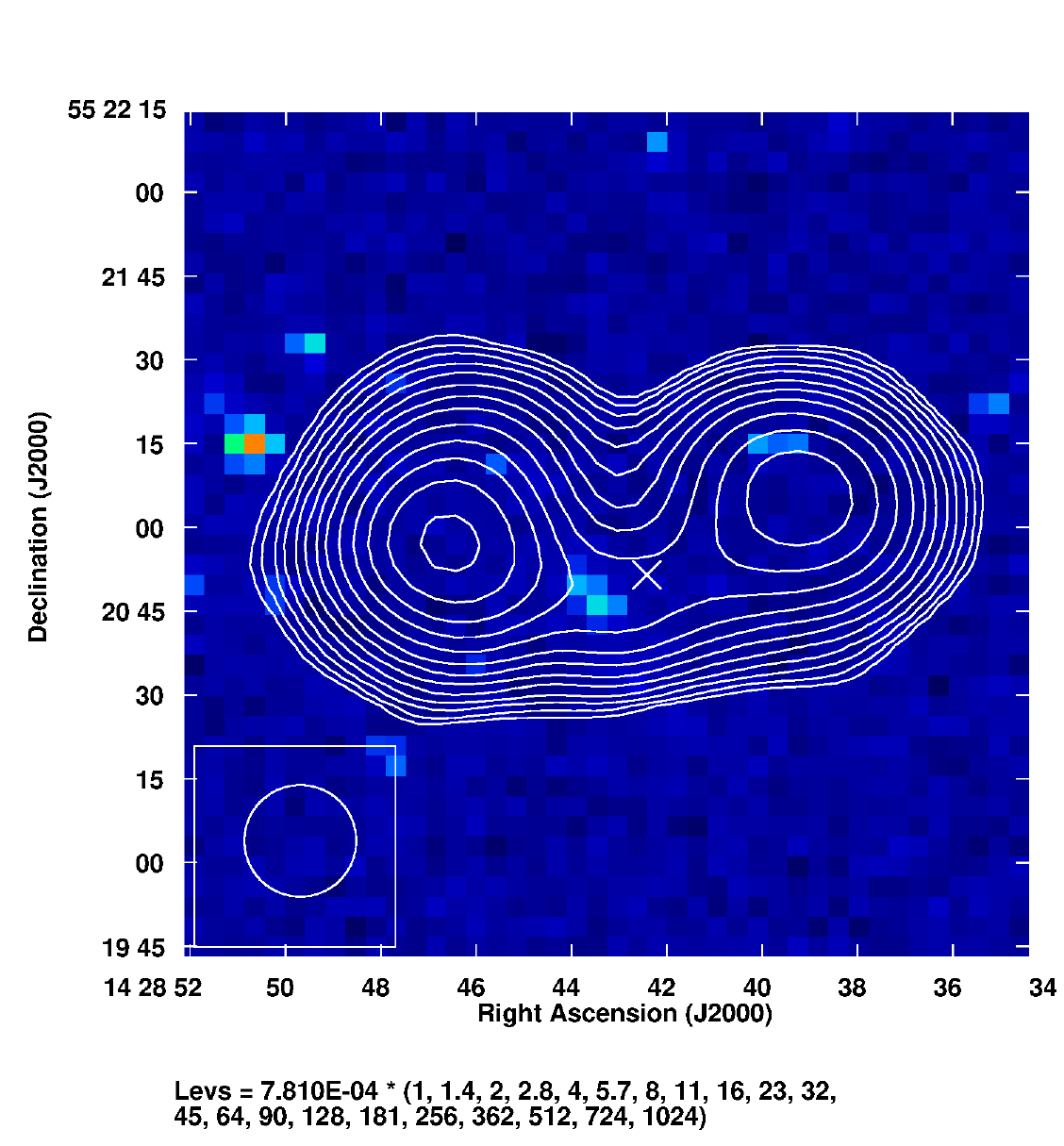}
\includegraphics[height=5.6cm,width=5.6cm]{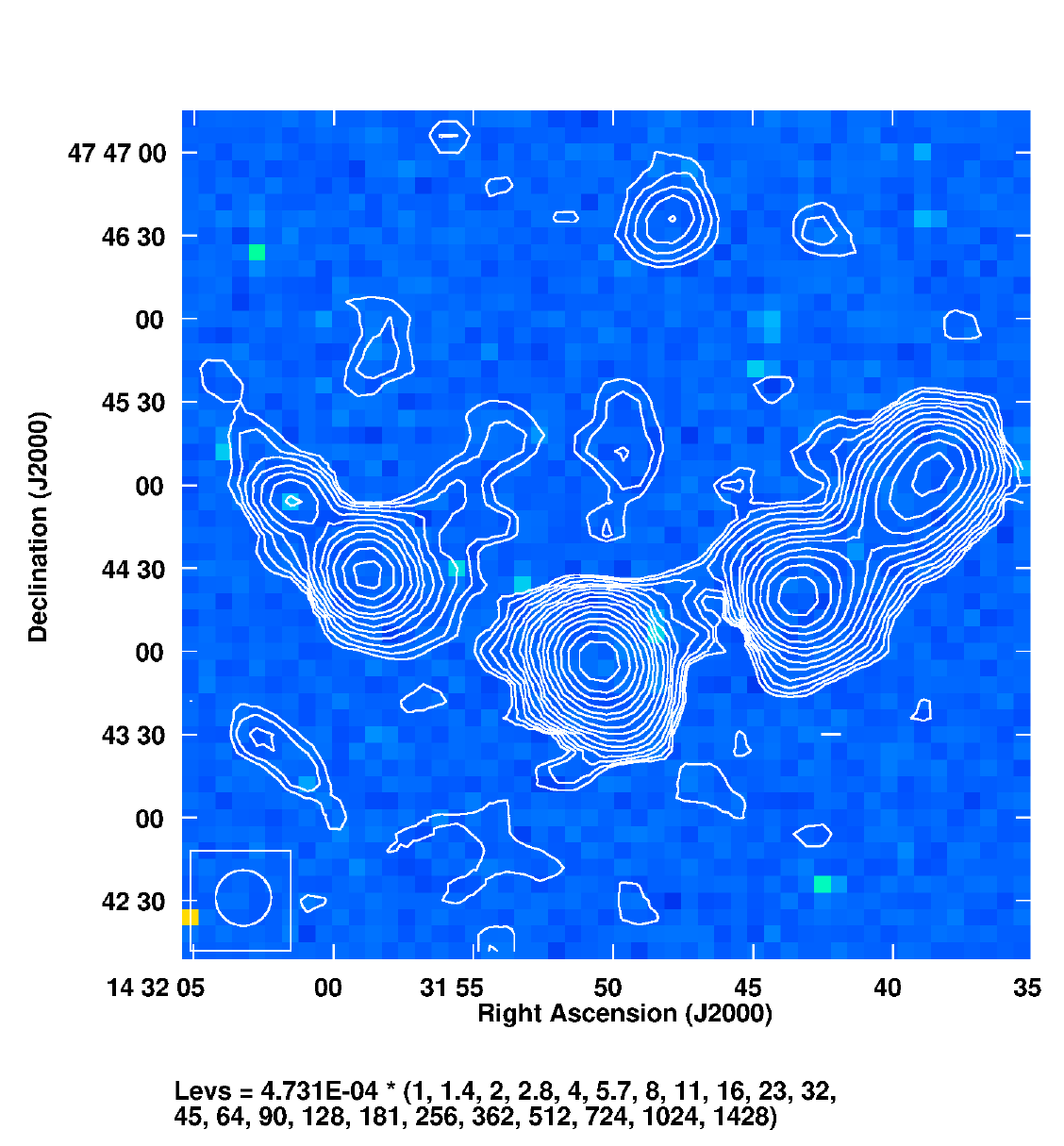}
}
}
\vbox{
\centerline{
\includegraphics[height=5.6cm,width=5.6cm]{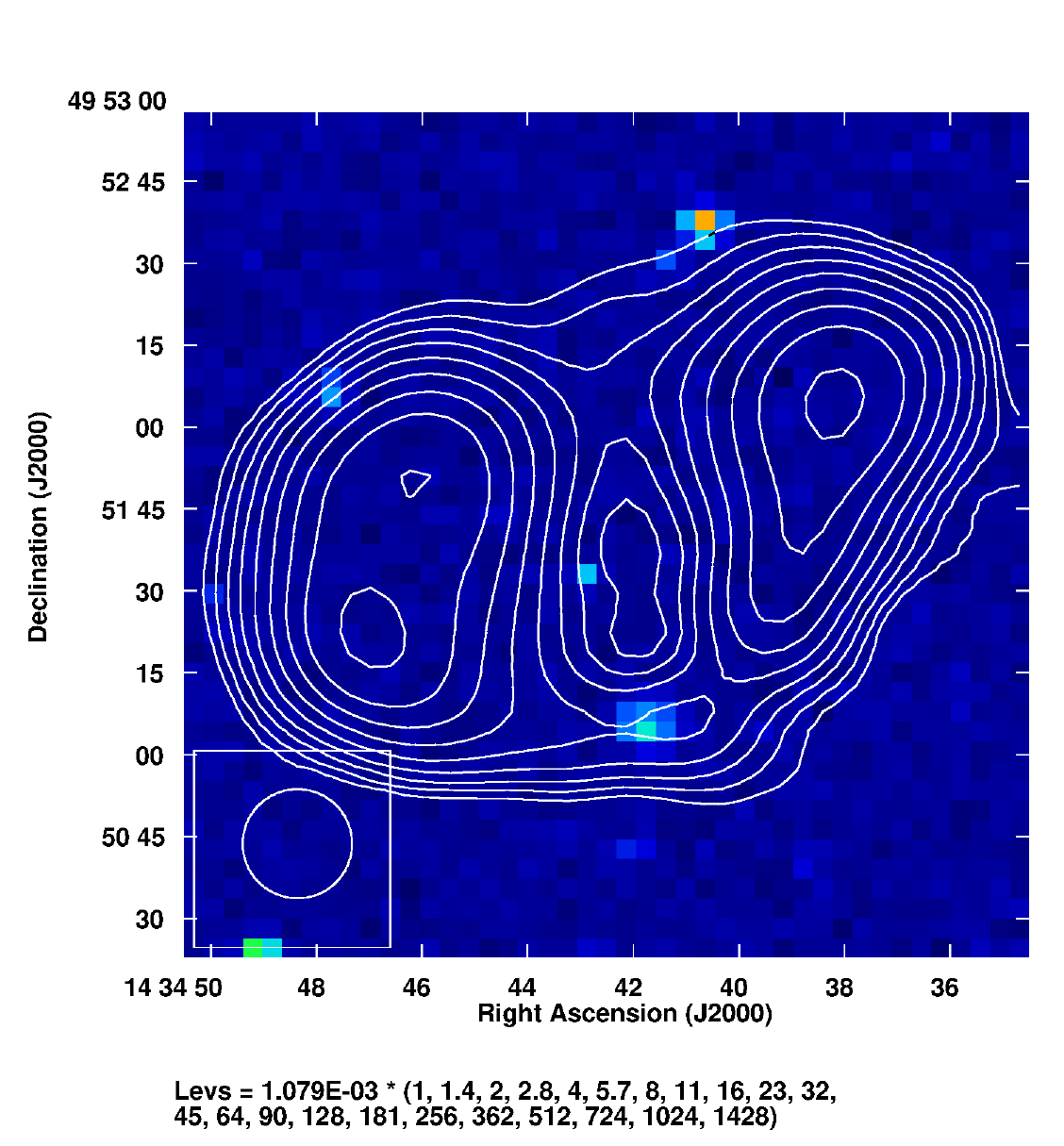}
\includegraphics[height=5.6cm,width=5.6cm]{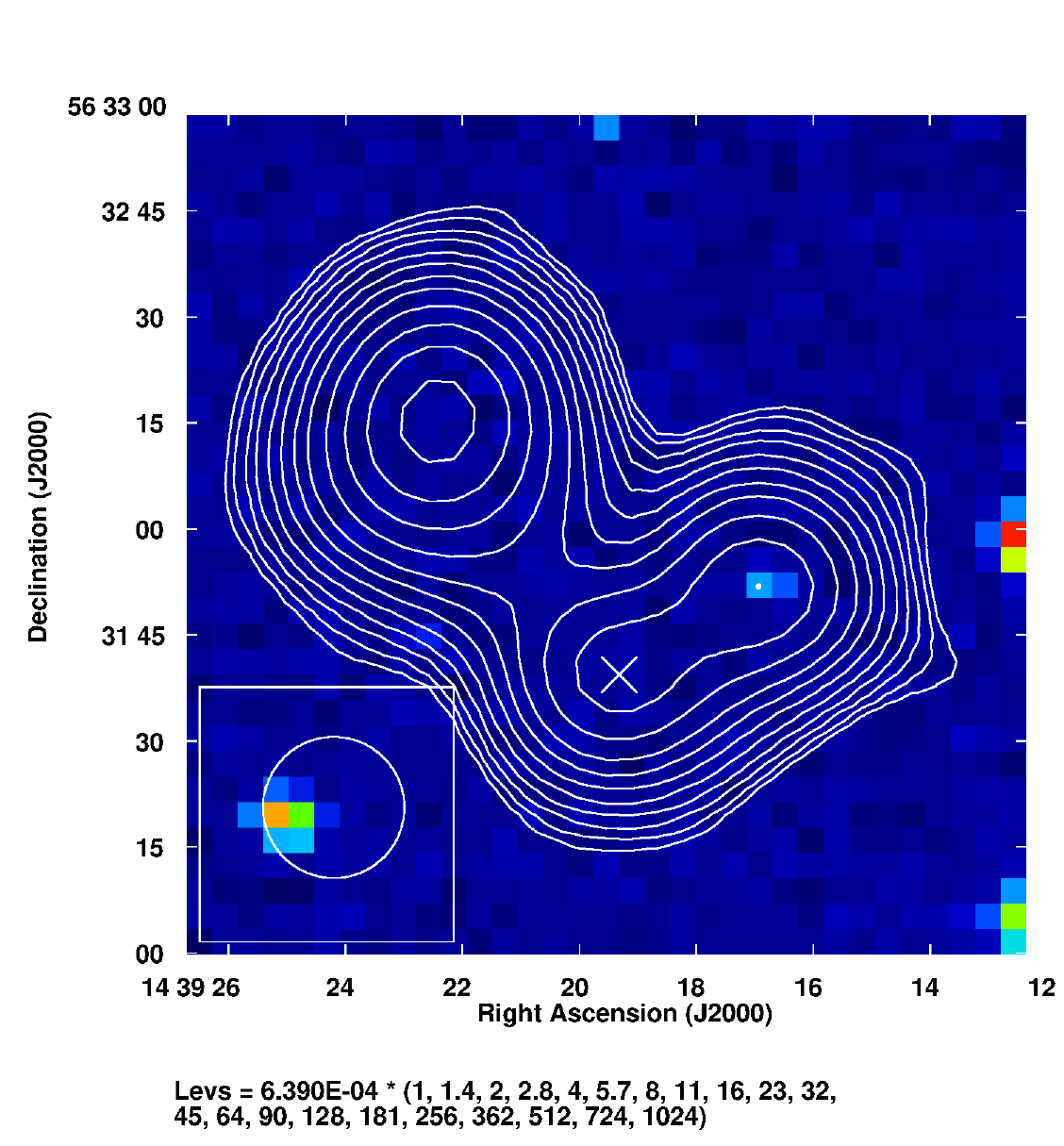}
\includegraphics[height=5.6cm,width=5.6cm]{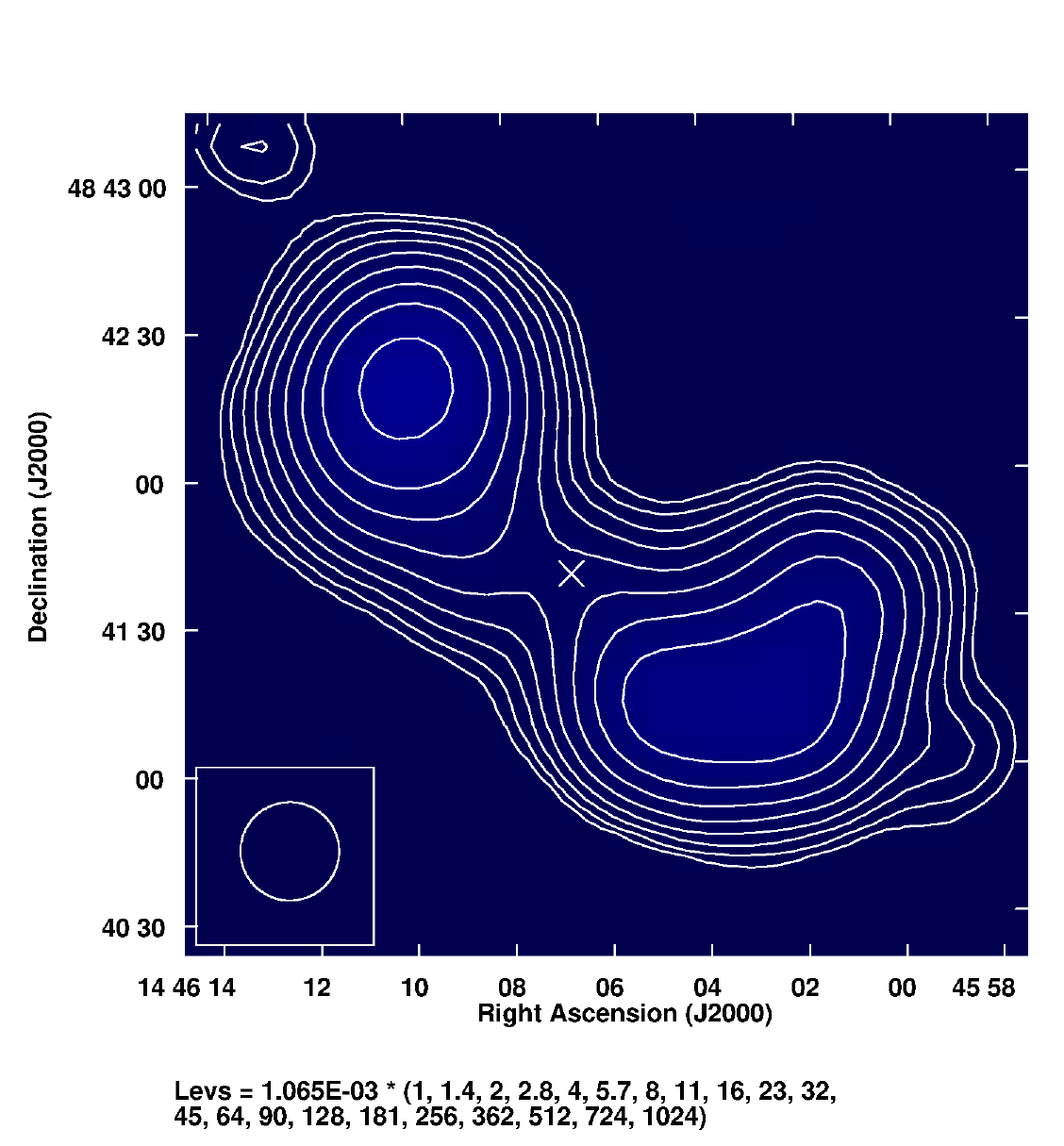}	
}
}
\vbox{
\centerline{
\includegraphics[height=5.6cm,width=5.6cm]{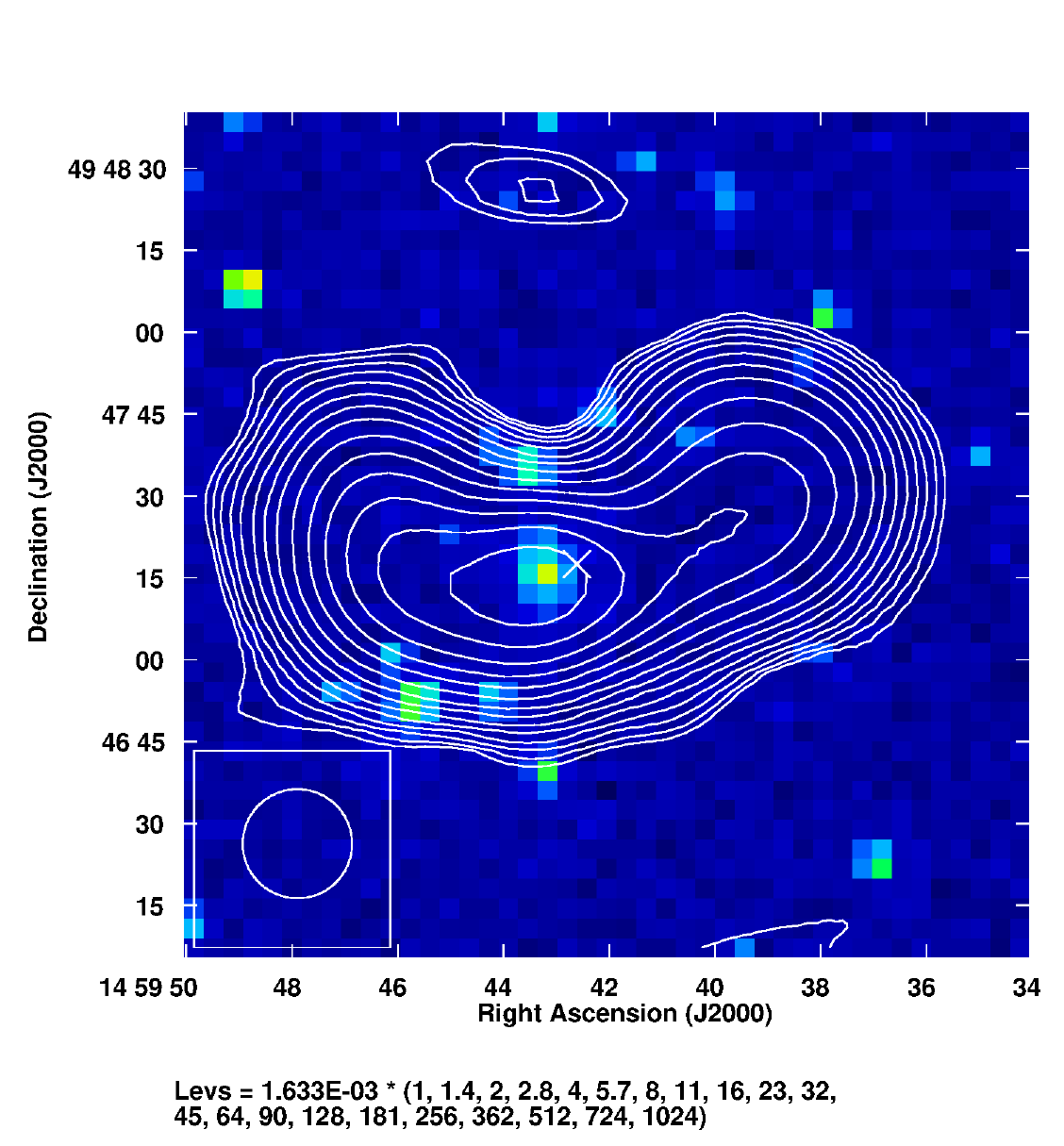}
\includegraphics[height=5.6cm,width=5.6cm]{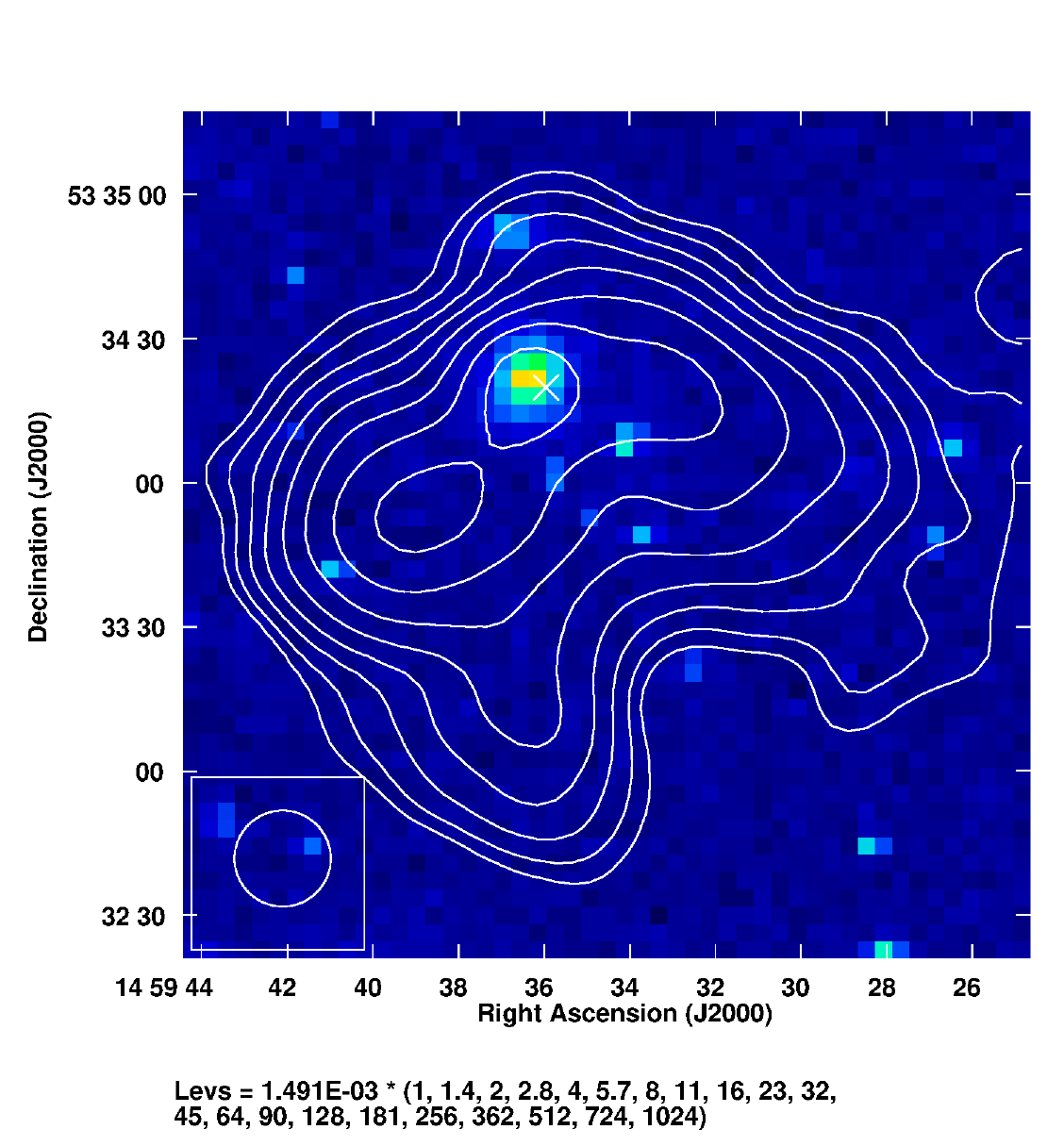}
\includegraphics[height=5.6cm,width=5.6cm]{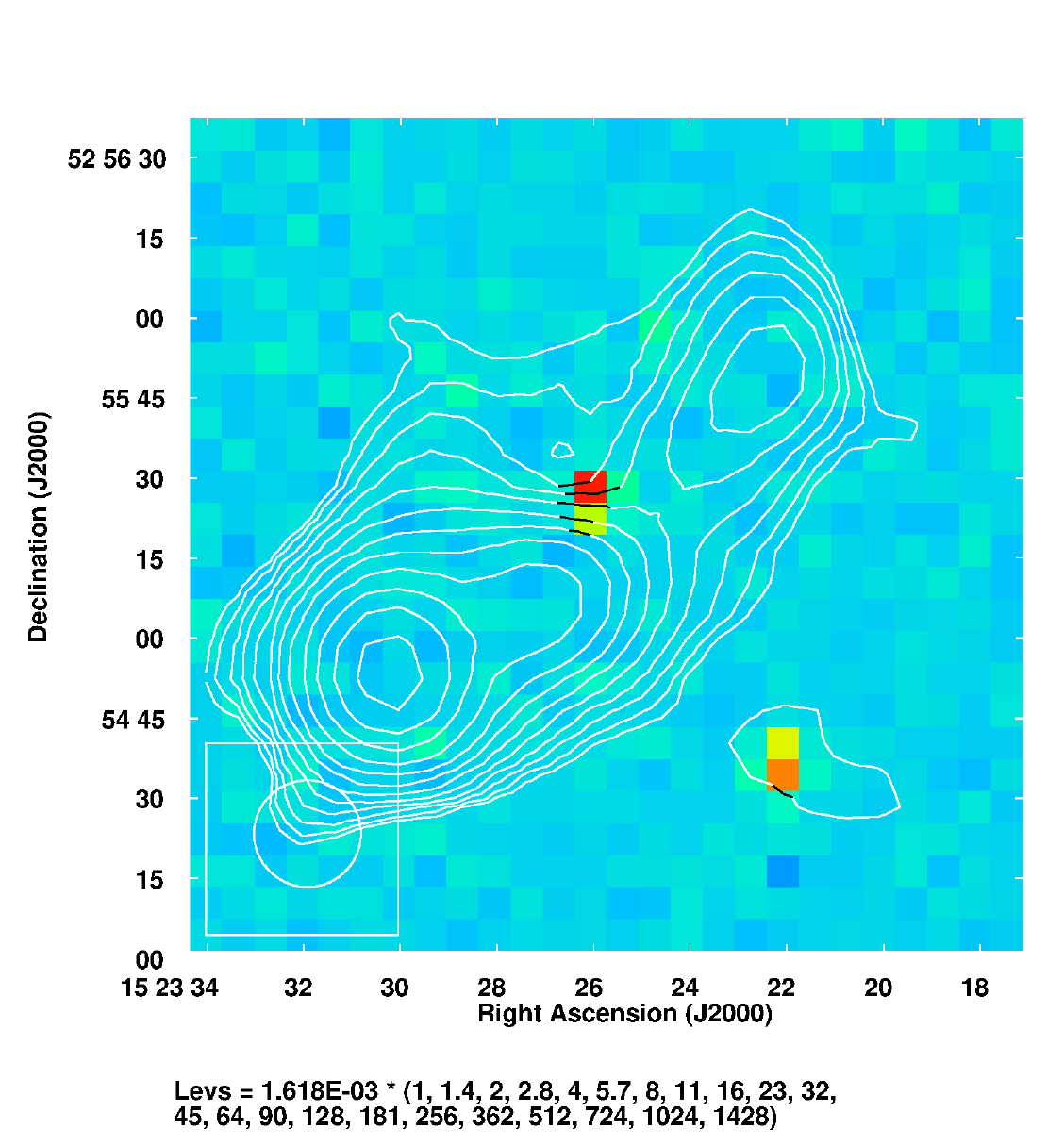}
}
}
\contcaption{LOFAR images of WAT radio galaxies (contours) overlaid on DSS2 red images (colour). Here white cross mark represent the optical counterpart for the sources, when available.}
\end{figure*}

\subsection{Radio luminosity}
\label{subsec:lum}
The radio luminosities ($L_{\textrm{rad}}$) of WAT and NAT sources are measured (when the value of $z$ is available) using the standard formula \citep{Do09}
\begin{equation}	
    L_{\textrm{rad}}=4\pi{D_{L}}^{2}S(1+z)^{\alpha-1}
\end{equation}
where $D_{L}$ is luminosity distance to the source in metre (m), $\alpha$ is the spectral index ($S \propto \nu^{-\alpha}$), $z$ is the redshift of the radio galaxy, and $S$ is the flux density (W m$^{-2}$ Hz$^{-1}$) at a given frequency. In column 9 of Table \ref{tab:-Head-tail-W} and \ref{tab:-Head-tail-N}, the luminosities of sources are shown. The range of $\log L_{\textrm{rad}}$ in the present sample is 24.47 to 26.35.
The most luminous NAT source in the present sample is J1207+4805 with a $\log L_{\textrm{rad}} = 26.31$ and the most luminous WAT source is J1049+4619 with a $\log L_{\textrm{rad}} =26.35$. The least luminous NAT source in the present sample is J1506+5311 with a $\log L_{\textrm{rad}} = 25.21$ and the least luminous WAT source is J1115+4834 with a $\log L_{\textrm{rad}} = 24.47$.

\subsection{Possible radio relics }
\label{sec:radio relics}
Radio relics are synchrotron-derived diffuse, elongated radio emitters that are frequently observed in galaxy clusters without an apparent host galaxy. It is believed that radio relics are traces of shock wave generation during the merger of galaxy clusters \citep{Fe21}. These objects may either have no AGN activity or activity that is so weak that outflowing jets can no longer be sustained or maybe at the final stage of radio source evolution \citep{Ta15}.

Radio relics have been observed in clusters like Coma, Abell 2255, and Abell 2256 with sizes $\geq$ 1 Mpc. It is found that in some of the cases (MACS J1752.0+4440 \citep{va12, Bo12}, PSZ1 G108.18–11.53 \citep{De15}, and CIZA J2242.8+5301 \citep{va10}), two radio relics were observed in the same cluster, usually running parallel to the distribution of ICM. The cluster Abell 3667, Abell 2345, Abell 1240, as well as ZwCl 0008.8+5215, and ZwCl 2341.1+0000, comprise two very luminous, virtually similar relics in a distance of larger than 5 Mpc \citep{Wi74, Co87, Ve98, Mu11, Ta15, Hu15}.

We reported four sources (J1147+5548, J1346+5250, J1347+5022, and J1510+5146),  with HT-like morphology (2 WATs and 2 NATs) which have no optical counterparts but are associated with known clusters (WHL J114703.7+554715, MaxBCG J206.50836+52.82084, WHL J134708.2+502222, and WHL J151003.4+514612, respectively). These extended bend sources could be radio relics.

\section{Discussion}
\label{sec:discuss} 
 
\subsection{Radio properties of HT radio galaxies}
\label{subsec:Rad-prop}
Radio galaxies can be distinguished by their radio properties such as spectral index, radio luminosity, angular size, and linear size with respect to their measured spectroscopic redshifts. Among 55 HT radio galaxies presented in the current paper, redshifts are found for 26 HT radio galaxies (48 per cent) from the SDSS catalogue, with 5 NATs and 21 WATs. All 26 HT radio galaxies have spectroscopic redshifts. The majority of detected sources have redshift $z$ $\leq0.5$. Two NATs (J1132+5459 and J1147+4805) have redshifts $z>0.5$. J1147+4805 is the furthest NAT with a redshift of $z = 0.70$. Only one WAT source, J1113+4952, has redshift $z>0.5$ ($z= 0.61$, the furthest source in the present paper). The nearest WAT source in our sample is J1115+4834 with $z= 0.07$. Previously, in the HT catalogue of \citet{Mis19}, sources had redshift $z\leq0.15$. Two sixty-five (204 WATs and 61 NATs) HT radio galaxies were presented by \citet{Bh21}, also had a redshift in the range of 0.01--0.68 in which only four sources had a redshift $z\geq0.5$. This can be concluded that the majority of HT radio galaxies are found in low redshift regions.

\begin{figure}
	\centering
\includegraphics[width=8.5cm,angle=0]{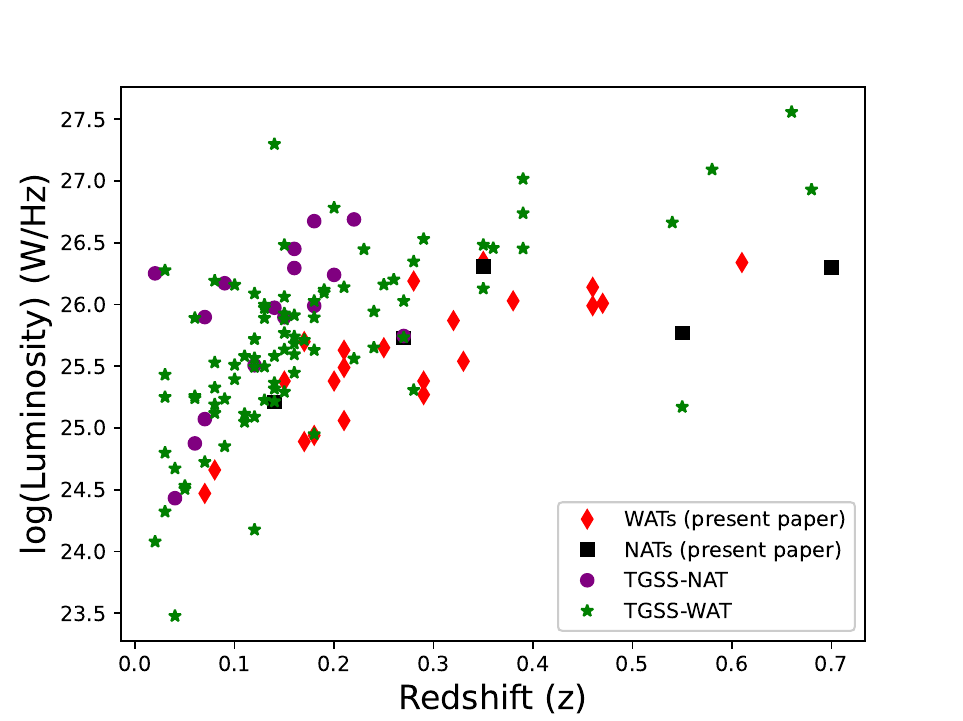}
\caption{Plot showing the $\log L$ distribution of WAT and NAT radio galaxies with the respected redshift presented in the current paper and Head-Tailed radio galaxies reported by \citet{Bh21}. Red diamond color points and black square color points represented WATs and NATs presented in the current paper while purple circle color points and green star color points represented TGSS-NATs and TGSS-WATs.}
\label{fig:z-logL}
\end{figure}

\begin{figure*}
	\centering
\includegraphics[width=8.5cm,angle=0,origin=c]{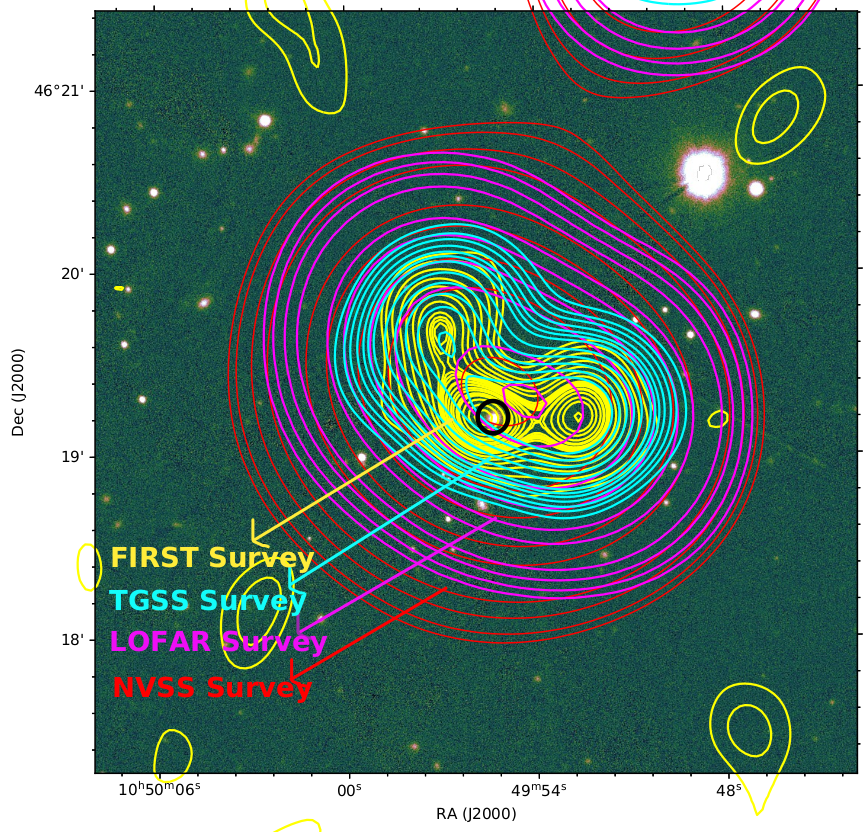}
\includegraphics[width=8.5cm,angle=0,origin=c]{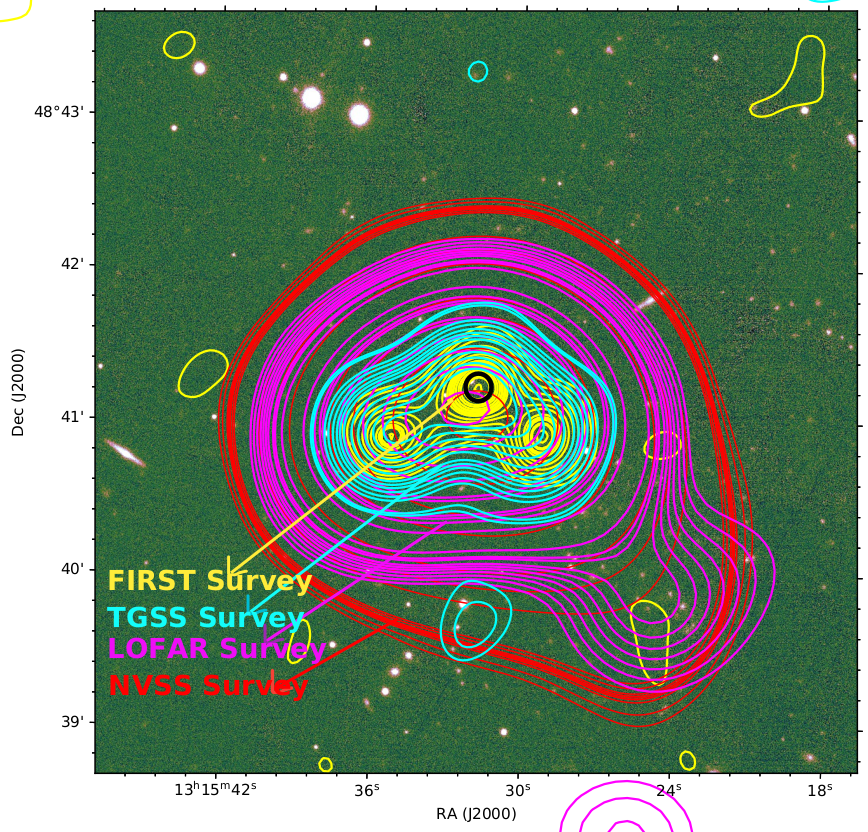}
\caption{Example of two WAT sources: Overlaid image of J1049+4619 and J1315+4841 with FIRST (yellow color), TGSS (cyan color), LOFAR (magenta color), and NVSS (red color) surveys. The background is in optical wavelength, taken from Pan-STARRS (Panoramic Survey Telescope and Rapid Response System). Black color circle represent the optical host galaxy core.}
\label{fig:1049}
\end{figure*}

The spectral indices for the presented sources in the current paper are listed in the tables \ref{tab:-Head-tail-W} and \ref{tab:-Head-tail-N}.

The average spectral index for WATs in the current paper is measured to be 0.78 with a median of 0.75, and the average spectral index for NATs is 0.56 with a median of 0.68. Using a large number (204 WATs and 61 NATs) of HT radio galaxies, \citet{Bh21} found the average spectral index for WATs was 0.77 with a median of 0.73 and for NATs, the average spectral index was 0.81 with a median of 0.73. The spectral index for normal radio galaxies typically lies in the range of 0.7--0.8 \citep{Oo88, Gr97, Ka98, Is10, Ma16}. The measured average spectral index for forty-five WAT sources in the present article and 265 sources reported by \citet{Bh21}, are close to this range. It can be concluded that in terms of spectral index, HT radio galaxies are similar to those of normal radio galaxies \citep{Mis19, Bh21}. Because of the small number of NATs reported in the current paper (which is not statistically significant), the average spectral index for NATs did not fall within this range.

Radio luminosity is one of the most important properties used to distinguish radio galaxies. In Fig. \ref{fig:z-logL}, the distribution of radio luminosity of NATs and WATs is shown for the sources presented in the current paper. 265 sources (204 WATs and 61 NATs) presented by \citet{Bh21} with known red-shift ($z$) are also added in the figure. It is found that for WAT sources, the mean $\log L_{\textrm{rad}}$ and median $\log L_{\textrm{rad}}$ are calculated as 25.60 and 25.62, respectively. Similarly, for NAT sources, the mean $\log L_{\textrm{rad}}$ and median $\log L_{\textrm{rad}}$ are 25.54 and 25.73, respectively. The mean $\log L_{\textrm{rad}}$ and median $\log L_{\textrm{rad}}$ for the HT radio galaxies presented by \citet{Mis19} were 25.40 and 25.35, respectively. The mean and median $\log L_{\textrm{rad}}$ for WAT sources in \citet{Bh21} were 25.62 and 25.63, respectively, and for NAT sources, the mean and median were 25.82 and 25.83, respectively. This result suggests that the average luminosity of WAT and NAT sources in the present paper is more or less the same as that reported by \citet{Bh21} (with the help of TGSS ADR 1 at 150 MHz) and \citet{Mis19}.

In Fig. \ref{fig:1049}, we showed the examples of two WAT (J1049+4619 and J1315+4841) sources. We made overlay images of these sources using LOTSS DR1), FIRST, TGSS, and NVSS surveys. Emissions from these sources for the LOTSS survey are visible with a magenta colour contour, and those of FIRST, TGSS, and NVSS are visible with yellow, cyan, and red colour contours, respectively. The optical background of these images is taken from the Pan-STARRS. The bending nature is not evident with NVSS contours due to the poor resolution of the survey. A black circle in the centre of the image represents the core of the host galaxy.

\subsection{Association with cluster}
\label{subsec:cluster}
A galaxy cluster is a structure composed of hundreds or thousands of gravitationally bound galaxies with masses ranging from 10$^{14}$--10$^{15}$ $M_{\odot}$ with enormous size \citep{We12}. We used the NASA Extragalactic Database (NED) to conduct a 3D search of galaxy clusters for each HT radio galaxy with a search radius of 3 arcmins. From the search results, we chose the clusters that are closest to the HT radio galaxies.  Using this method of searching, we found 30 HT radio galaxies (25 are WATs and 5 are NATs) that match known galaxy clusters, as listed in Table \ref{tab:-Head-tail2W} and \ref{tab:-Head-tail2N}. Twenty-three clusters are from the Wen+Han+Liu (WHL) cluster catalog \citep{We15}, which consists of 158,103 clusters (X-ray frequency). Two clusters are from the Gaussian Mixture Brightest Cluster Galaxy (GMBCG) \citep{Ha10} catalog, which consist of 55,880 clusters (optical), 1 from the NSC \citep{Sm12} cluster catalog (optical), 1 from MSPM \citep{Sm12} cluster catalog (optical), 2 from SDSS--C4--DR3 \citep{Vo07} cluster catalog (optical), and 1 from MaxBCG \citep{Ko07} cluster catalog (optical).

We computed linear distances between HT radio galaxies and their associated known cluster centres. The majority of cluster centres are located within a few kpc and the shortest and longest distances from the host galaxies are 3.84 kpc and 936.80 kpc, respectively, with a median of 41.27 kpc.

\subsection{Cluster mass}
\label{subsec:clus-mass}
The density of the inter-cluster medium is correlated with cluster mass, with heavier clusters having a denser inter-cluster medium. The velocity dispersion of HT radio galaxies is also correlated with the mass of the cluster. It is believed that HT radio galaxies show rapid movement in high mass clusters \citep{Ma10}. We calculated cluster mass $M_{500}$ (10$^{14} M_{\odot})$ for detected associated clusters of the HT sources described in the present paper with the help of cluster richness $R_{L}$ using,  
\begin{equation}
    \log M_{500}=(1.08\pm0.02)\log R_{L}-(1.37\pm0.02)
\end{equation}
The cluster richness ($R_{L}$) and cluster radius from optical luminosity ($r_{500}$) of 22 WHL clusters are listed in Table \ref{tab:-Head-tail2W} and \ref{tab:-Head-tail2N} (taken from \citep{We15}).

The mass of the ICM gas that is contained inside a radius of $r_{500}$ is $M {500}$, where $r_{500}$ is the radius containing the mean over-density of $\Delta c= 500\rho_{cr}$, where $\rho_{cr}(z)= 3H(z)^2/8 \pi G$ is the critical mean density of the universe as defined in terms of the Hubble function $H(z)$. The relative velocities of galaxies are consistent with the temperature of the ICM, which indicates that both galaxies and gas are nearly in equilibrium within a common gravitational potential well. The depth of the potential well cannot be explained by the mass of galaxies and hot gases, suggesting that the majority of the mass in clusters is made up of dark matter. 

We reported that 30 out of 55 HT radio galaxies are associated with groups of relatively low mass clusters (in the order of 10$^{14} M_\odot$). In our catalogue, the most massive cluster associated with a WAT is WHL J145943.2+494716 (with mass ($M_{500}$) 5.09$\times10^{14} M_\odot$) of which host galaxy is J1459+5334. The most massive cluster associated with a NAT is WHL J145943.2+494716 (with mass ($M_{500}$) 4.47$\times10^{14} M_\odot$) of which host galaxy is J1505+4706. 

In the classification of HT radio galaxies, projection effects may have an impact. Due to projection effects, the large jets (up to a few degrees) of HT radio galaxies that would appear straight in the line of sight are excluded from our sample. The number of observed HT radio galaxies at any given time is also highly dependent on the AGN duty cycle in the formation of bent jets.

The redshifts of the associated cluster are mentioned in Table \ref{tab:-Head-tail2W} and \ref{tab:-Head-tail2N}. These redshifts are taken from NED and SDSS.

\section{Summary and Conclusions}
\label{subsec:summ}
In this paper, we report the identification of 55 HT radio galaxies (45 WATs and 10 NATs). The current catalogue of 55 newly detected HT radio galaxies will help to increase the number of known HT radio galaxies. Among all 55 HT radio galaxies, SDSS optical counterparts are found for 44 HT radio galaxies. Four sources are reported as possible radio relic sources. Seven sources (four WATs (J1115+4729, J1344+5552, J1431+4743, and J1523+5255) and three NATs (J1329+5246, J1431+5518, and J1325+5544)) have no optical counterparts, and they are also not associated with any galaxy clusters. More in-depth optical follow-up observations are needed to search for possible optical counterparts of these sources. We found that thirty out of the fifty-five HT are linked to known galaxy clusters. 

The detection of 55 HT radio galaxies in a 424 square degree area suggests that these types of sources are not uncommon, particularly in cluster-rich areas. Future multi-wavelength follow-up observations with high resolution and high sensitivity surveys, especially using SKA, will be able to detect a large number of HT radio galaxies and will help to understand the nature of these sources in more detail.

\section*{Acknowledgments}
The LOFAR Two-metre Sky Servey (LoTSS) is being conducted with the high-band antennas (HBA) of LOFAR which is funded by the EU, European Fund for Regional Development and the Northern Netherlands Provinces (SNN), and EZ/KOMPAS and managed by ASTRON Netherlands Institute for Radio Astronomy. This research has made use of the NASA/IPAC Extra-galactic Database (NED) which is operated by the Jet Propulsion Laboratory, California Institute of Technology, under contract with the National Aeronautics and Space Administration.



\begin{theunbibliography}{}
\vspace{-1.5em}

\bibitem[{Aghanim {\em et al.}}(2020)]{Ag20}
Aghanim N. {\em et al.}, 2020, A\&A, 641, 67

\bibitem[{Ag\"ueros {\em et al.}}(2005)]{Ag05}
Ag\"ueros M.A. {\em et al.}, 2005, AJ, 130, 1022


\bibitem[{Baan \& McKee }(1985)]{Ba85}
Baan W.A., McKee M.R., 1985, A\&A, 143, 136

\bibitem[{Becker {\em et al.}}(1995)]{Be95}
Becker R.H. {\em et al.}, 1995, ApJ, 450, 559

\bibitem[{Benn {\em et al.}}(1982)]{Be82}
Benn C. R., Grueff G., Vigotti M., Wall J.V., 1982, MNRAS, 200, 747

\bibitem[{Benn {\em et al.}}(1988)]{Be88}
Benn C. R., Grueff G., Vigotti M., Wall J.V., 1988, MNRAS, 230, 1

\bibitem[{Bhukta, Pal \& Mondal}(2022)]{Bh21}
Bhukta N., Pal S., Mondal S., 2022, MNRAS, 516, 372

\bibitem[{Blanton {\em et al.}}(2000)]{Bl00}
Blanton E.L., Gregg M.D., Helfand D.J., Becker R.H., White R.L., 2000, ApJ, 531, 118

\bibitem[{Blanton {\em et al.}}(2003)]{Bl03}
Blanton E.L., Gregg M.D., Helfand D.J., Becker R.H., White R.L., 2003, AJ, 125, 1635

\bibitem[{Bolton, Gardner \& Mackey}(1964)]{Bo64}
Bolton J.G., Gardner F.F., Mackey M.B., 1964, AuJPh, 17, 340

\bibitem[{Bonafede {\em et al.}}(2012)]{Bo12}
Bonafede A. {\em et al.}, 2012, MNRAS, 426, 40

\bibitem[{Burns}(1998)]{Bu98}
Burns J. O., 1998, Sci, 280, 400


\bibitem[{Chung {\em et al.}}(2011)]{Ch11}
Chung S.M., Eisenhardt P.R., Gonzalez A.H., Stanford S.A., Brodwin M., Stern D., Jarrett T., 2011, ApJ, 743, 34

\bibitem[{Cohen {\em et al.}}(2007)]{Co07}
Cohen A.S., Lane W.M., Cotton W.D., Kassim N.E., Lazio T.J.W., Perley R.A. Condon J.J., Erickson W.C., 2007, AJ, 134, 1245


\bibitem[{Cordey}(1987)]{Co87}
Cordey R. A., 1987, Monthly Notices of the RAS, 227, 695

\bibitem[{Condon {\em et al.}}(1998)]{Co98}
Condon J.J., Cotton W.D., Greisen E.W., Yin Q.F., Perley R.A., Taylor G.B., Broderick J.J., 1998, AJ, 115, 1693

\bibitem[{de Gasperin {\em et al.}}(2015)]{De15}
de Gasperin F. {\em et al.}, 2015, MNRAS, 453, 3483

\bibitem[{Dehghan {\em et al.}}(2014)]{De14}
Dehghan S., Johnston-Hollitt M., Franzen T.M.O., Norris R.P., Miller N.A., 2014, AJ, 148, 75


\bibitem[{Donoso, Best \& Kauffmann}(2009)]{Do09}
Donoso E., Best P. N., Kauffmann G., 2009, MNRAS, 392, 617

\bibitem[{Douglas {\em et al.}}(1996)]{Do96}
Douglas J.N., Bash F.N., Bozyan F.A., Torrence G.W., Wolfe C., 1996, AJ, 111, 1945

\bibitem[{Dreyer}(1888)]{Dr88}
Dreyer J.L.E., 1888, MmRAS, 49, 1


\bibitem[{Eilek {\em et al.}}(1984)]{Ei84}
Eilek J.A., Burns J.O., O'Dea C.P., Owen F.N., 1984, ApJ, 278, 37


\bibitem[{Eilek \& Owen}(2002)]{Ei02}
Eilek J.A., Owen F.N., 2002, ApJ, 567, 202



\bibitem[{Fernánde {\em et al.}}(2021)]{Fe21}
Fernández P.D. et al., 2021, MNRAS, 507, 2714

\bibitem[{Ficarra, Grueff \& Tomassetti}(1985)]{Fi85}
Ficarra A., Grueff G., Tomassetti G., 1985, A\&AS, 59, 255


\bibitem[{Gregory \& Condon}(1991)]{Gr91}
Gregory P.C., Condon J.J., 1991, ApJS, 75, 1011

\bibitem[{Gruppioni {\em et al.}}(1997)]{Gr97}
Gruppioni C., Zamorani G., de Ruiter H.R., Parma P., Mignoli M., Lari C., 1997, MNRAS, 286, 470

\bibitem[{Gunn \& Gott}(1972)]{Gu72}
Gunn J. E., Gott J. R. I., 1972, ApJ, 176, 1

\bibitem[{Gunn {\em et al.}}(2006)]{Gu06}
Gunn J.E. {\em et al.}, 2006, AJ, 131, 2332


\bibitem[{Hales, Baldwin \& Warner}(1988)]{Ha88}
Hales S.E.G., Baldwin J.E., Warner P.J., 1988, MNRAS, 234, 919

\bibitem[{Hales {\em et al.}}(1990)]{Ha90}
Hales S.E.G., Masson C.R., Warner P.J., Baldwin J.E., 1990, MNRAS, 246, 256

\bibitem[{Hales {\em et al.}}(1991)]{Ha91}
Hales S.E.G., Mayer C.J., Warner P.J., Baldwin J.E., 1991, MNRAS, 251, 46

\bibitem[{Hales {\em et al.}}(1993a)]{Ha93a}
Hales S.E.G., Baldwin J.E., Warner P.J., 1993a, MNRAS, 263, 25

\bibitem[{Hales {\em et al.}}(1993b)]{Ha93b}
Hales S.E.G., Masson C.R., Warner P.J., Baldwin J.E., Green D.A., 1993b, MNRAS, 262, 1057

\bibitem[{Hao {\em et al.}}(2010)]{Ha10}
Hao J. {\em et al.}, 2010, ApJS, 191, 254

\bibitem[{Hardcastle, Sakelliou \& Worrall}(2005)]{Ha05}
Hardcastle M.J., Sakelliou I., Worrall D.M., 2005, MNRAS, 359, 1007

\bibitem[{Heald {\em et al.}}(2015)]{He15}
Heald G. H. {\em et al.} 2015, A\&A, 582, A123


\bibitem[{Hurley-Walker {\em et al.}}(2015)]{Hu15}
Hurley-Walker N., {\em et al.}, 2015, MNRAS, 447, 2468

\bibitem[{Intema {\em et al.}}(2017)]{In17}
Intema H. T., Jagannathan P., Mooley K.P., Frail D.A., 2017, A\&A, 598, A78

\bibitem[{Ishwara-Chandra {\em et al.}}(2010)]{Is10}
Ishwara-Chandra C.H., Sirothia S.K., Wadadekar Y., Pal S., Windhorst R., 2010, MNRAS, 405, 436

\bibitem[{Kapahi {\em et al.}}(1998)]{Ka98}
Kapahi V.K., Athreya R.M., van Breugel W., McCarthy P.J., Subrahmanya C.R., 1998, ApJS 118, 275

\bibitem[{Kenderdine, Ryle \& Pooley}(1966)]{Ke66}
Kenderdine S., Ryle M.S., Pooley G.G., 1966, MNRAS, 134, 189

\bibitem[{Klamer, Subrahmanyan \& Hunstead}(2004)]{Kl04}
Klamer I., Subrahmanyan R., Hunstead R. W., 2004, MNRAS, 351, 101

\bibitem[{Kollgaard {\em et al.}}(1994)]{Ko94}
Kollgaard R.I., Brinkmann W., Chester M.M., Feigelson E.D., Hertz P., Reich P., Wielebinski R., 1994, ApJS, 93, 145

\bibitem[{Koester {\em et al.}}(2007)]{Ko07}
Koester B. P., McKay T. A. Annis J.,  Wechsler R. H {\em et al.}, 2007, ApJ, 660, 239


\bibitem[{Lane {\em et al.}}(2014)]{La14}
Lane W.M., Cotton W.D., van Velzen S., Clarke T.E., Kassim N.E., Helmboldt J.F., Lazio T.J.W., Cohen A.S., 2014, MNRAS, 440, 327




\bibitem[{Mao {\em et al.}}(2010)]{Ma10}
Mao M. Y., Sharp R., Saikia D. J., Norris R. P., Hollitt M. J., Middelberg E., Lovell J. E. J., 2010, MNRAS, 406, 2578

\bibitem[{Mahony {\em et al.}}(2016)]{Ma16}
Mahony E.K. {\em et al.}, 2016, MNRAS, 463, 2997

\bibitem[{McGilchrist {\em et al.}}(1990)]{Mc90}
McGilchrist M.M., Baldwin J.E., Riley J.M., Titterington D.J. Waldram E.M., Warner P.J., 1990, MNRAS, 246, 110


\bibitem[{Miley {\em et al.}}(1972)]{Mi72}
Miley G. K., Perola G. C., van der Kruit P. C., van der Laan H., 1972, Nat, 237, 269

\bibitem[{Mingo {\em et al.}}(2019)]{Mi19}
Mingo {\em et al.}, 2019, MNRAS, 488, 2701

\bibitem[{Missaglia {\em et al.}}(2019)]{Mis19}
Missaglia V., Massaro F., Capetti A., Paolillo M., Kraft R. P., Baldi R. D., Paggi A., 2019, A\&A, 626, A8

\bibitem[{Murgia {\em et al.}}(2011)]{Mu11}
Murgia M., {\em et al.}, 2011, a\&a, 526, A148


\bibitem[{O'Brien {\em et al.}}(2018)]{O'B18}
O'Brien A.N. {\em et al.}, 2018 MNRAS, 481, 5247


\bibitem[{O'Dea \& Owen}(1986)]{O'D86}
O'Dea C.P., Owen F.N., 1986, ApJ, 301, 841



\bibitem[{Oort, Steemers \& Windhorst}(1988)]{Oo88}
Oort M.J.A., Steemers W.J.G., Windhorst R. A., 1988, A\&AS, 73, 103

\bibitem[{Owen \& Rudnick}(1976)]{Ow76}
Owen F.N., Rudnick L., 1976, ApJ, 205, L1

\bibitem[{Pal \& Kumari}(2021)]{Pa21}
Pal S., Kumari S., 2021, astro-ph:2104.00410

\bibitem[{Patra {\em et al.}}(2019)]{Pa19}
Patra D., Pal S., Konar C., Chakrabarti S.K., 2019, Astrophys Space Sci, 364, 72

\bibitem[{Pearson}(1975)]{Pe75}
Pearson T.J., 1975, MNRAS, 171, 475

\bibitem[{Pearson \& Kus}(1978)]{Pe78}
Pearson T.J., Kus A.J., 1978, MNRAS, 182, 273


\bibitem[{Piffaretti {\em et al.}}(2011)]{Pi11}
Piffaretti R., Arnaud M., Pratt G. W., Pointecouteau E., Melin J. B., 2011, A\&A, 534, 109

\bibitem[{Pooley \& Kenderdine}(1968)]{Po68}
Pooley G.G., Kenderdine S., 1968, MNRAS, 139, 529

\bibitem [{Pooley}(1969)]{Po69}
Pooley G.G., 1969, MNRAS, 144, 101

\bibitem[{Proctor}(2011)]{Pr11}
Proctor D.D., 2011, ApJS, 194, 31

\bibitem[{Rebull {\em et al.}}(2011)]{Re11}
Rebull L.M. {\em et al.}, 2011, ApJS, 196, 4


\bibitem[{Rudnick \& Owen}(1976)]{Ru76}
Rudnick L., Owen F.N., 1976, AJ, 203, L107

\bibitem[{Rudnick \& Owen}(1977)]{Ru77}
Rudnick L., Owen F. N. 1977, AJ, 82, 1

\bibitem[{Ryle \& Windram}(1968)]{Ry68}
Ryle M., Windram M.D., 1968, MNRAS, 138, 1

\bibitem[{S\'anchez {\em et al.}}(2011)]{Sa11}
S\'anchez A.J., Aguerri J.A.L., Mu\~{n}oz-Tu\~{n}\'on C., Huertas-Company M., 2011, ApJ, 735, 125

\bibitem[{Sasmal {\em et al.}}(2022)]{Sa22}
Sasmal T.K., Bera S., Pal S., Mondal S., 2022, ApJS, 259, 9

\bibitem[{Schuch}(1981)]{Sc81}
Schuch N.J., 1981, MNRAS, 196, 695

\bibitem[{Skrutskie {\em et al.}}(2006)]{Sk06}
Skrutskie M.F. {\em et al.}, 2006, AJ, 131, 1163

\bibitem[{Shimwell {\em et al.}}(2019)]{Sh19}
Shimwell T.W. {\em et al.}, 2019, A\&A, 622, A1

\bibitem[{Smith {\em et al.}}(2012)]{Sm12}
Smith A.G. {\em et al.}, 2012, MNRAS, 422, 25

\bibitem[{Tamhane {\em et al.}}(2015)]{Ta15}
Tamhane P., Wadadekar Y., Basu A., Singh V., Ishwara-Chandra C. H., Beelen A., Sirothia
S., 2015, MNRAS, 453, 2438

\bibitem[{van Weeren {\em et al.}}(2010)]{va10}
van Weeren R. J., Röttgering H. J. A., Brüggen M., Hoeft M., 2010, Science, 330, 347

\bibitem[{van Weeren {\em et al.}}(2012)]{va12}
van Weeren, R. J. {\em et al.}, 2012, MNRAS, 425, L36 

\bibitem[{Venkatesan {\em et al.}}(1994)]{Ve94}
Venkatesan T. C. A., Batuski D. J., Hanisch R. J., Burns J. O., 1994, ApJ, 436, 67

\bibitem[{Vessey \& Green}(1998)]{Ve98}
Vessey S.J., Green D.A., 1998, MNRAS, 294, 607

\bibitem[{Von {\em et al.}}(2007)]{Vo07}
Von D.L.A., Best P.N., Kauffmann G., White S.D.M., 2007, MNRAS, 379, 867

\bibitem[{Waggett}(1977)]{Wa77}
Waggett P.C., 1977, MNRAS, 181, 547

\bibitem[{Waldram {\em et al.}}(1996)]{Wa96}
Waldram E.M., Yates J.A., Riley J.M., Warner P.J., 1996, MNRAS, 282, 779

\bibitem[{Wayth {\em et al.}}(2015)]{Wa15}
Wayth R.B. {\em et al.}, 2015, PASA, 32, e025

\bibitem[{Wen, Han \& Liu}(2012)]{We12}
Wen Z. L., Han J. L., Liu F. S., 2012, ApJS, 199, 34

\bibitem[{Wen \& Han}(2015)]{We15}
Wen Z.L., Han J.L., 2015, ApJ, 807, 178

\bibitem[{Willson}(1970)]{Wi70}
Willson M.A.G., 1970, MNRAS, 151, 1

\bibitem[{Willis, Strom \& Wilson}(1974)]{Wi74}
Willis A. G., Strom R. G., Wilson A. S., 1974, Nature, 250, 625

\bibitem[{Yu-Xing {\em et al.}}(2019)]{Yu19}
Yu-Xing Liu, Hai-Guang Xu, Dong-Chao Zheng, {\em et al.},
2019, RAA, 19, 127

\end{theunbibliography}
\end{document}